\documentclass{mn2e}

\usepackage[dvips]{graphicx}

\begin{document}

\title[Dark matter response to galaxy formation]{Dark matter response to galaxy formation}

\author[Tissera et al. ]{Patricia  B. Tissera$^{1,2}$\thanks{E-mail:patricia@iafe.uba.ar},
Simon D. M. White$^{3}$, Susana Pedrosa$^{1,2}$ and Cecilia Scannapieco$^{4}$\\
$^1$  Consejo Nacional de Investigaciones Cient\'{\i}ficas y T\'ecnicas, CONICET, Argentina.\\
$^2$ Instituto de Astronom\'{\i}a y F\'{\i}sica del Espacio, Casilla de Correos 67, Suc. 28, 1428, Buenos Aires, Argentina. CONICET-UBA\\
$^3$ Max-Planck Institute for Astrophysics, Karl-Schwarzchild Str. 1, D85748, Garching, Germany\\
$^4$ Astrophysikalisches Institut Potsdam, An der Sternwarte 16, D-14482
Potsdam, Germany
}

\maketitle

\begin{abstract}
We have resimulated the six galaxy-sized haloes of the Aquarius Project
including metal-dependent cooling, star formation and supernova feedback. This
allows us to study not only how dark matter haloes respond to galaxy formation,
but also how this response is affected by details of halo assembly history. In
agreement with previous work, we find baryon condensation to lead to increased
dark matter concentration. Dark matter density profiles differ substantially
in shape from halo to halo when baryons are included, but in all cases the
velocity dispersion decreases monotonically with radius. Some haloes show an
approximately constant dark matter velocity anisotropy with $ \beta \approx
0.1-02$, while others retain the anisotropy structure of their baryon-free
versions. Most of our haloes become approximately oblate in their inner
regions, although a few retain the shape of their dissipationless
counterparts.  Pseudo-phase-space densities are described by a power law in
radius of altered slope when baryons are included. The shape and concentration
of the dark matter density profiles are not well reproduced by published
adiabatic contraction models.  The significant spread we find in the density
and kinematic structure of our haloes appears related to differences in their
formation histories.  Such differences already affect the final structure in
baryon-free simulations, but they are reinforced by the inclusion of baryons,
and new features are produced. The details of galaxy formation need to be
better understood before the inner dark matter structure of galaxies can be
used to constrain cosmological models or the nature of dark matter.
\end{abstract}

\section{Introduction}

Dissipationless cosmological simulations have contributed substantially to our
understanding of the structure and evolution of cold dark matter (CDM) haloes,
showing them to have triaxial shapes (Frenk et al. 1998; Dubinski \& Carlberg
1991; Jing \& Suto 2002; Hayashi, Navarro \& Springel 2007) and density
profiles with inner cusps (Dubinski \& Carlberg 1991, Navarro, Frenk \& White
1996; Moore et al. 1999; Diemand et al. 2005) and an approximately universal
shape, independent of mass or cosmological parameters (Navarro, Frenk \& White
1997).  This universal cuspy profile appears in conflict with several
observations, for example, the slow increase of rotation velocity with radius
at the centre of low surface brightness galaxies (Flores \& Primack 1994;
McGaugh \& de Blok 1998; Moore et al 1999; Salucci, Yegorova \& Drory 2008) and the
relatively weak concentration of galaxy clusters inferred from strong lensing
(Sand et al. 2008). Additional evidence supporting central dark matter
densities lower than predicted by CDM models comes from the difficulty in
simultaneously matching observed luminosity functions and the zero-point of
the Tully-Fisher relation (e.g.  Dutton, van den Bosch \&
Courteau 2008).  Recently, Navarro et al. (2008, hereafter N08) analysed in detail the
density profiles of very high-resolution dark matter-only simulations of six
Milky Way mass haloes from the Aquarius Project (Springel et al. 2008). They
found the innermost cusps in these haloes to be weaker than claimed in some
earlier work, and showed how halo-to-halo variations in radial structure
reflect the detailed formation histories of individual haloes (see also
Vogelsberger et al. 2009).

Galaxy formation might reinforce such history-specific features in the dark
matter distribution, since baryons are subject to dissipative processes such
as cooling, star formation and feedback in addition to gravity.  The
condensation of baryons within dark matter haloes modifies both their dynamics
and their structure, but exactly how this happens is still quite
uncertain. The simplest model assumes that the dark halo is compressed
radially and adiabatically by the added mass at its centre (Blumenthal et
al. 1986, hereafter B86; see Eggen et al. 1962, Zeldovich et al. 1980 and
Barnes \& White 1984 for previous applications of this formalism). Recent
simulations have found the very simple adiabatic compression (AC) scheme of
B86 to overestimate the effect (e.g. Gnedin et al. 2004 and Abadi et al. 2009,
hereafter G04 and A09, respectively; Pedrosa, Tissera \&   Scannapieco 2010) but it is nevertheless often used in
simplified modelling of galaxy formation (e.g. Mo, Mao \& White 1998).  More
sophisticated AC models, based, for example, on the formalism of Young (1980)
have claimed to reproduce better the contraction and the final shape of haloes
(G04; Sellwood \& McGaugh 2005). The common outcome of these schemes is
nevertheless an increase in the dark matter density in the central regions,
exacerbating the problem of reconciling $\Lambda$CDM models with observation.

Simulations of galaxy formation which follow both dark matter and baryons in
their proper cosmological context are the primary tool for studying how galaxy
assembly affects dark matter haloes.  There are many papers devoted to this
subject. Improving numerical algorithms and increasing computer power have led
to continual progress in understanding the complex interplay between these two
components.  Previous analyses of the evolution of dark haloes in
hydrodynamical simulations have reported an increase both of the central mass
concentration and of the central velocity dispersion, which no longer shows
the 'temperature inversion' characteristic of dissipationless CDM haloes
(e.g. Katz \& Gunn 1991; Evrard, Summers \& Davis 1994; Navarro \& White 1994; Tissera \&
Dominguez-Tenreiro 1998; O\~norbe et al. 2008; Romano-Diaz et al. 2008).
Recently, Pedrosa, Tissera \& Scannapieco (2009, 2010) investigated how baryonic
assembly history affects the dark matter distribution. They found the SN
feedback process to play a key role by regulating star formation activity and
ejecting material not only from the main system, but also from infalling
satellites.  It seems clear that the final distribution of baryons at the
centre of a halo is insufficient to determine halo response to galaxy
assembly, thus contradicting the AC hypothesis.

Typical dark matter haloes have long been reported to be triaxial with major
to minor axis ratios often exceeding two in CDM-only simulations (e.g. Barnes
\& Efstathiou 1987; Frenk et al. 1988; Dubinski \& Carlberg 1991; Jing \& Suto
2002; Hayashi, Navarro \& Springel 2007). However, their shape changes,
becoming more nearly oblate as baryons condense within them (Katz \& Gunn
1991; Evrard et al. 1994; Tissera \& Dom\'{\i}nguez-Teneiro 1998; Kazantzidis
et al 2004; Debattista et al. 2008; A09). These results suggest the need for a
comprehensive analysis of halo structure in simulations which include both the
appropriate cosmological context and a realistic description of the physics of
baryon condensation. It is clearly necessary to analyse a number of different
galaxy-sized haloes in order to explore how differing assembly histories
affect the structure of the final haloes.

In this paper, we study a set of high-resolution resimulations of the six
galaxy-mass haloes of the Aquarius Project (Springel et al. 2008). These were
carried out with a version of {\small GADGET-3} which includes a multi-phase
treatment of metal-dependent cooling, star formation and SN feedback
(Scannapieco et al. 2005, 2006). The original haloes were selected from
a cosmological CDM-only simulation with no restriction on merger history,
except that implied by eliminating objects with high-mass close neighbours.
This Aquarius halo set is well-suited to study how formation history and
baryonic condensation together determine the final structure of dark matter
haloes.  Our ability to isolate these effects is aided by comparing results
from our hydrodynamical simulations (hereafter SPH runs) with corresponding
results from the CDM-only simulations (hereafter DM runs) as reported by N08.
Properties of the galaxies in these haloes are studied in Scannapieco et
al. (2009) and will be further analysed in a forthcoming paper.

This paper is organized as follows. In Section 2, we describe the numerical
experiments. In Section 3, we analyse the dark matter density
profiles. Section 4, describes the velocity dispersion structure of our
haloes. In Section 5, we study the effects of baryons on the
pseudo-phase-space density profile, while Section 6 discusses halo
shapes. Section 7 compares the change in halo circular velocity profile
between the SPH and the DM runs with the predictions of AC models. Finally, in
Section 8, we summarize our main findings.

\section{The Simulated Haloes}

The six haloes studied in this paper were taken from the Aquarius Project
(Springel et al. 2008). They were selected at random from a lower resolution
version of the $(100\ h^{-1}{ \rm Mpc })^3$ Millennium-II Simulation
(Boylan-Kolchin et al. 2009) subject only to the requirements that their mass
should be similar to that inferred for the Milky Way's halo and that they
should have no close massive neighbour. Dark matter only versions of these
haloes were then resimulated at a variety of much higher resolutions as part
of the Aquarius Project itself (Springel et al. 2008; N08; Vogelsberger et al. 2009). 
The lowest resolution version of
each halo (designated as resolution level 5) was also resimulated with
detailed modelling of baryonic processes by Scannapieco et al. (2009,
hereafter S09).  Details of how the initial conditions were created can be
found in these two papers.  In the following we will refer to the original
Aquarius simulations at the higher resolution level 2 as the DM simulations,
and to the simulations of S09, which have roughly 200 times worse dark matter
mass resolution but include the baryonic physics, as the SPH simulations. The
analysis in N08 shows that the properties of the dark haloes which concern us
in this paper are extremely well numerically converged all the way down to
resolution level 5. The individual haloes are labelled Aq-A-$n$, Aq-B-$n$,
etc., with $n$ corresponding to the resolution level in to be consistent with
the convention in the earlier papers.

The Millennium and Millennium-II Simulations, and thus also these halo
simulations, were carried out assuming a $\Lambda$CDM cosmology with
parameters $\Omega_{\rm m}=0.25,\Omega_{\Lambda}=0.75, \sigma_{8}=0.9,
n_{s}=1$ and $H_0 = 100 \ h \ { \rm km s^{-1} Mpc^{-1}}$ with $h
=0.73$. Evolution in both the DM and the SPH simulations was followed from $z
= 127$ to $z=0$ using versions of {\small GADGET-3}, an update of {\small
  GADGET-2} (Springel et al. 2001; Springel 2005) optimized for massively
parallel simulation of highly inhomogeneous systems such as individual dark
haloes.

The version of {\small GADGET-3} used for the SPH runs includes a multiphase
model for the gas component with metal-dependent cooling, star formation and
phase-dependent treatments of SN feedback and chemical enrichment, as set out
in Scannapieco et al. (2005, 2006) and employed for studying $\Lambda$CDM
galaxy formation in Scannapieco et al. (2008, 2009). The model describes
chemical enrichment by SNII and SNIa separately, using appropriate yields and
delays (Mosconi et al. 2001; Scannapieco et al. 2005). The multiphase
treatment of both hydrodynamics and feedback is quite effective in reproducing
the observed phenomenology of  star formation and wind-generation in both
quiescent and starburst galaxies. Without introducing system-specific
parameters, the scheme produces substantial mass-loaded galactic winds in
rapidly star-forming systems with speeds that reflect the escape velocity much
as in observed galaxies (Scannapieco et al. 2006, 2008). This freedom from {\it ad
  hoc} scale-dependent parameters, makes the algorithm particularly suited for
studying galaxy formation in its cosmological context, since the simultaneous
formation of systems of widely differing mass is the norm in this situation.

The SPH runs have maximum gravitational softenings in the range $\epsilon_{G}
=0.5 - 1\ h^{-1}$kpc.  To prevent spurious results due to limited numerical
resolution in the innermost regions, we analyse halo properties only outside
$2  h^{-1}$kpc in all our systems.  
 Haloes are considered bounded by their virial
radius $r_{200}$ defined as the largest radius within which the mean enclosed
density exceeds $\approx 200$ times the critical density.  The simulated SPH
haloes have $\approx 1$ million particles in total within this virial radius
while the DM simulations have approximately $10^8$ particles in the same
region.  The virial masses of the six systems are in the range $5$ to $11
\times 10^{11} h^{-1} {\rm M_\odot}$. Hence, dark matter particles in the SPH
runs have masses of the order $10^{6}h^{-1}{\rm M_\odot }$ while gas particles
initially have $\approx 2 \times 10^{5} h^{-1}{\rm M_\odot}$. In Table
~\ref{tab1} we summarize the principal characteristics of our SPH simulations.
 We also include the corresponding information for two lower resolution versions of
halo Aq-E. We find convergent results between Aq-E-5 and Aq-E-6 for their characteristic properties.
However, for the lowest resolution run, Aq-E-7,  larger differences are found.
 In Section 3, we discuss the effects of numerical resolution on the dark matter distributions. 

The Aquarius haloes have varied assembly histories and these produce a variety
of structures and star formation histories for their central galaxies, even
though the simulations were all run with similar SF and SN feedback
parameters.  A detailed description of the code and the parameters adopted can
be found in S09, together with images of the central galaxies and considerable
analysis of their structure.  Here we provide a brief summary so that the
reader has a general picture of the central galaxies of the haloes we analyse
here.

In all six galaxies star formation peaked between 13 Gyr (Aq-C-5) and 10 Gyr
(Aq-B-5) ago and there has been rather little star formation over the last 8
Gyr. The total stellar masses at $z=0$ are in the range $2.5$ to $6 \times
10^{10} h^{-1} {\rm M_\odot}$ and are listed in Table ~\ref{tab1}, along with
the total masses and baryon masses of the $z=0$ haloes and the numbers of
simulation particles representing these masses.

The analysis of S09 showed that most of the central galaxies contain
centrifugally supported disks, although none of these accounts for more than
about a fifth of the total stellar mass. Aq-F-5 is an exception in that it has
no disk, only a spheroidal component.  This can be traced to the fact that the
system underwent a major merger at $z \sim 0.6$.  In some cases the stellar
spheroid has substantial nett rotation (Aq-E-5) while in others it retains
very little angular momentum (Aq-A-5). All the galaxies have been
substantially affected by SN feedback, which drives winds which limit their
mass.  In all cases the baryon fraction within the virial radius is around
10\%, significantly smaller than the global value, $f_{\rm b}=0.16$.

\begin{table*}
  \begin{center}
  \caption{General characteristics of our Aquarius halo simulations
    and their central galaxies. The first column show the halo name,
    taken from Springel et al. (2008). $\epsilon_G$ is the
    gravitational softening.  $r_{\rm 200}$ and $M_{\rm 200}$ are the
    radius and mass of the sphere enclosing total mean mass density
    200 times the critical value.  $M_{\rm dm}$ and $M_{\rm bar}$ are
    dark matter and baryonic masses within this same sphere. $M_{\rm
     s}$ is the stellar mass of the main galaxy hosted by each halo,
    measured within an optical radius defined to contain 83\% of the
    galaxy's baryonic mass. $N_{\rm dm}$ and $N_{\rm bar}$ are the
    total number of particle representing each mass component.}
\label{tab1}
\begin{tabular}{|l|c|c|c|c|c|c|c|c}\hline
 {Halo}  &$\epsilon_G$ & $r_{\rm 200}$& $M_{\rm 200}$&$M_{\rm dm}$ &$M_{\rm bar}$ &$M_{\rm s}$
&$N_{\rm dm}$&$N_{\rm bar}$ \\ 
   &kpc h$^{-1}$&  kpc h$^{-1}$ &${10^{12}\rm M_{\odot}h^{-1}} $&$ 10^{11}{\rm M_{\odot}}h^{-1} $&$ 10^{10} {\rm M_{\odot}h^{-1}}$ &$10^{10} {\rm M_{\odot}h^{-1}} $&& \\\hline
Aq-A-5 &0.5&	169.42 &  1.10 &  9.95 & 10.17  & 5.92 &529110&425737 \\
Aq-B-5& 0.5&	132.10 &  0.52 &  4.79 & 4.15 &  2.53 &435330&354976\\
Aq-C-5&	1.0&    173.19 &  1.18 &  10.70 &  10.68& 5.93& 681143&647325\\
Aq-D-5&	0.5&    170.63 &  1.09 &  10.10 &  8.29 & 4.41 &599438&460845\\
Aq-E-5&	1.0&    149.93 &  0.79 &  7.08 &  8.05& 4.97&554245&606136\\
Aq-F-5& 1.0&    142.74&   0.67&   5.99&  6.92& 5.39& 680129&759456\\\hline
Aq-E-6& 2.0&    150.79&   0.79&   7.20&  7.60&   4.89  & 132662 &132698\\
Aq-E-7& 4.0&    146.39&   0.73&   6.75&  5.41&   3.06  &44429 &32214\\
\end{tabular} 
 \end{center}
\vspace{1mm}
\end{table*}

\begin{table*}
  \begin{center}
  \caption{Characteristics of the density profiles of the haloes in our
    Aquarius galaxy formation simulations. Column (1) gives the name
    of each halo.  Columns (2),(3) and (4) list $\alpha$, $r_{-2}$ and $\rho_{-2}$ (in units
    of kpc $h^{-1}$ and  ${\rm M_\odot h^2 kpc^{-3}}$), the parameters of the best fitting Einasto
    model. Column (5) gives the {\it rms} scatter in  $\log \rho$ around
    this fit. Column (6) lists exponents
    for the best power law fits to $\rho / \sigma^3$ as a function of
    $r$. Columns (7), (8) (9)  (10) and (11) list  the corresponding fitting parameters  for
    the corresponding DM runs. Note that the fitting parameters for the profiles in the DM runs differ slightly from those given by N08 because we  have  re-calculated them over the same radial
    range used for the SPH runs. }
\label{tab2}
  \begin{tabular}{|l|c|c|c|c|c|c|c|c|c|c|}\hline
 { Run}  & { $\alpha$} & { $r_{-2}$}  &  {$\log \rho_{-2}$}  &{\it rms }& $\chi$ & { $\alpha^{\rm DM}$} & { $r_{-2}^{\rm DM}$} &{$\log \rho_{-2}^{\rm DM}$}& {\it rms }&$\chi^{\rm DM}$ \\\hline
Aq-A-5 &  0.065 &  3.68 & 7.81 &   0.016    & -1.67 & 0.108 &13.13 & 6.71  & 0.001&-2.05\\	
Aq-B-5 &  0.145 &  10.95& 6.59 &   0.015   &-1.63  & 0.192 &16.64 & 6.28&0.007&-1.87\\
Aq-C-5 &  0.115&  7.17  & 7.28 &   0.014    &-1.63  & 0.179 &13.58 &6.76 &0.010&-1.95\\
Aq-D-5 &  0.102 & 10.35 & 6.85 &   0.018   &-1.66  & 0.178 &21.23 &6.28&0.008&-1.86\\
Aq-E-5 &  0.098 &  7.79 & 6.99 &   0.017    &-1.62  & 0.149 &14.97 & 6.47 &0.010&-1.91\\
Aq-F-5 &  0.112 &  10.89& 6.62 &   0.011   &-1.60  & 0.164&18.96  &6.22&0.016&-1.98\\ \hline     

Aq-E-6 &  0.091 &  7.77 & 6.99 &   0.012    &-1.68  & -- & -- & -- & -- & --\\
Aq-E-7 &  0.15 &  14.69 & 6.41 &   0.013    &-1.96  & -- & -- & -- & --& --\\
\end{tabular}
 \end{center}
\vspace{1mm}
\end{table*}

\section{Dark matter density profiles}

We have constructed spherically-averaged dark matter density profiles from
$2\ h^{-1}$kpc to the virial radius $r_{200}$. The inner radius is four times
the gravitational softening for three of our simulations (A, B and D) and
twice the gravitational softening in the other three. It typically contains
more than  a thousand dark matter particles.
 Finding an appropriate centre is crucial to obtaining
accurate profiles. Here we use the shrinking sphere technique as laid out in
of Power et al. (2003). In practise, this gives a centre very close to the minimum
of the gravitational potential in all our systems.
 We have measured profiles both including and excluding
substructures, but this only causes rather minor effects in the outer
regions. In the following we use the full DM mass including substructures in
order to be able to compare consistently with the results of N08. Note that we
re-normalize the dark matter profiles of N08 by the global DM fraction of
$0.84$ adopted in the SPH simulations when comparing profiles from the two
set-ups.

To characterise our dark matter profiles we adopt the three-parameter Einasto
model (Einasto 1965) which N08 found to give a relatively good fit to the profiles of the DM
runs. This model has also been used by earlier authors to fit the dark matter
distribution when baryons are included (e.g. O\~norbe et al. 2008; Gao et
al. 2008). N08 parametrise the Einasto profile using $\alpha$, $r_{-2}$ and
$\rho_{-2}$, which indicate its curvature in a log-log plot, and the radius
and density at the point where its logarithmic slope is $-2$, the isothermal
value. We fit this formula, 
leaving all three parameters free, to our measured
density profile at a set of points spaced logarithmically from $2\ h^{-1}$kpc
to $r_{200}$, minimising the rms residual in $\log\rho$.
 In Table~\ref{tab2}
we give the results of these fits, together with the {\it rms} residual in
$\log\rho$. To facilitate the comparison with the DM runs, we also include
the corresponding parameters given by N08; note that these were obtained for
fits over a slightly wider radial range. We see that in all cases $r{-2}$
and $\alpha$ are both smaller for the simulations which include baryons,
showing that the condensation of the galaxy has increased the concentration
of the halo and given it a density profile which is approximately isothermal
over a wider radial range than in the DM-only case. The residuals in
$\log\rho$ are only slightly larger, showing that the Einasto profile
is  a good  fit to both the  SPH and the  DM runs.

\begin{figure*}
\resizebox{7cm}{!}{\includegraphics{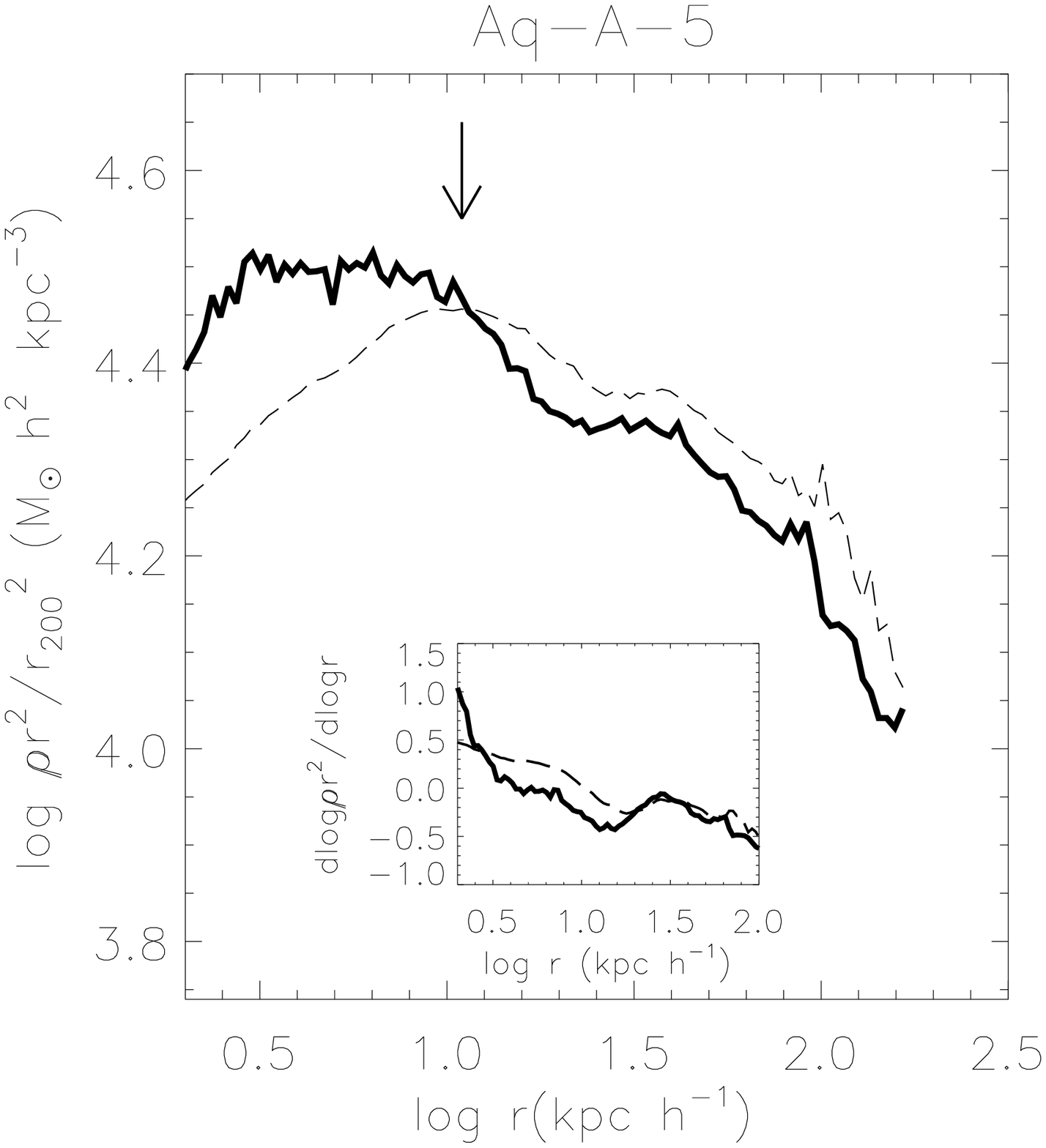}}
\resizebox{7cm}{!}{\includegraphics{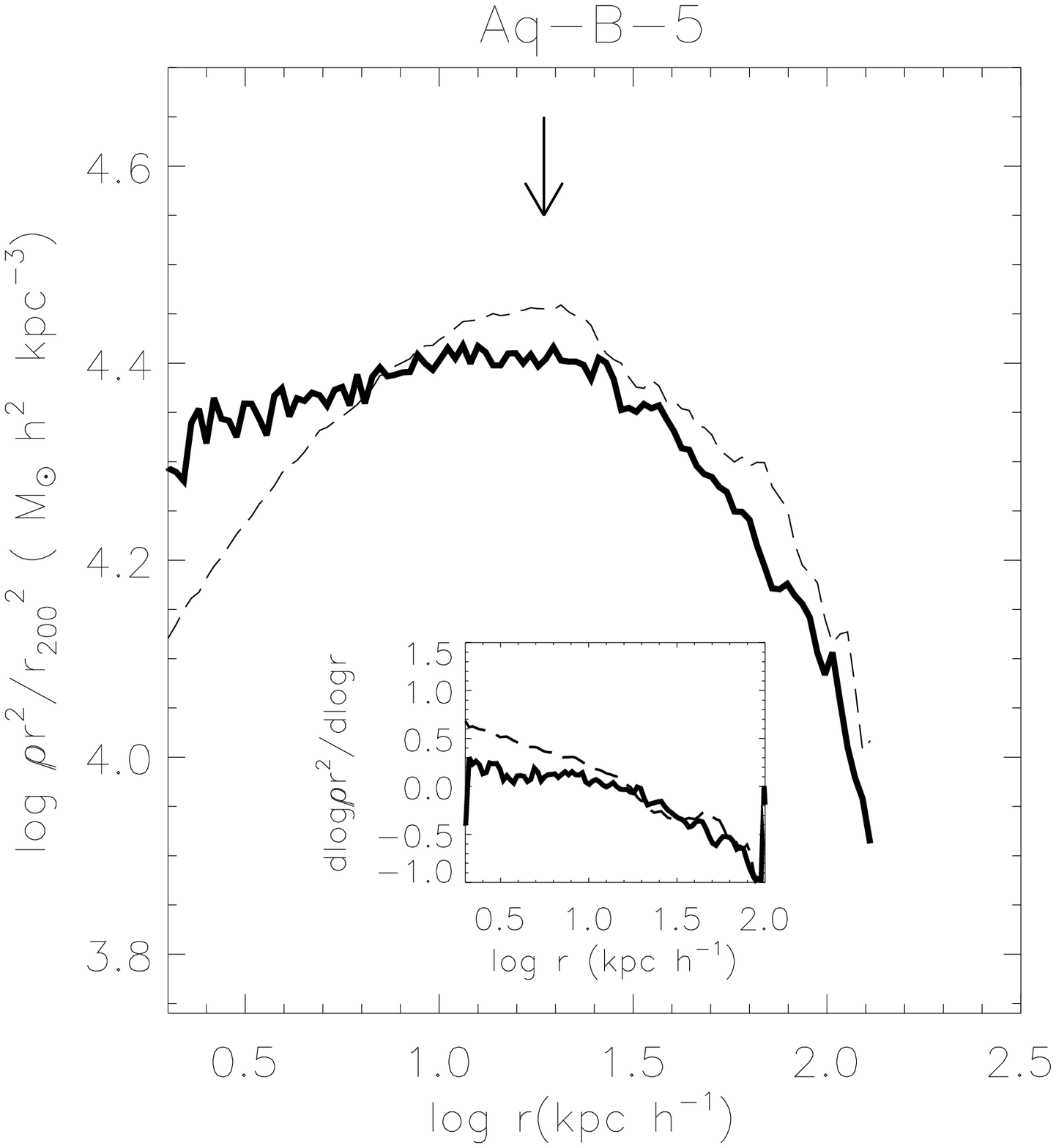}}\\
\resizebox{7cm}{!}{\includegraphics{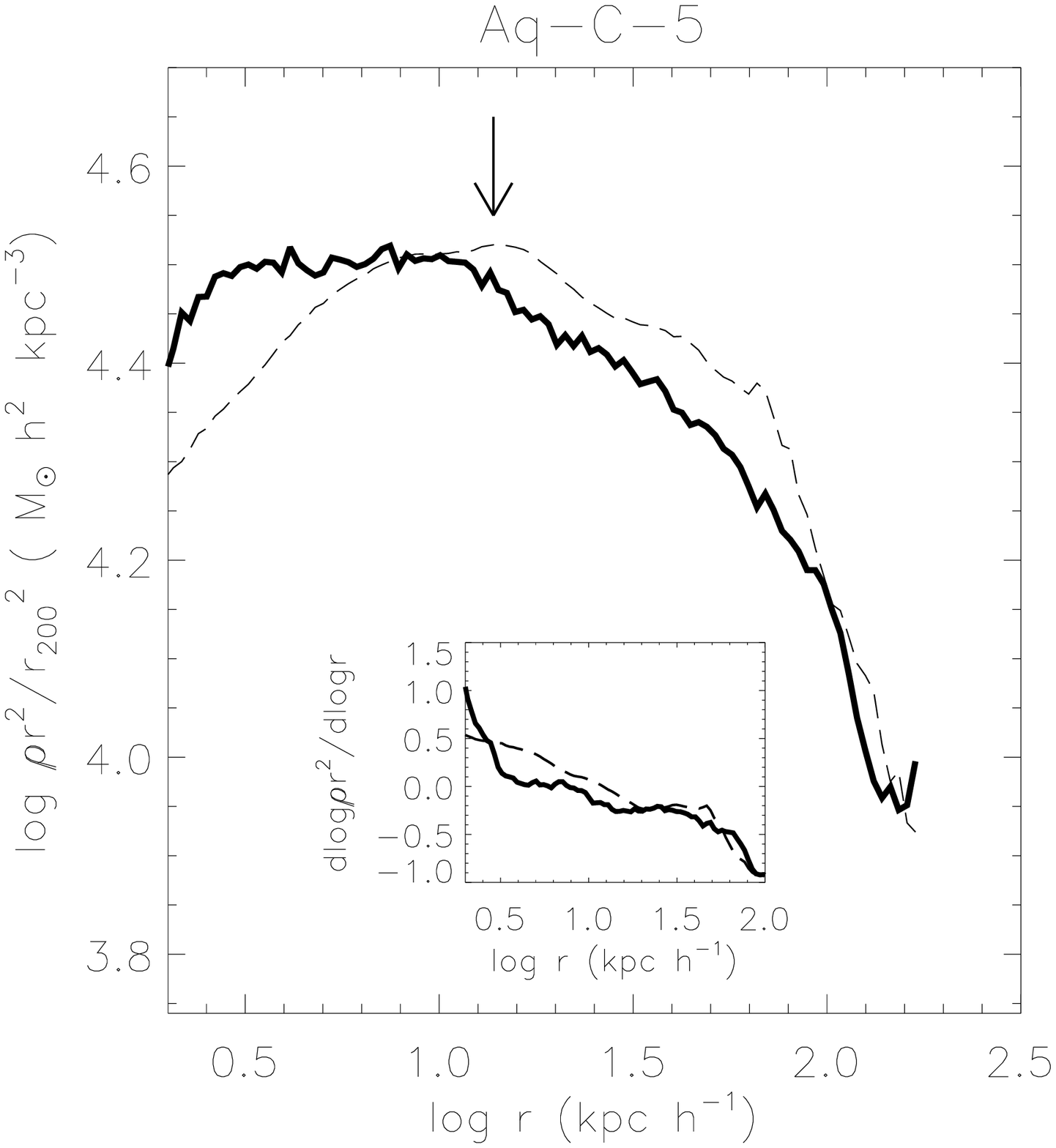}}
\resizebox{7cm}{!}{\includegraphics{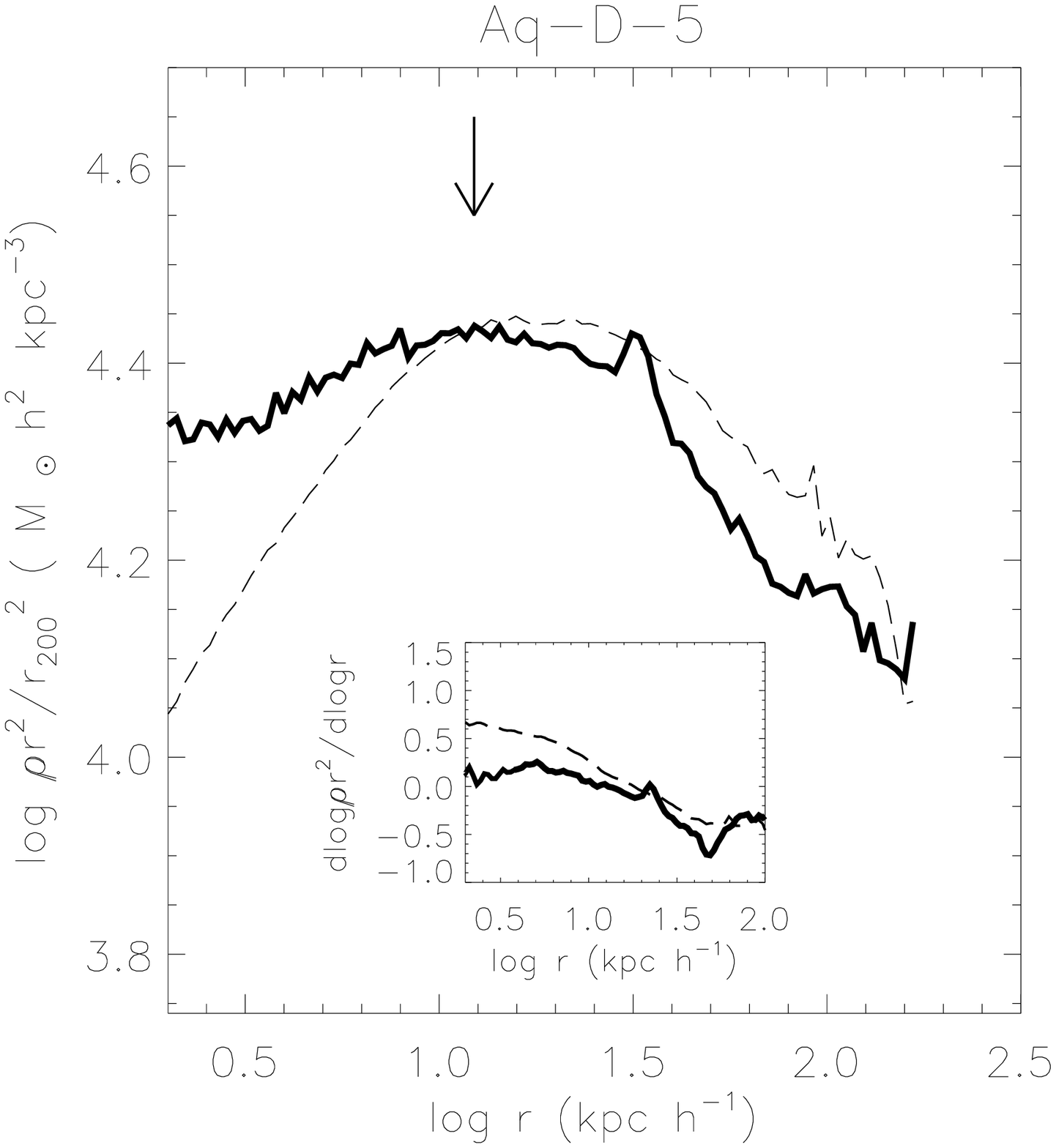}}\\
\resizebox{7cm}{!}{\includegraphics{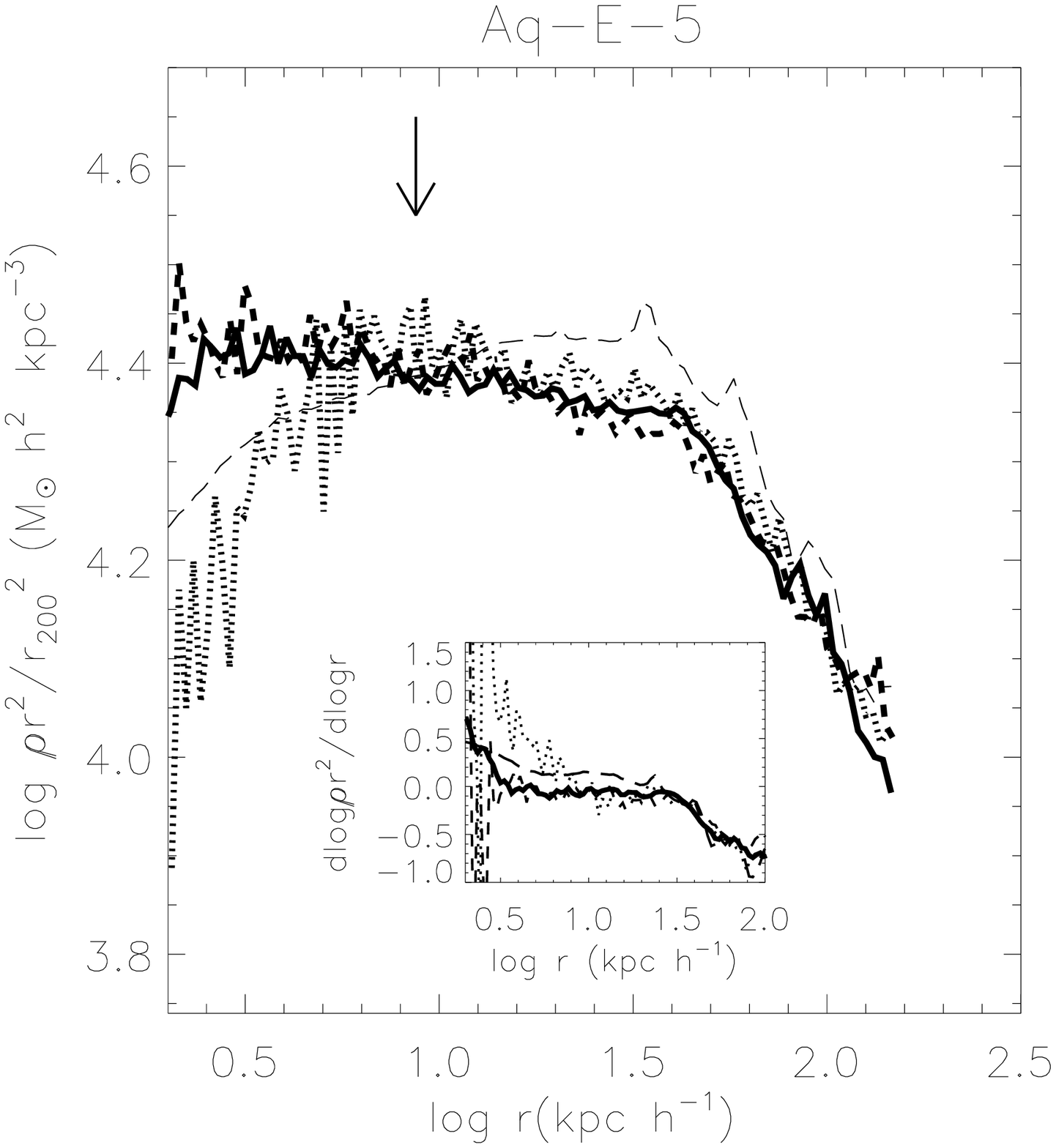}}
\resizebox{7cm}{!}{\includegraphics{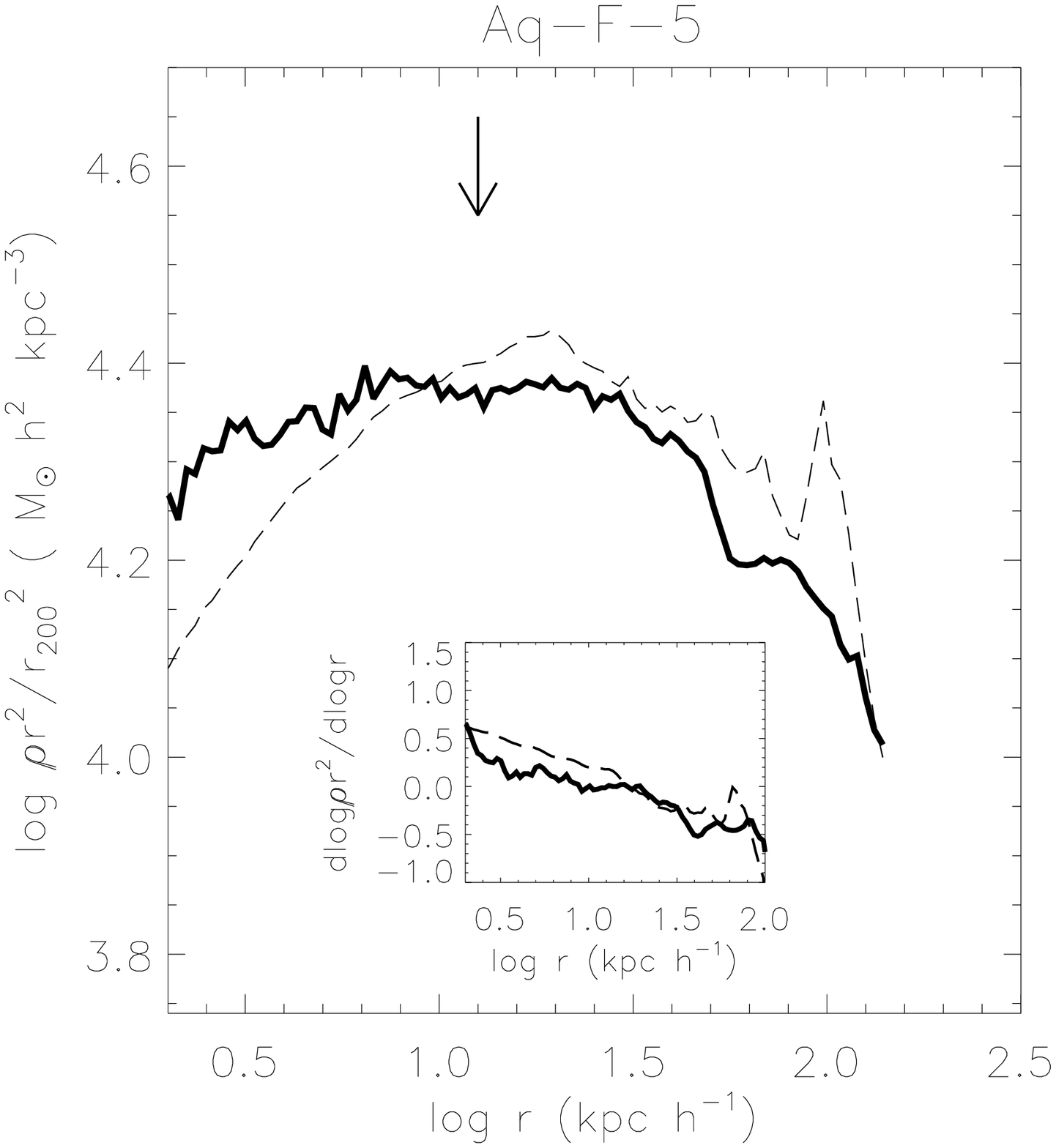}}
\hspace*{-0.2cm}
\caption{Spherically-averaged dark matter profiles for six haloes from the
  Aquarius Project simulated with (solid lines) and without (long dashed lines)
  baryons. We plot the profiles from $2\ h^{-1}$kpc to the virial radius
  $r_{200}$. The DM runs are the resolution level 2 simulations studied by
  Navarro et al. (2008).The arrow indicates the baryonic radius of the central
  galaxy defined as the radius enclosing $83\%$ of its stars and cold gas. The
  inset boxes show the logarithmic derivatives of the profiles.
 We include the lower resolution versions of Aq-E: Aq-E-6 (short-dashed lines) and Aq-E-7 (dotted lines). }
\label{haloes}
\end{figure*}

This is evident in Fig.~\ref{haloes} where we compare the spherically-averaged
dark matter density profiles of the haloes in our SPH runs (solid lines) to
the corresponding profiles for the DM runs from N08 (dashed lines).  Every SPH
profile lies below its corresponding DM profile at large radii and above it at
small radii.  In all cases the two profiles cross at, or slightly inside, the
baryonic radius of the central galaxy, which we define to be the radius
containing 83\% of its stars and cold gas. Inside the baryonic radius,
the SPH profiles are all relatively flat, and so
approximately ``isothermal''. This is indicated more explicitly in the insets
in Fig.~\ref{haloes} which plot the logarithmic derivatives of the two
profiles as a function of radius. This slope changes smoothly with radius in
most of the DM haloes, but shows a clear change in behaviour in the inner
regions in many of the SPH haloes. 
 For Aq-A-5, Aq-C-5 and Aq-E-5, we detect a 
change in the logarithmic derivatives of the SPH profiles at the smallest radii suggesting a flattening of the profiles. However, this occurs very close to our resolution limit and therefore, should be viewed with caution.
At radii well outside the central galaxy the
SPH profiles parallel the DM profiles quite closely, and indeed the offset
between them is close to the renormalisation factor that we introduced to
account for the difference in the total amount of dark matter in the two kinds
of simulation.

We have looked for correlations between the changes in $\alpha$ and $r_{-2}$
and various other parameters of our haloes, in particular, their spin
parameter, the fraction of the baryons gathered in the central galaxy, and their 
disk-to-total mass ratio. However, in no case did we find a clear trend.

  To study the robustness of our results against numerical resolution, we analysed the dark matter
 profiles of the two lower resolution versions of halo Aq-E. The results show  excellent convergence 
 for the dark matter profile and its characteristic properties, as  can be seen in  Fig.~\ref{haloes}. 
Runs Aq-E-5 and Aq-E-6 are in very good agreement over the whole analysed region. The same is true for the
 lowest resolution Aq-E-7 but only over the region resolved with at least a thousand 
dark matter particles. A lower resolution seems to produce an underestimation of the
dark matter density in the innermost regions.
In Table 2, we include the fitting parameters for these two additional simulations 
(carried out from twice their gravitational softening).
Based on these findings, we are confident that our main set of simulations have been run with adequate
 numerical resolution to assure convergent results for the density profiles 
over the analysed radial interval.

\section{Velocity dispersion distributions}

The velocity dispersion structure of the dark matter haloes in the DM runs has
been analysed by N08, who show that they all have a temperature inversion in
the central region (i.e. the velocity dispersion drops towards the centre at
small radii) but that there is no simple regularity among the haloes.  Rather
there is considerable diversity which seems to reflect the variety of
formation paths possible for these galaxy-sized haloes in the $\Lambda$-CDM
cosmology.
 
\begin{figure*}
\resizebox{5cm}{!}{\includegraphics{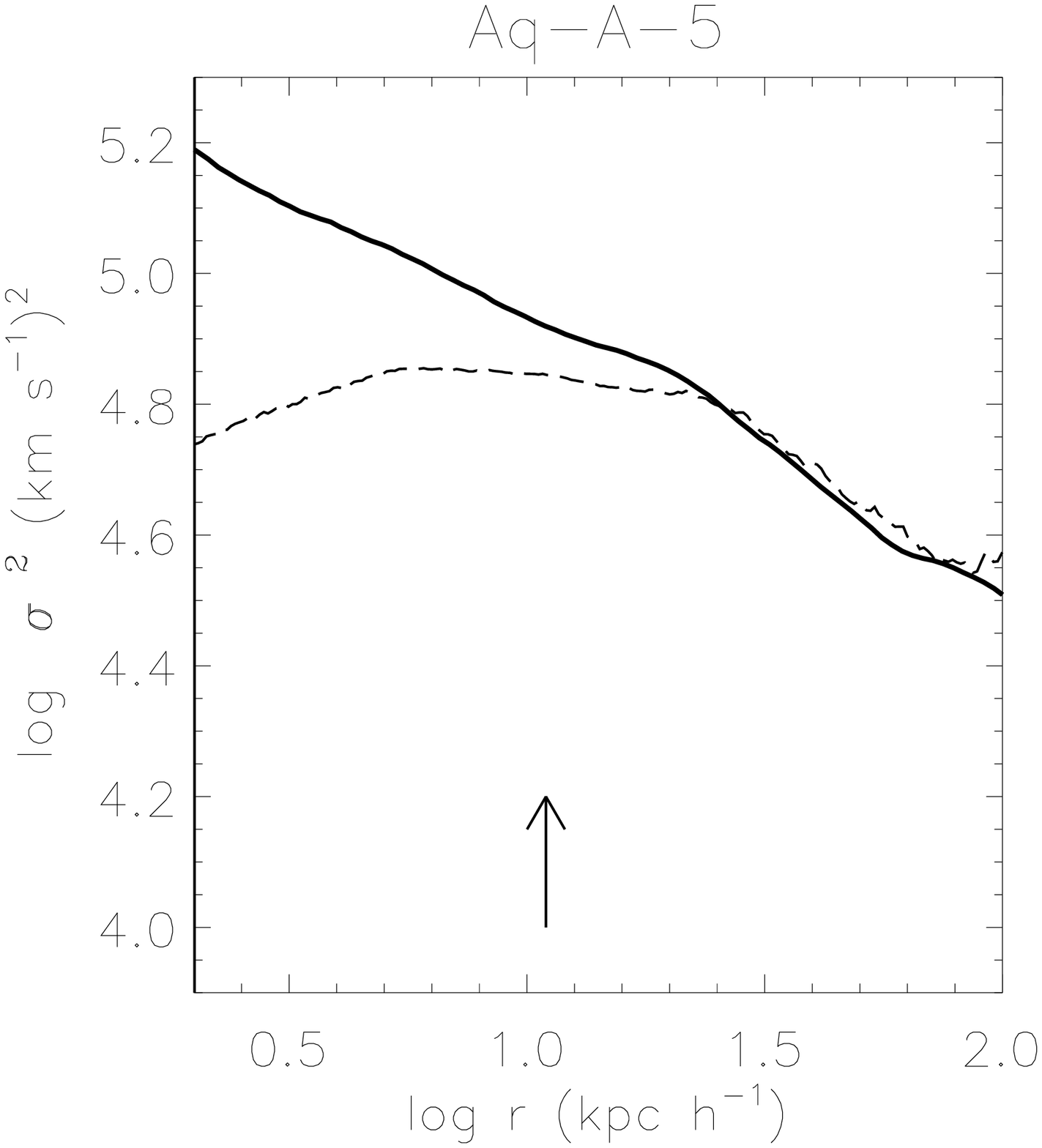}}
\resizebox{5cm}{!}{\includegraphics{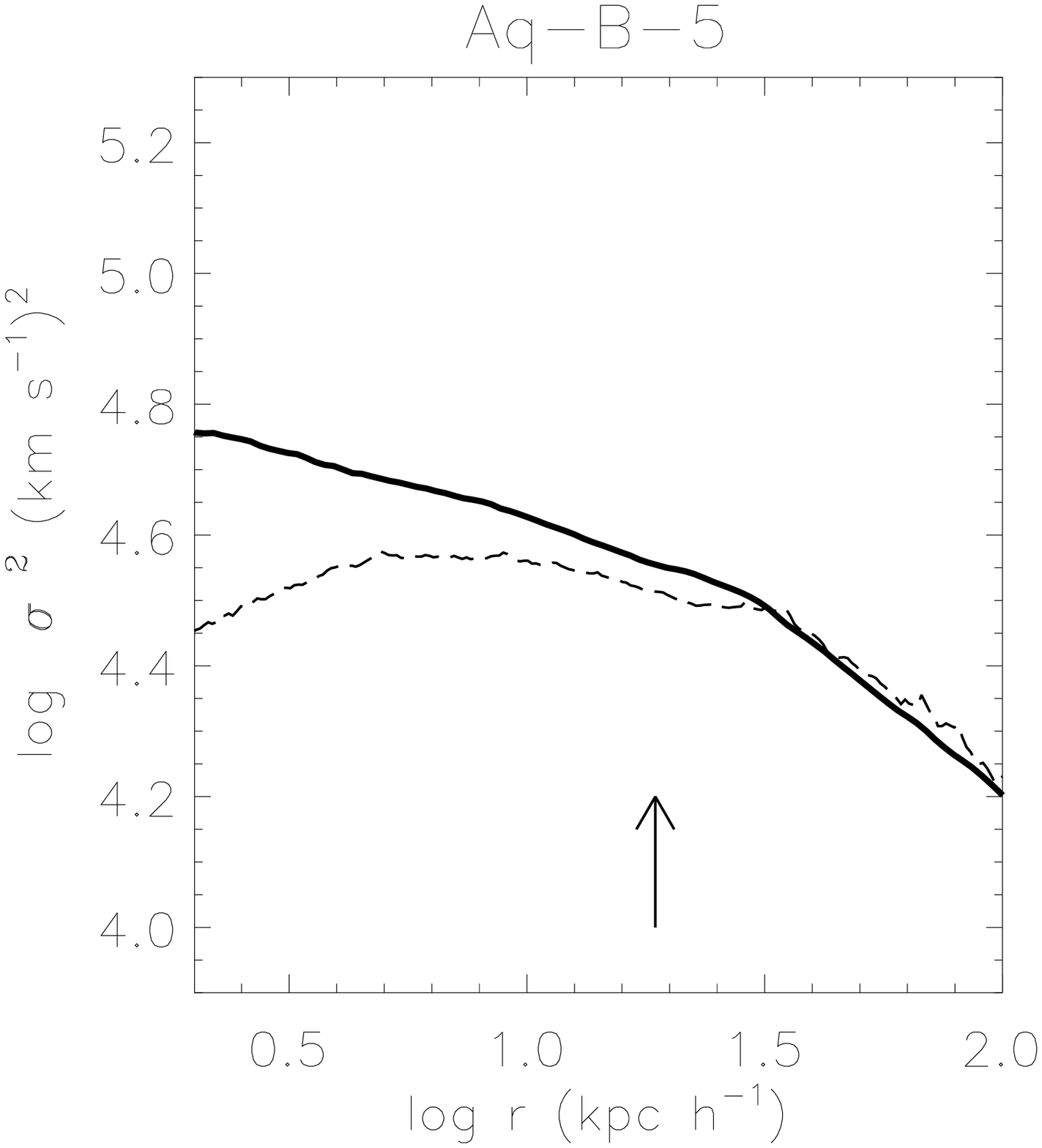}}
\resizebox{5cm}{!}{\includegraphics{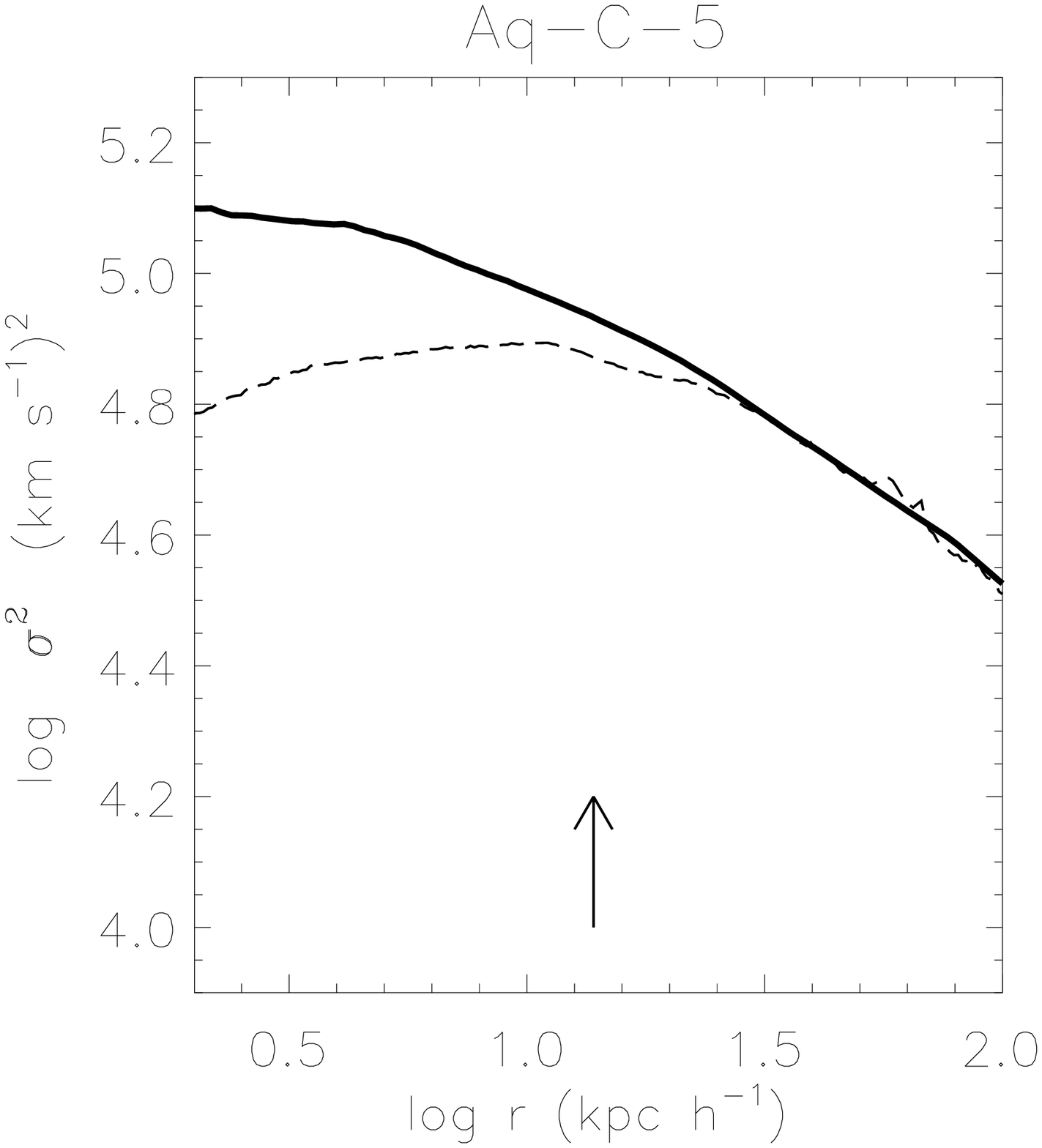}}\\
\resizebox{5cm}{!}{\includegraphics{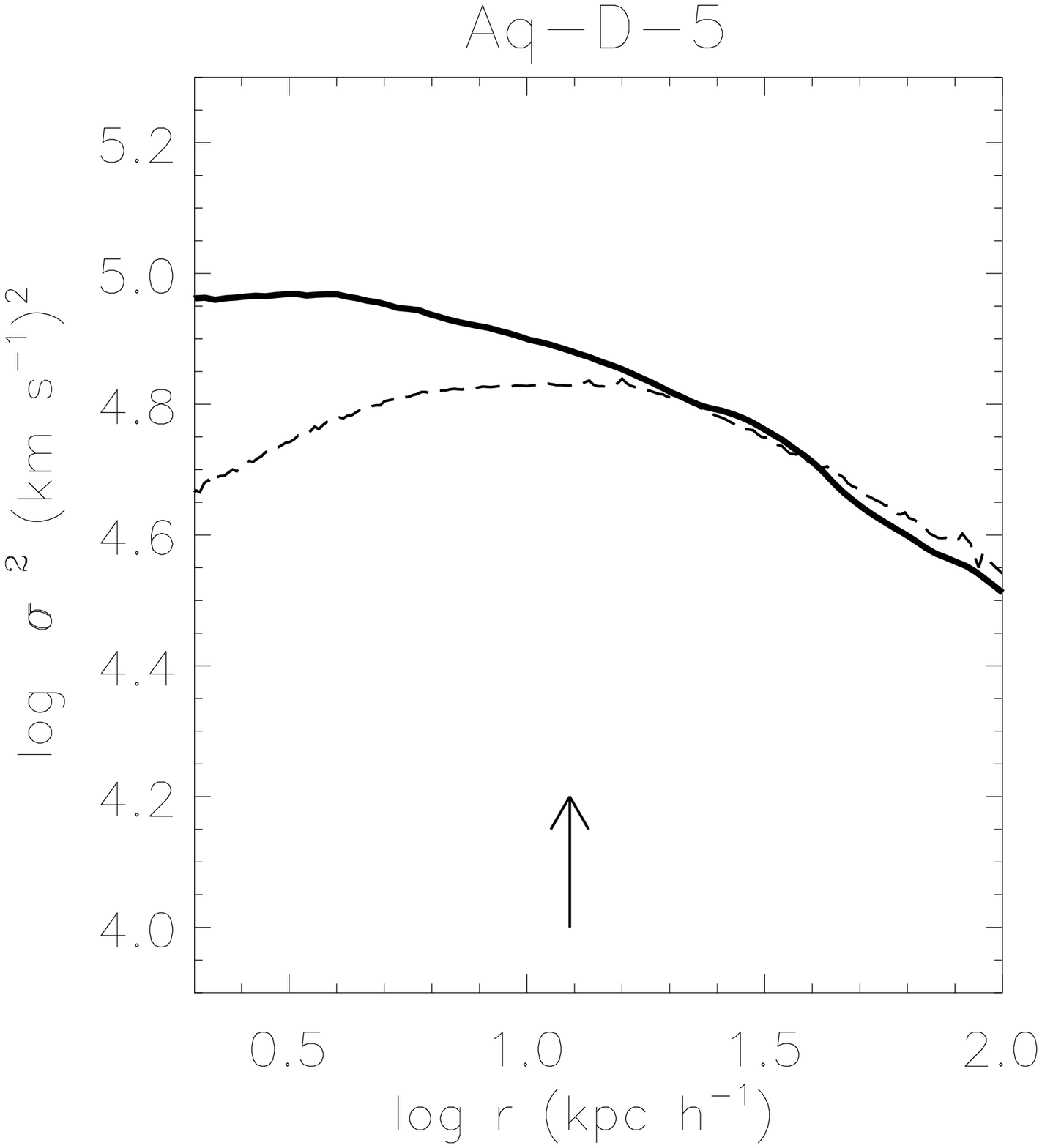}}
\resizebox{5cm}{!}{\includegraphics{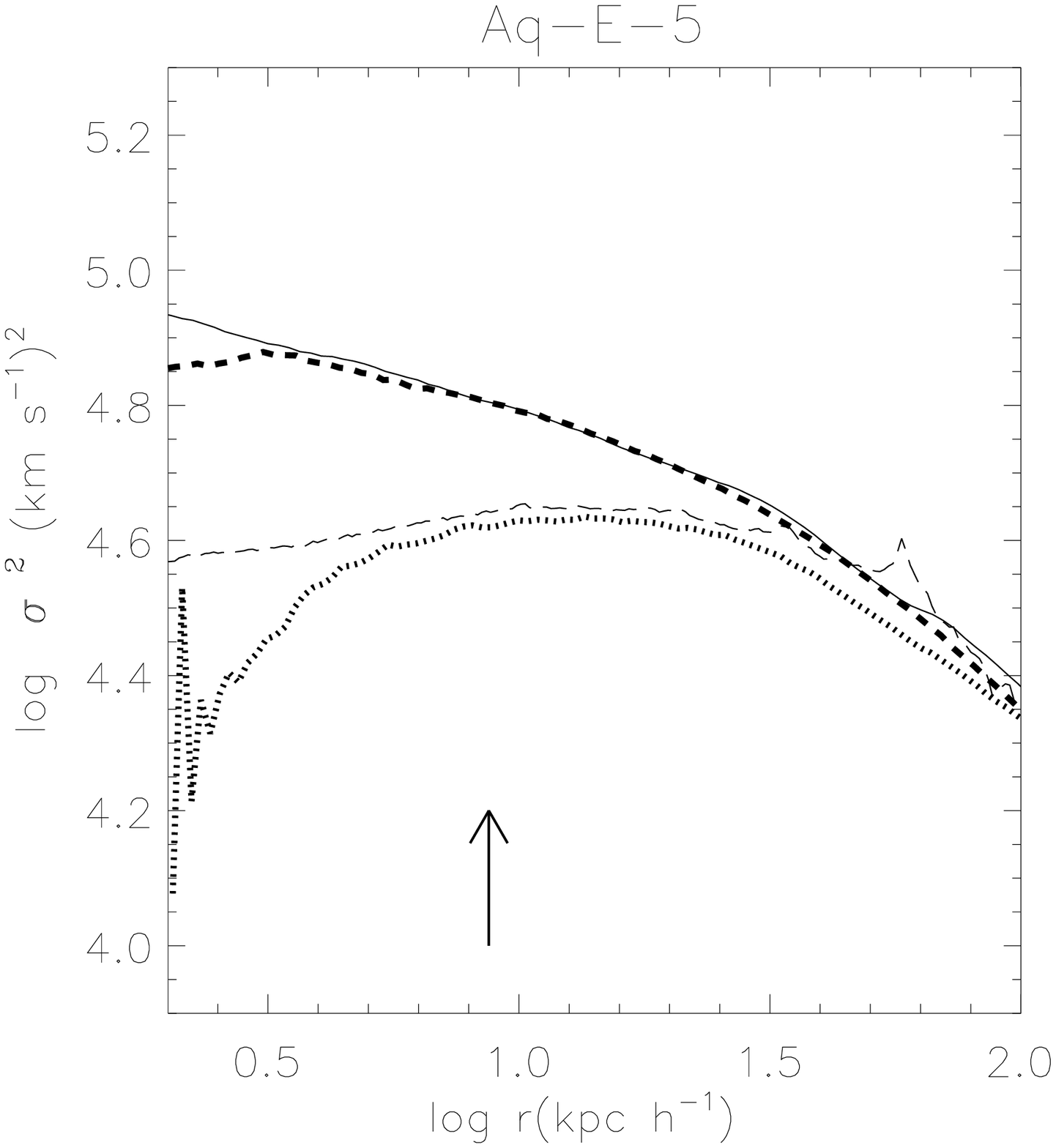}}
\resizebox{5cm}{!}{\includegraphics{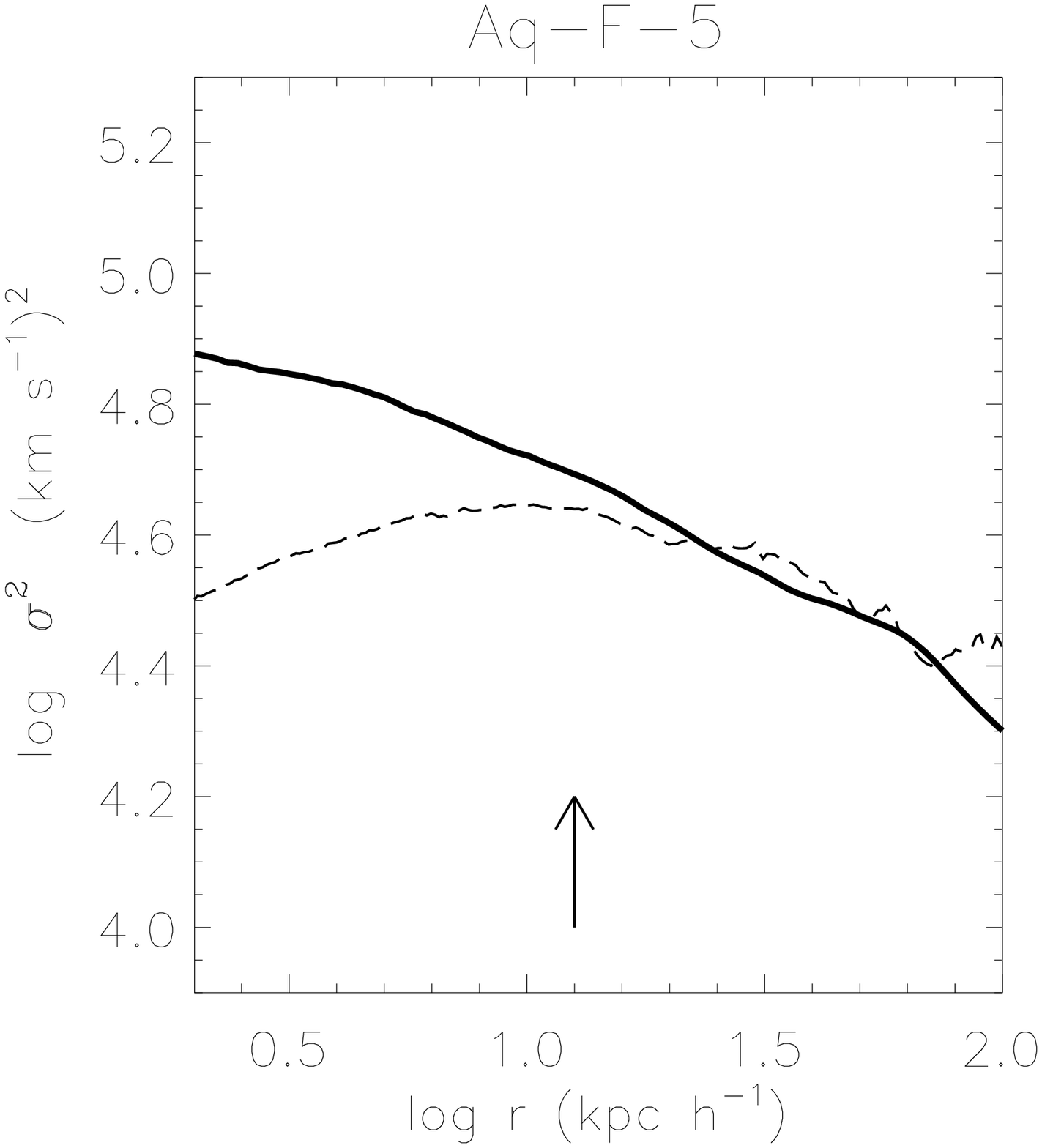}}
\hspace*{-0.2cm}
\caption{Velocity dispersion as a function of radius for the SPH (solid lines)
  and DM (dashed lines) haloes. The arrows indicate the baryonic radii of the
  central galaxies.  We include the lower resolution versions of Aq-E: Aq-E-6 (short dashed lines) and Aq-E-7 (dotted lines).}
\label{sigmas}
\end{figure*}

We here carry out a similar analysis of the velocity dispersion structure of
our set of SPH runs. In Fig. ~\ref{sigmas} we compare their total velocity
dispersion profiles to those of the DM runs. It is clear that the baryons
have substantially affected the velocities of dark matter particles within the
baryonic radius of the central galaxy. As reported in previous work
(e.g. Tissera \& Dominguez-Tenreiro 1998; Romano-Diaz et al. 2008; Pedrosa et al. 2010) the
increased velocity dispersion in the central regions results in a final profile
that decreases monotonically with radius, in contrast to the behaviour seen in
DM-only simulations.  For Aq-E-5, we have included its lower resolution versions (Aq-E-6 and Aq-E-7). 
As can be seen, the lowest resolution version provides a poor representation of the velocity
distribution obtained with the higher resolution runs. However, Aq-E-6 and Aq-E-5 agree
very well except in the very central region. 

\begin{figure*}
\resizebox{5cm}{!}{\includegraphics{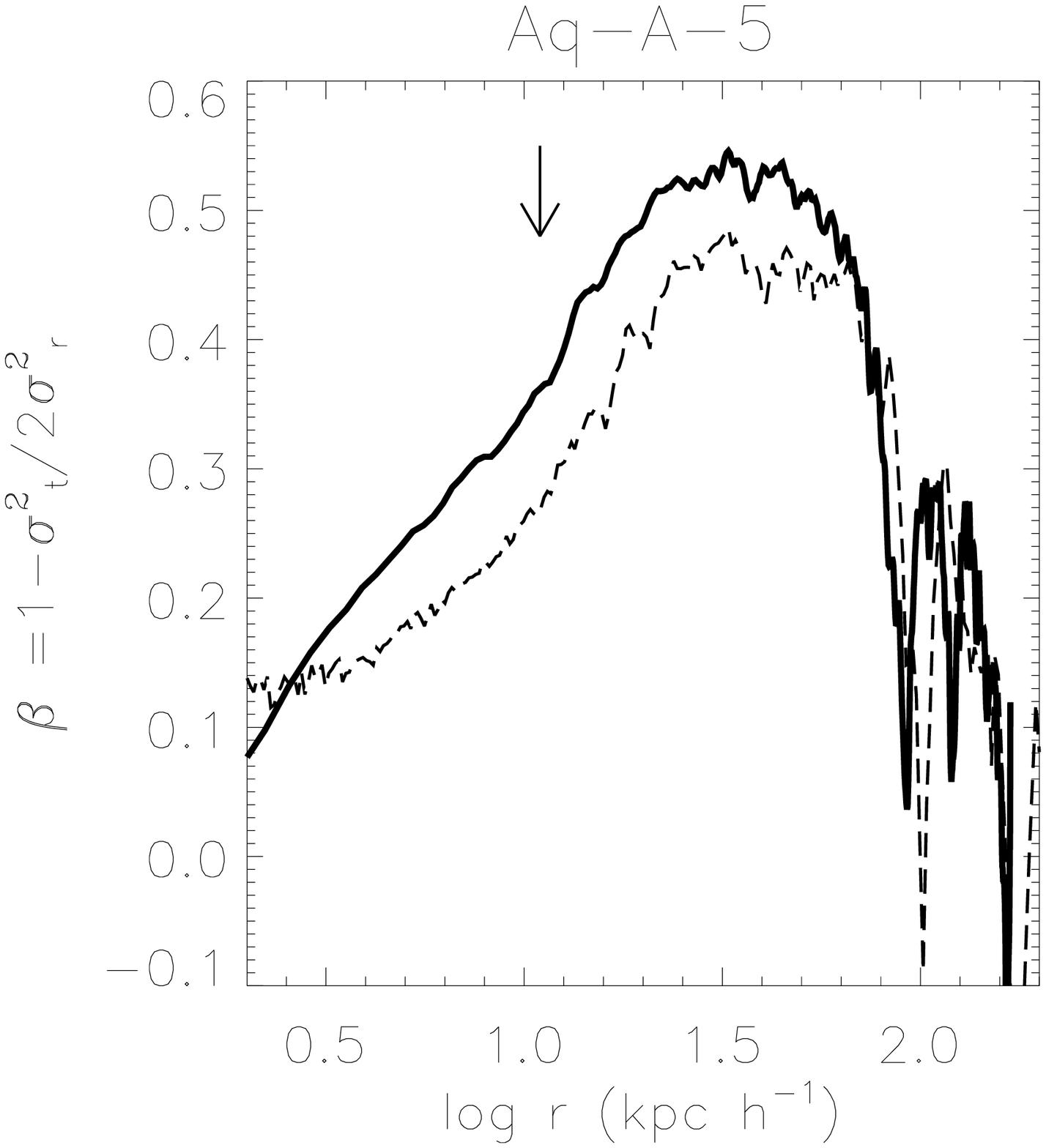}}
\resizebox{5cm}{!}{\includegraphics{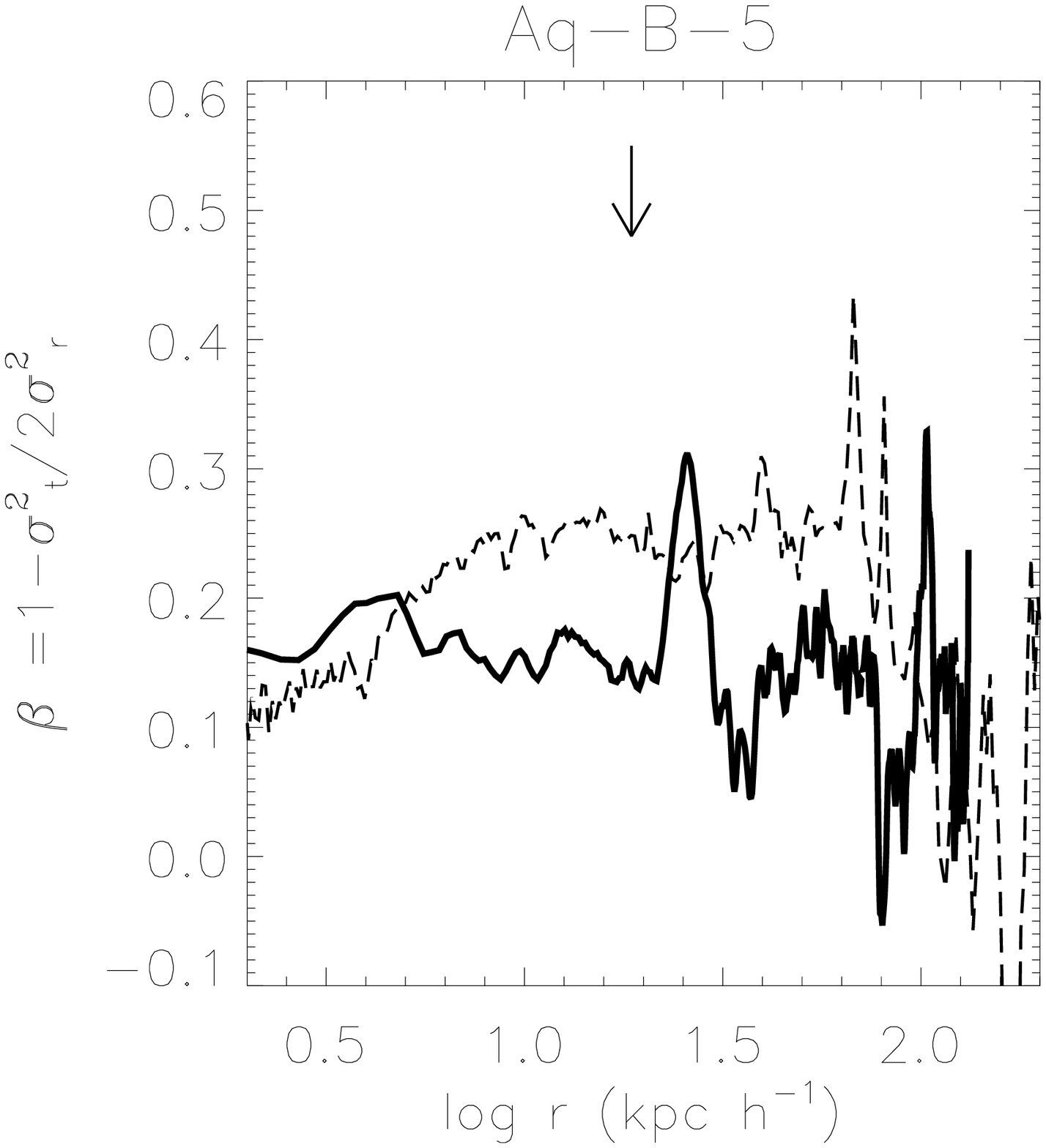}}
\resizebox{5cm}{!}{\includegraphics{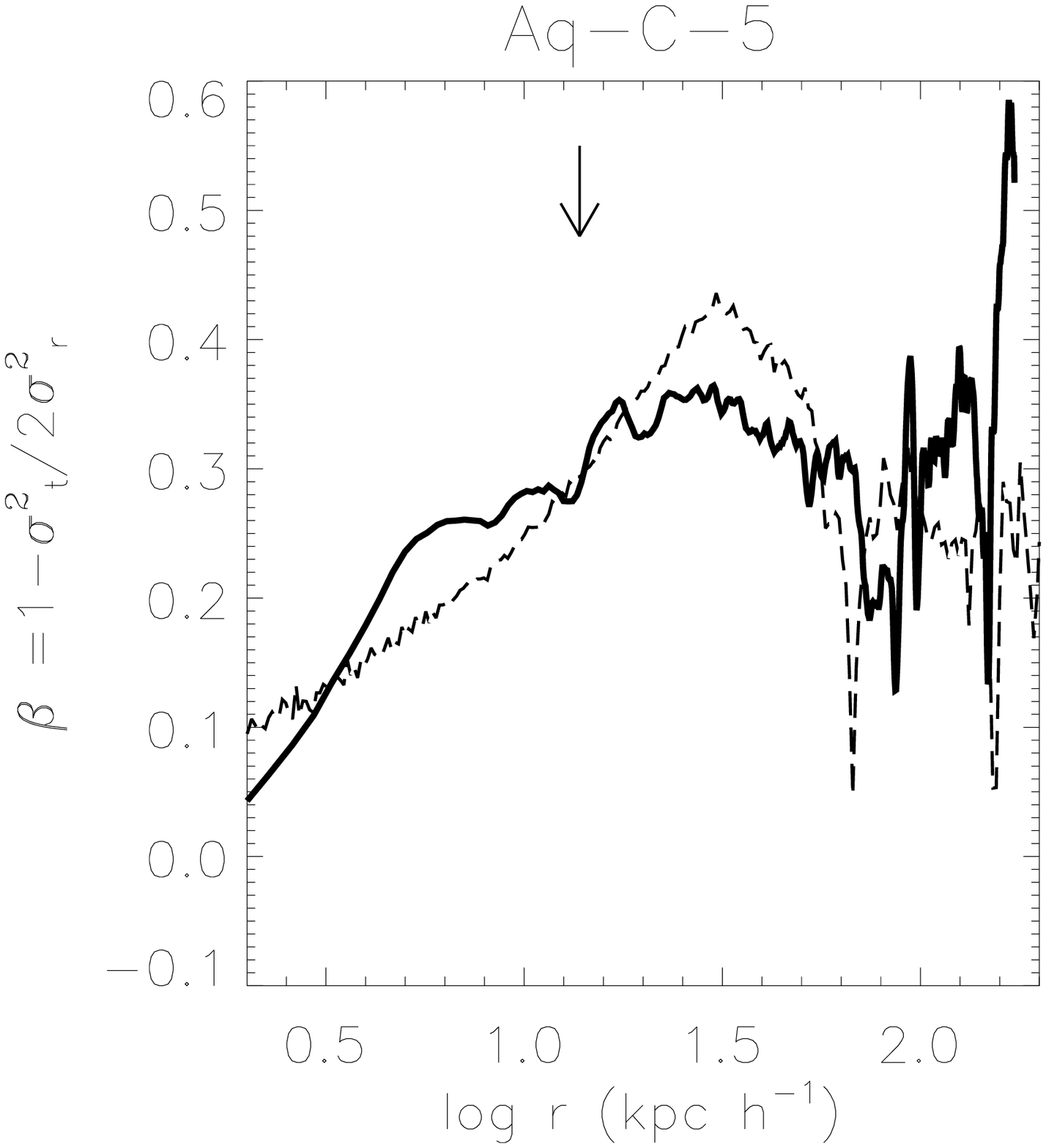}}\\
\resizebox{5cm}{!}{\includegraphics{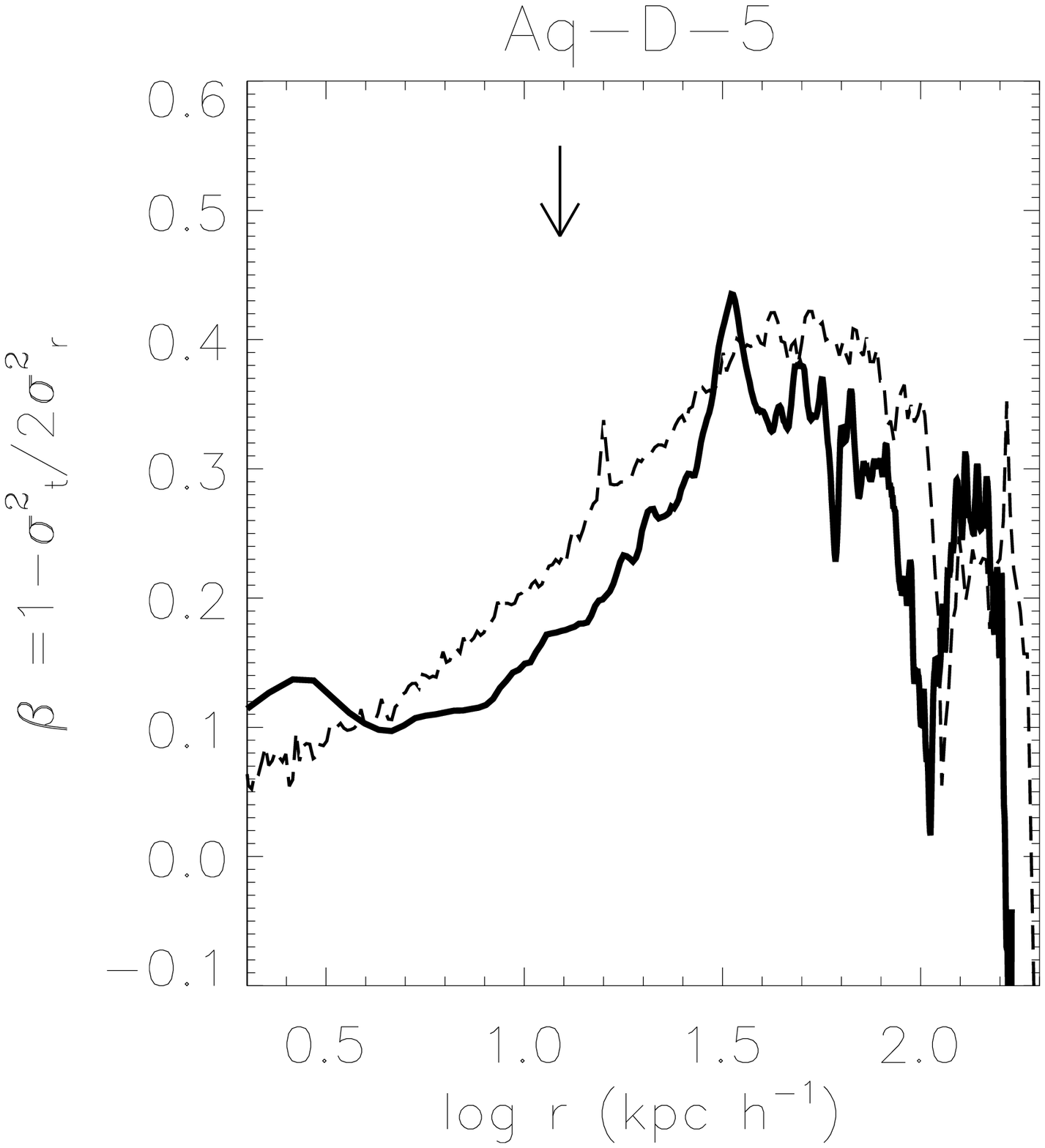}}
\resizebox{5cm}{!}{\includegraphics{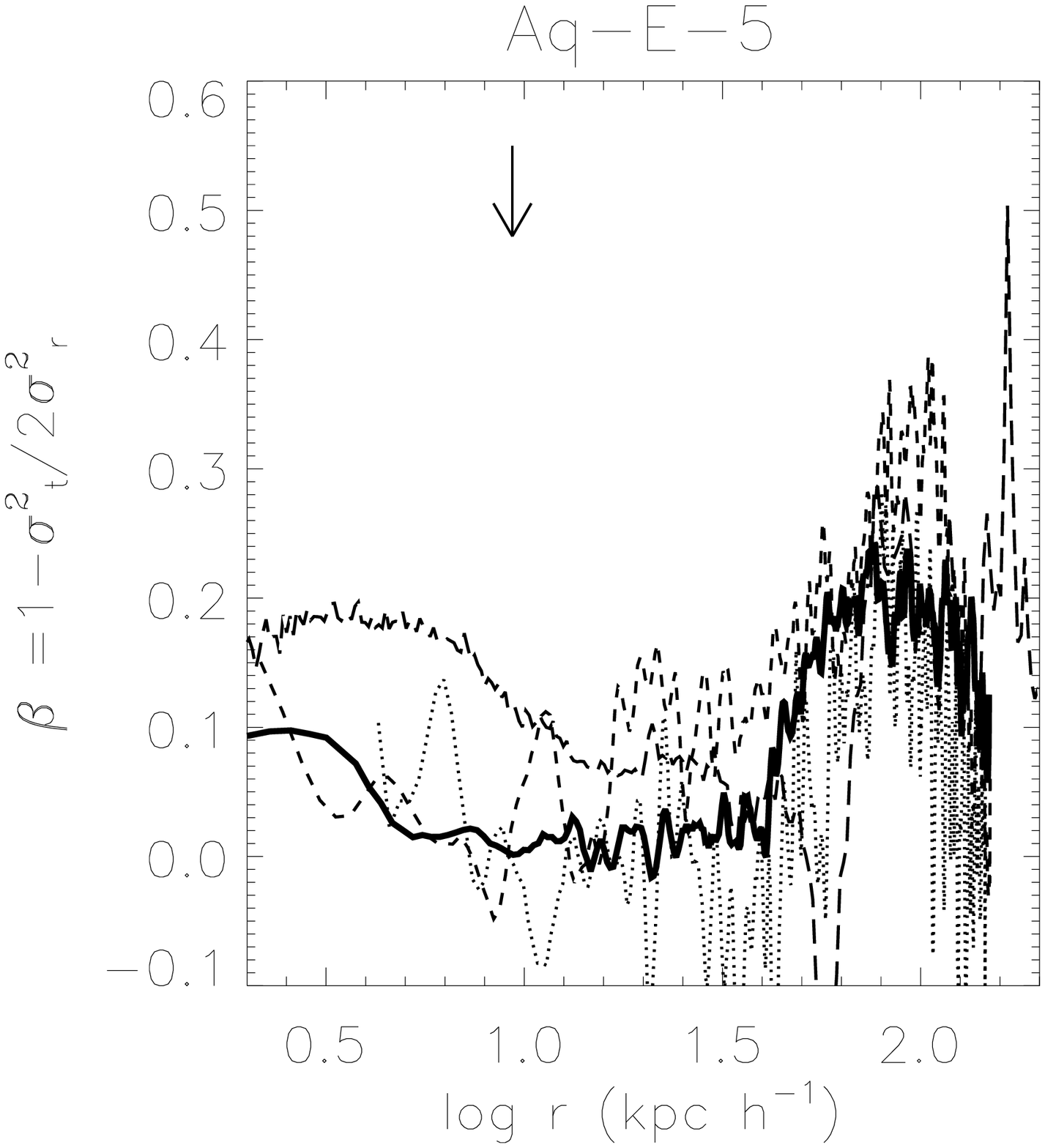}}
\resizebox{5cm}{!}{\includegraphics{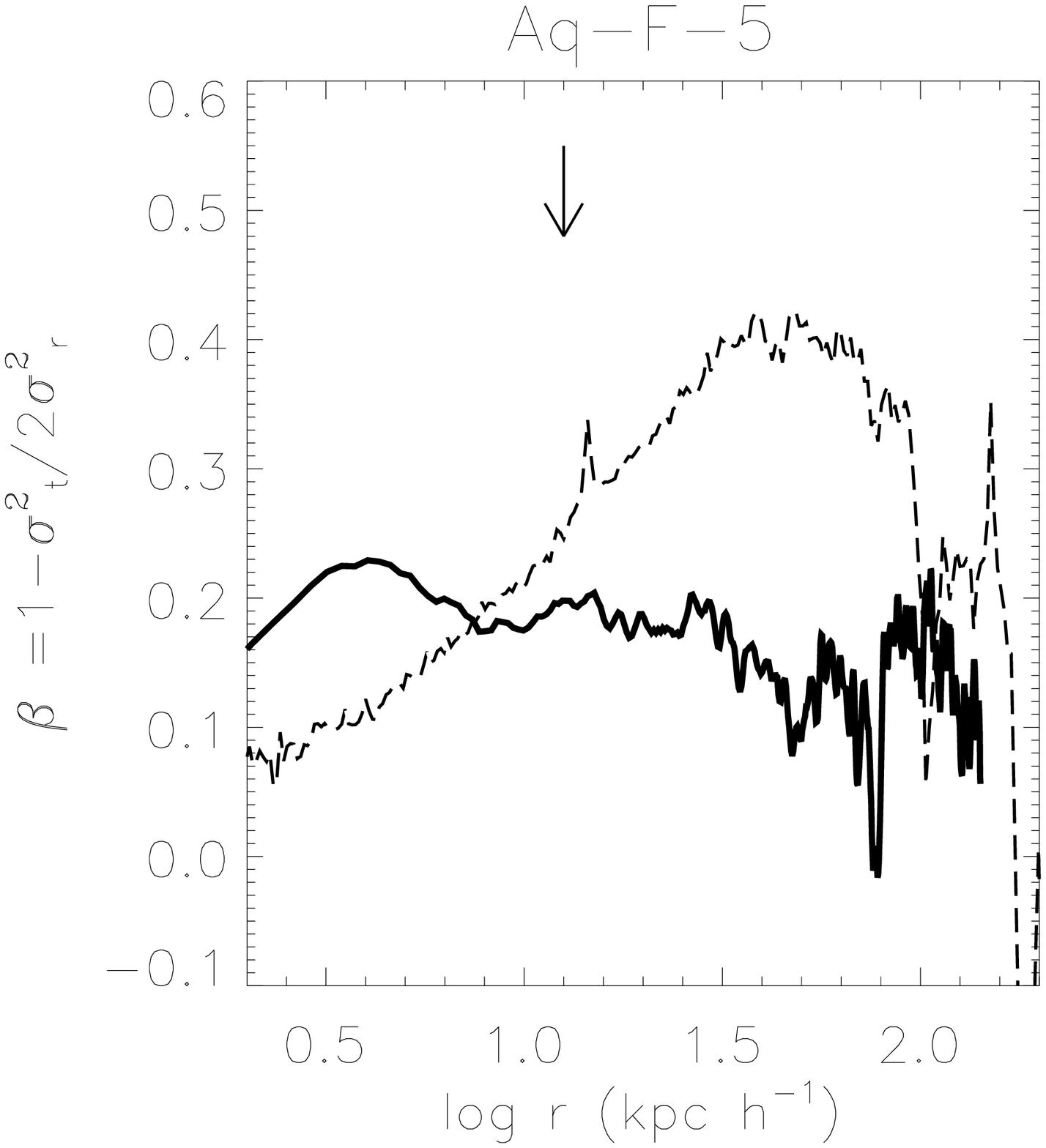}}
\hspace*{-0.2cm}
\caption{Anisotropy parameter ($\beta$) as a function of radius for the SPH
  (solid lines) and DM (dashed lines) runs. The arrows indicate the baryonic
  radii of the central galaxies.  We include the lower resolution versions of Aq-E: Aq-E-6 (short dashed lines) and Aq-E-7 (dotted lines).}
\label{betas}
\end{figure*}

\begin{figure*}
\resizebox{7.5cm}{!}{\includegraphics{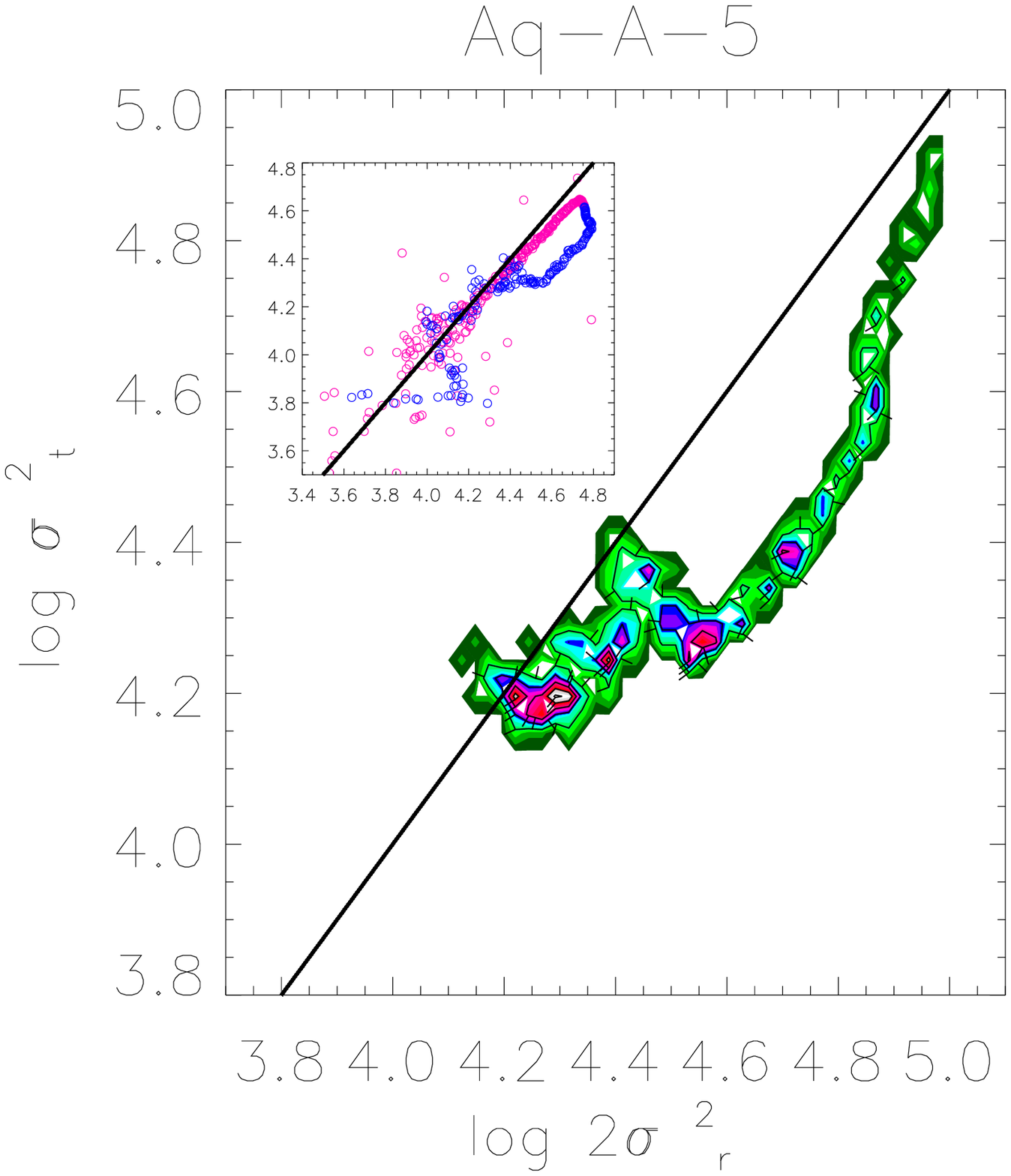}}
\resizebox{7.5cm}{!}{\includegraphics{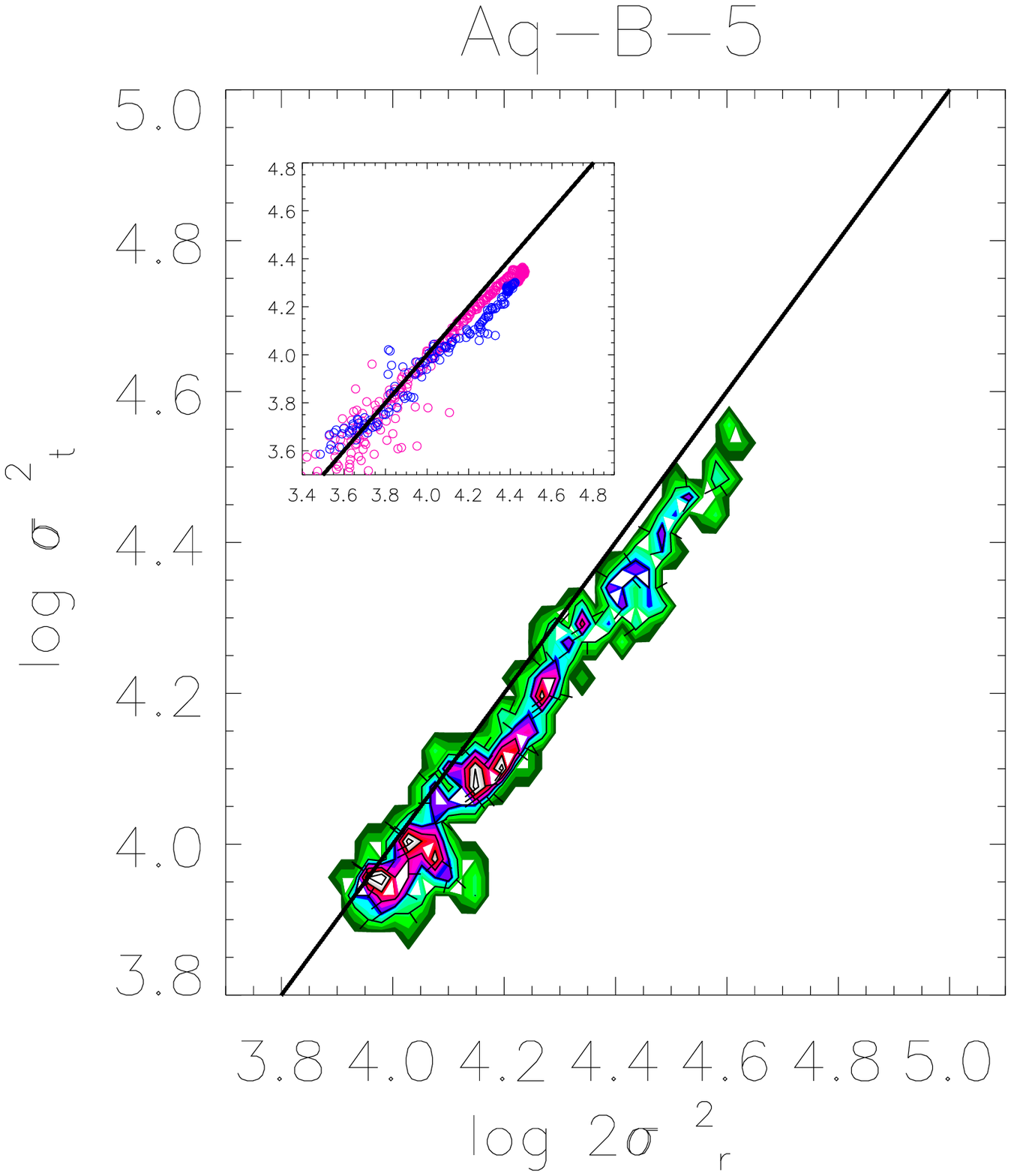}}\\
\resizebox{7.5cm}{!}{\includegraphics{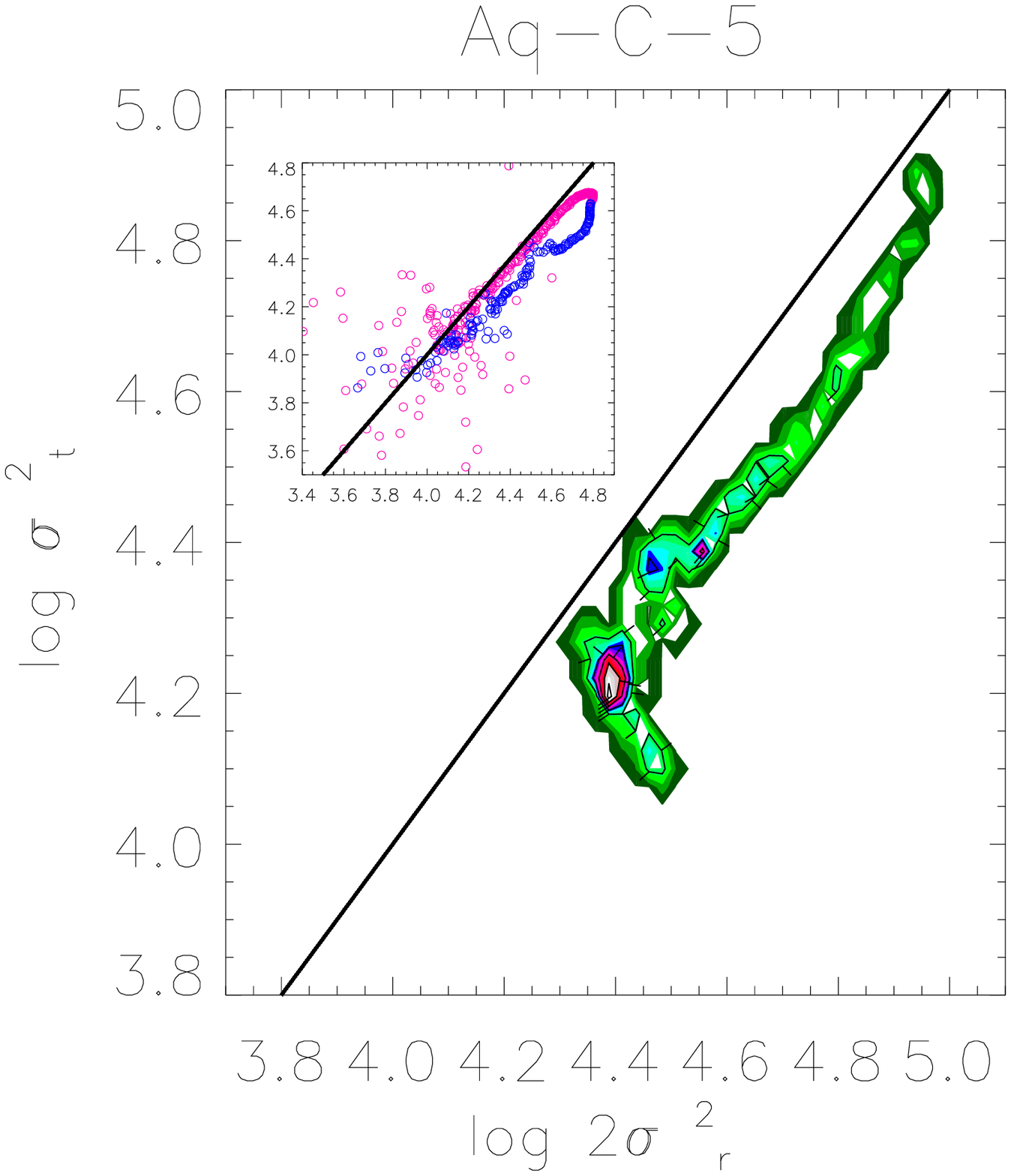}}
\resizebox{7.5cm}{!}{\includegraphics{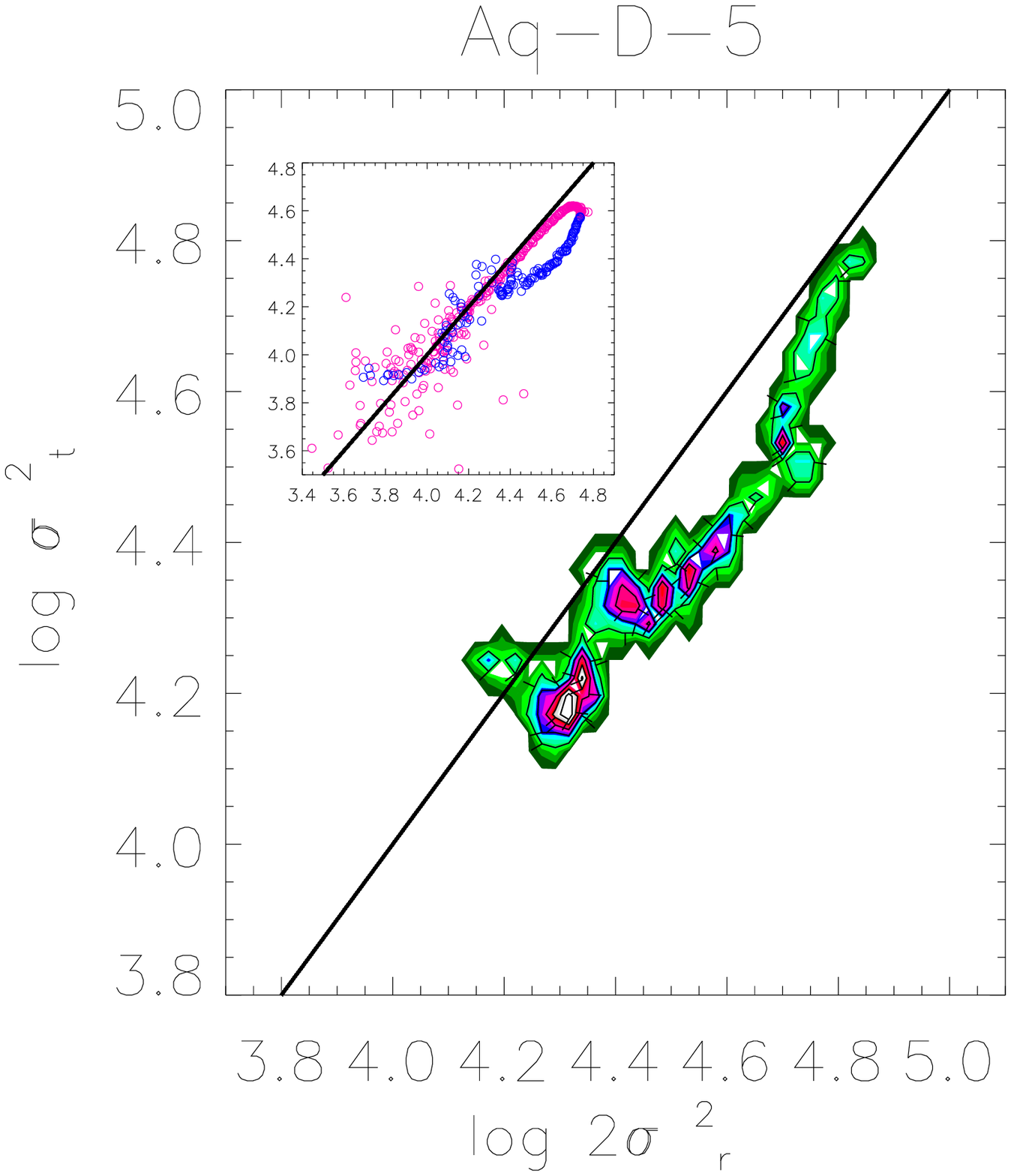}}\\
\resizebox{7.5cm}{!}{\includegraphics{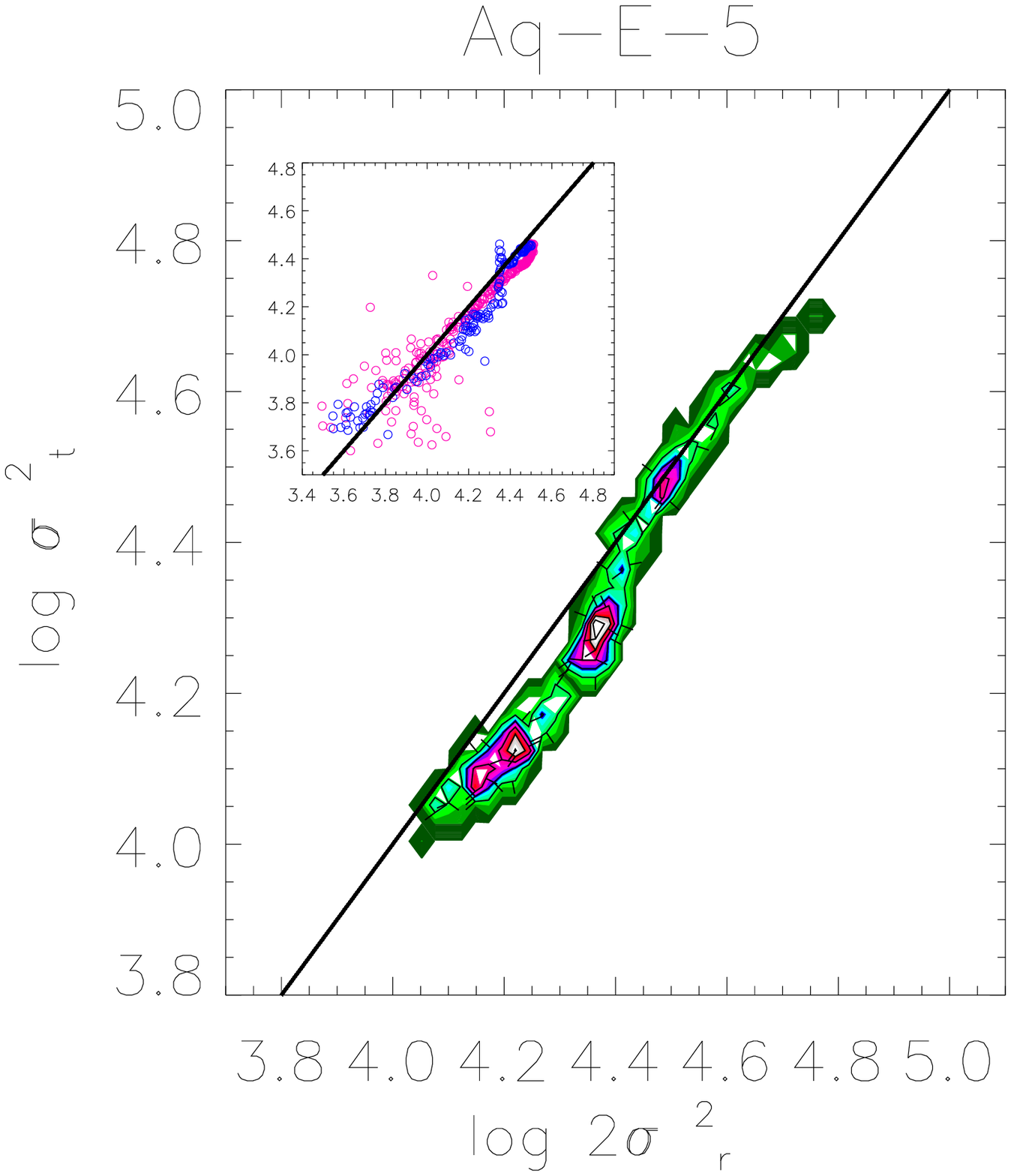}}
\resizebox{7.5cm}{!}{\includegraphics{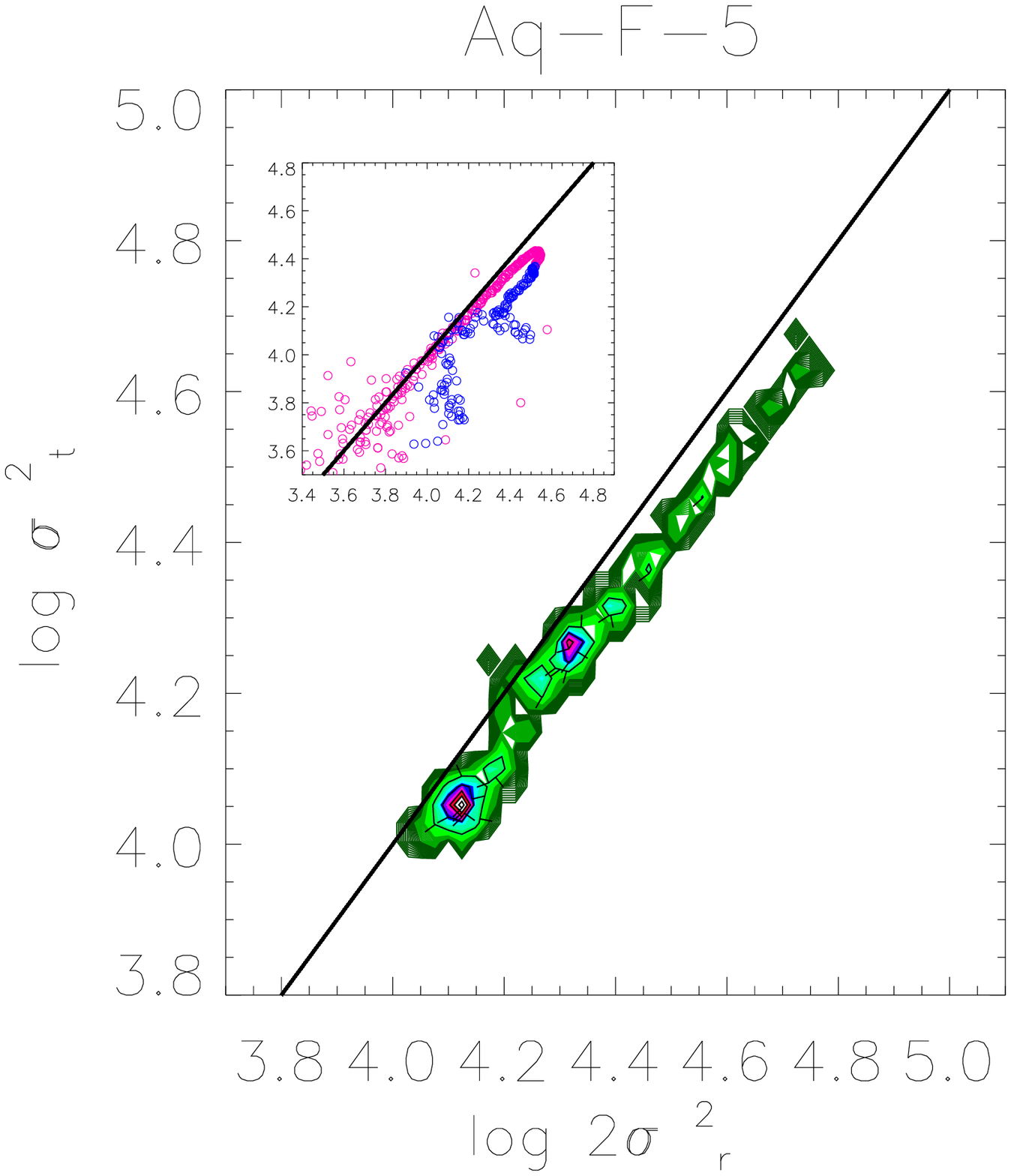}}
\hspace*{-0.2cm}
\caption{Tangential dispersion versus radial dispersion for the SPH
  haloes (velocities are in units of ${\rm km \ s^{-1}}$). The corresponding relations for the DM haloes of N08 are shown in
  magenta for $ 2\ h^{-1}{\rm kpc} < r < r_{-2}$ and in blue for $ r_{-2} < r
  < r_{200}$ in the insets. The solid black line depicts equality.  The SPH
  haloes show a monotonic relation with radius for both components with the
  higher dispersion values for smaller radius as shown in Fig.~\ref{sigmas}.}
\label{scatter}
\end{figure*}

The conventional velocity anisotropy parameter, defined as $\beta(r) \equiv 1-
{\sigma_{t}^2/(2\sigma^2_{r})}$ where $\sigma_{t}$ and $\sigma_{r}$ are
tangential and radial dispersions averaged over spherical shells, gives some
indication of the internal orbital structure of the haloes.  As shown by N08,
the DM runs are almost isotropic ($\beta\approx 0$) in their inner regions,
then in most cases become progressively more radially biased ($\beta > 0$)
with increasing radius out to about $r_{-2}$ before becoming more isotropic
again at even larger radii.  There are however, substantial differences
between the individual haloes.  We find that baryonic effects modify the
dispersion structure within these haloes in a complicated way, as can be seen
in Fig.~\ref{betas}. While some of the SPH haloes have similar anisotropy
structure to their DM counterparts (Aq-A-5, Aq-C-5 and Aq-D-5), others have
become less dominated by radial motions (Aq-F-5, Aq-B-5 and Aq-E-5).  All our  haloes
remain slightly radially biased, $\beta \approx 0.1-0.2 $,  in the inner resolved regions.
 The lower resolution versions of Eq-E (Aq-E-6 and Aq-E-7) 
 show a large level of 
noise around the trend determined by   the highest resolution run (Aq-E-5).  
In order to get further insight into the effects of baryons, we plot
$\sigma_{t}^{2}$ against $\sigma_{r}^{2}$ in Fig.~\ref{scatter} both for the
SPH runs and for their DM counterparts.  For the DM runs, nearly isotropic
behaviour is evident in the cool central regions ($r < r_{-2}$; magenta
circles, radius increasing with dispersion) with an inversion of the relation
in the outer regions where radial motions dominate (blue circles radius
increasing with decreasing dispersion).  Baryon condensation (either by infall
or mergers) increases both radial and tangential dispersions and changes the
shapes of orbits in the central regions.  (Note that the axis ranges of the
main plots and of their insets are different).  However, each halo has its own
particular distribution.  Haloes Aq-B-5, Aq-E-5 and Aq-F-5, which have
approximately constant velocity anisotropy, are also those that have an almost
isothermal density profile to radii larger than the baryonic radius of their
central galaxy, as is evident from the insets of Fig. ~\ref{haloes}.

\begin{figure*}
\resizebox{5cm}{!}{\includegraphics{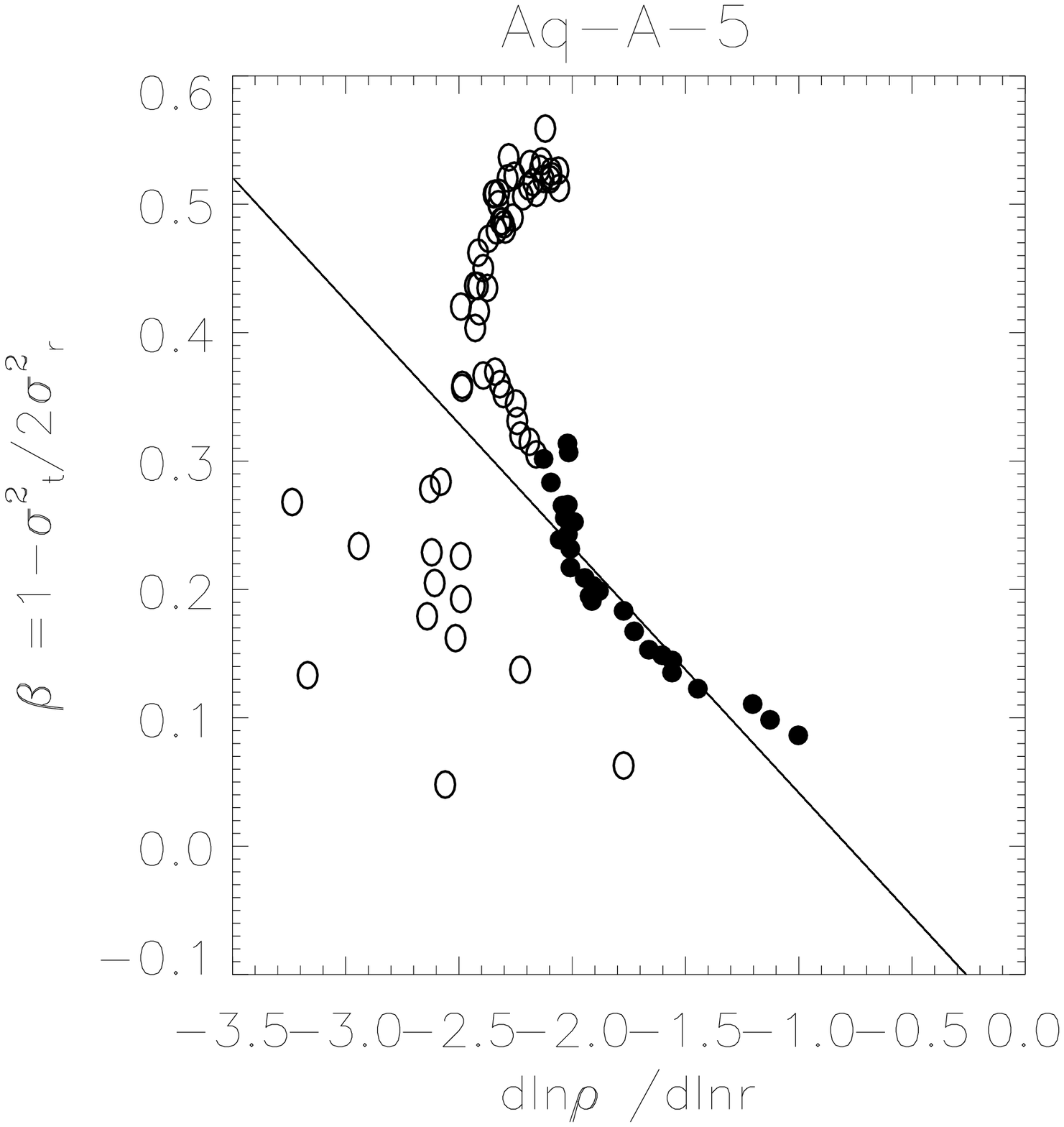}}
\resizebox{5cm}{!}{\includegraphics{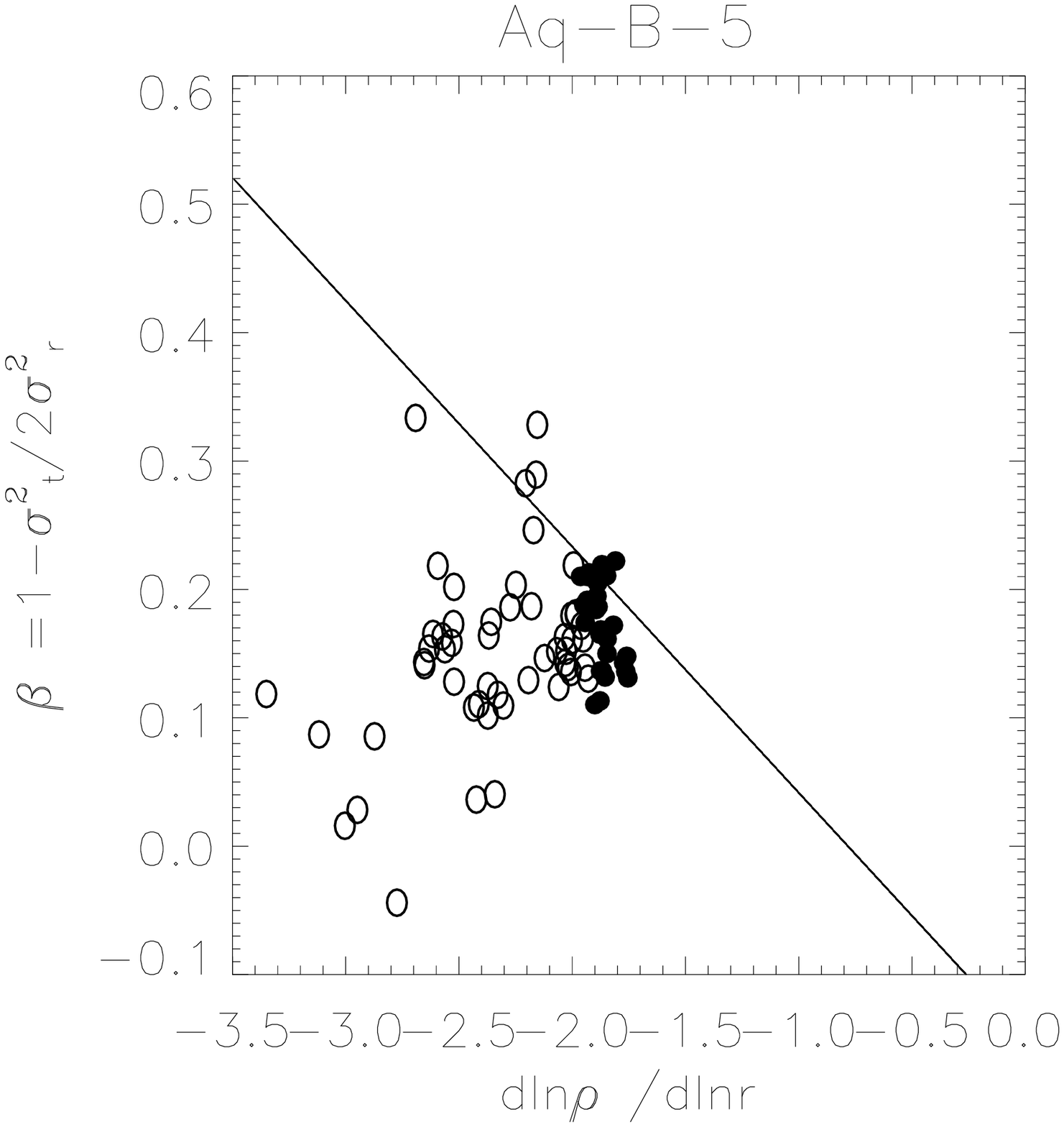}}
\resizebox{5cm}{!}{\includegraphics{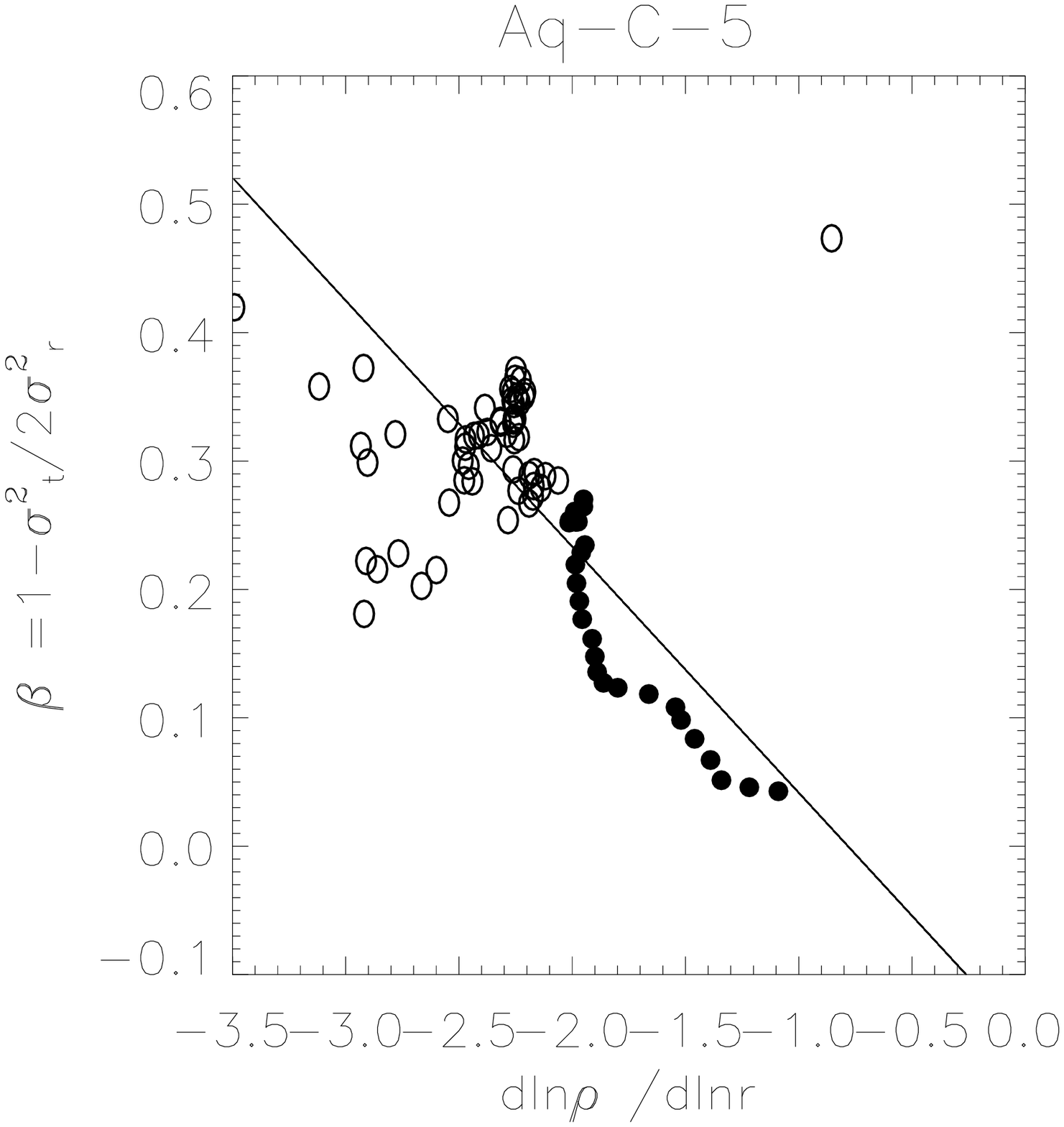}}\\
\resizebox{5cm}{!}{\includegraphics{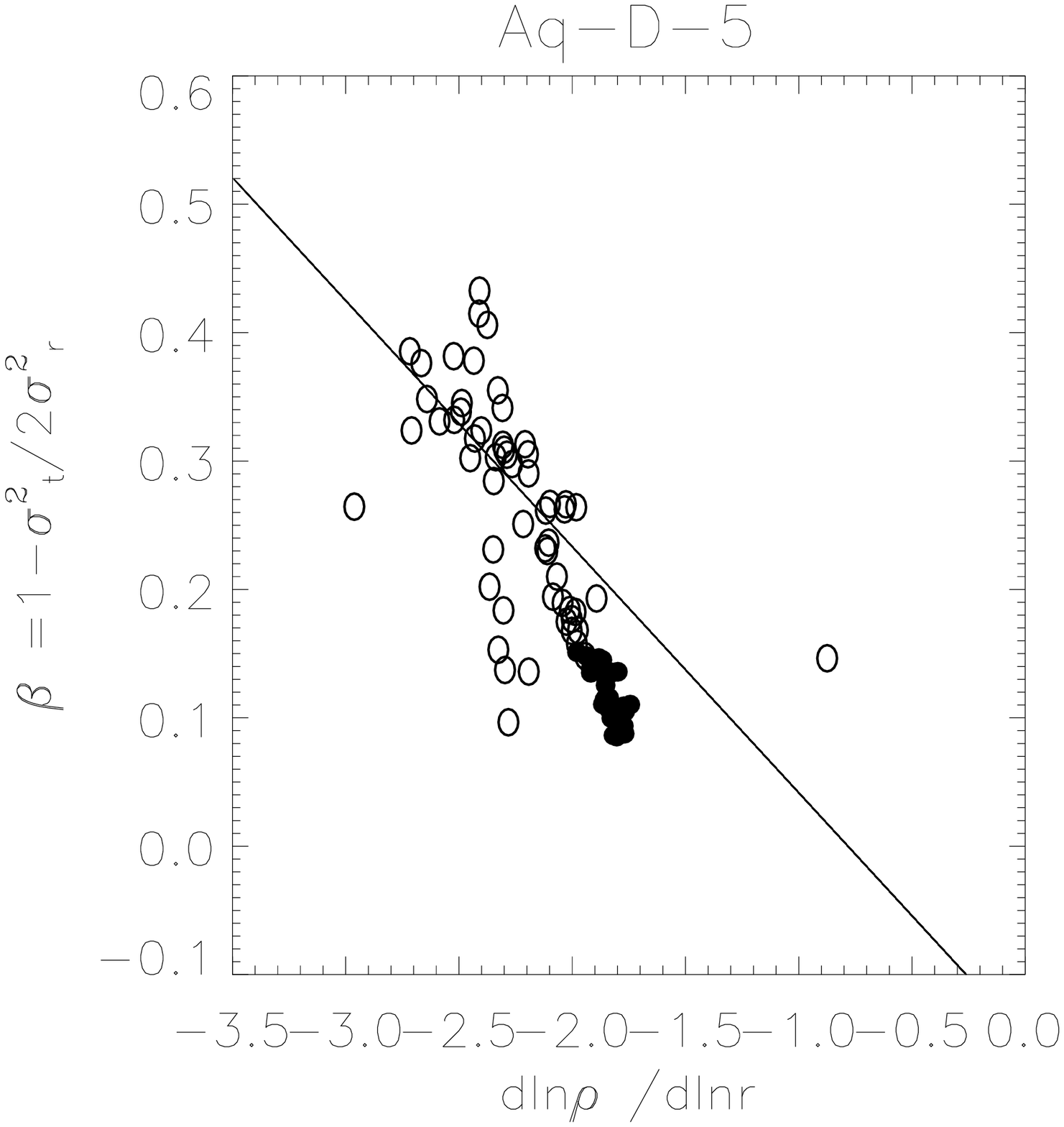}}
\resizebox{5cm}{!}{\includegraphics{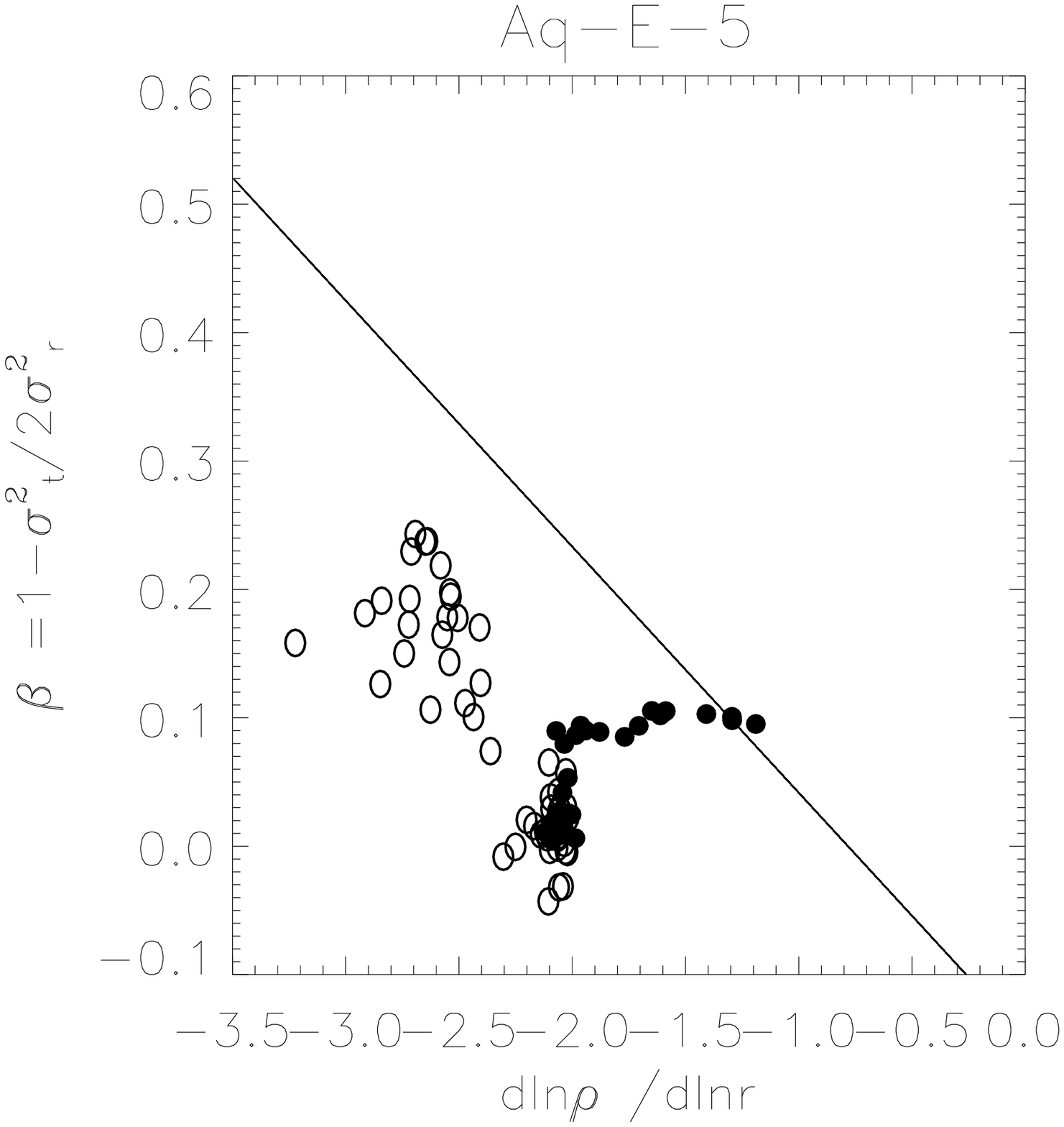}}
\resizebox{5cm}{!}{\includegraphics{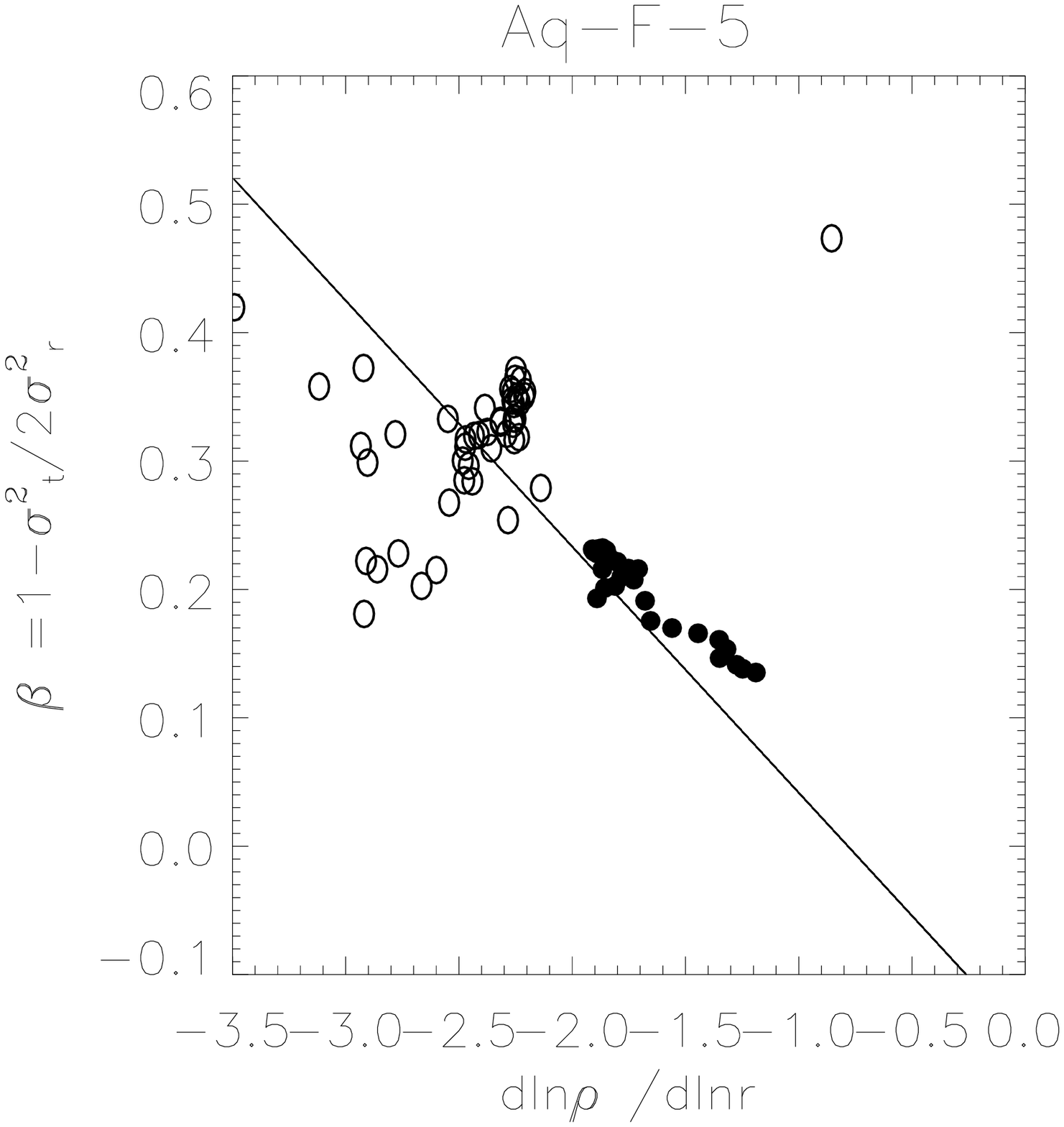}}
\hspace*{-0.2cm}
\caption{Velocity anisotropy $\beta$ as a function of the local value $\gamma$
  of the logarithmic slope of the density profile in the SPH runs. Filled
  symbols correspond the radial range $ 2\ h^{-1}{\rm kpc} < r < r_{-2}$ and
  open circles to the range $ r_{-2} < r < r_{200}$. The Hansen \& Moore
  (2006) relation in shown as a solid line. }
\label{moore}
\end{figure*}

Finally, we look at the relation between the logarithmic slope of the density
profile ($\gamma(r)$) and the velocity anisotropy parameter ($\beta(r)$).
Hansen \& Moore (2006) suggested that there might be a ``universal'' relation
between these parameters in dissipationaless haloes.  N08 found that their DM
haloes follow a well-defined relation only in the central regions where both
parameters are monotonic functions of $r$.  In Fig. ~\ref{moore} we show the
anisotropy-slope relation for our SPH runs. As can be seen, there are two
different behaviours. Three  of the haloes (Aq-A-5, Aq-C-5 and Aq-F-5) show
similar and well-defined relations over the rough range $ 2\ h^{-1}{\rm kpc} <
r < r_{-2}$, which are fairly close to the HM06 relation. However, for $r >
r_{-2}$, each halo behaves differently. In general, they show large variations
in $\beta$ (which differ from halo to halo) but relatively minor variations in
$\gamma$ (compare with Fig. ~\ref{haloes}).  For the other haloes (Aq-B-5,
Aq-D-5 and  Aq-E-5) the logarithmic slope shows only a weak dependence on
radius in the region dominated by baryons, consistent with a nearly isothermal
profile, while  the anisotropy varies from  $ \sim  0.1$ to $ \sim 0.4$.
As a result the HM06 relation fails dramatically in these cases.


\section{The phase-space density profile}

Taylor \& Navarro (2001) noticed that the quantity $Q(r)=\rho(r)/\sigma(r)^3$
is well approximated by a power law $Q(r) \propto r^{-\alpha}$ with $\alpha
\approx 1.875$ over more than 2.5 orders of magnitude in radius in simulated
dark matter haloes of any mass in a CDM cosmology (see also, Rasia, Tormen \& Moscardini 2004; Ascsibar et al. 2004; Vass et al. 2008, N08). This is the behaviour
predicted in the spherical similarity solution of Bertschinger (1985). This
$Q(r)$ is related to the entropy distribution within haloes, but the origin of
its power law behaviour and its relation to structure formation are not yet
understood. The apparent universality is established only for pure dark matter
haloes and may be broken by the effects of baryons which modify both the
velocity dispersion and the density profiles of the dark matter.

\begin{figure*}
\resizebox{5cm}{!}{\includegraphics{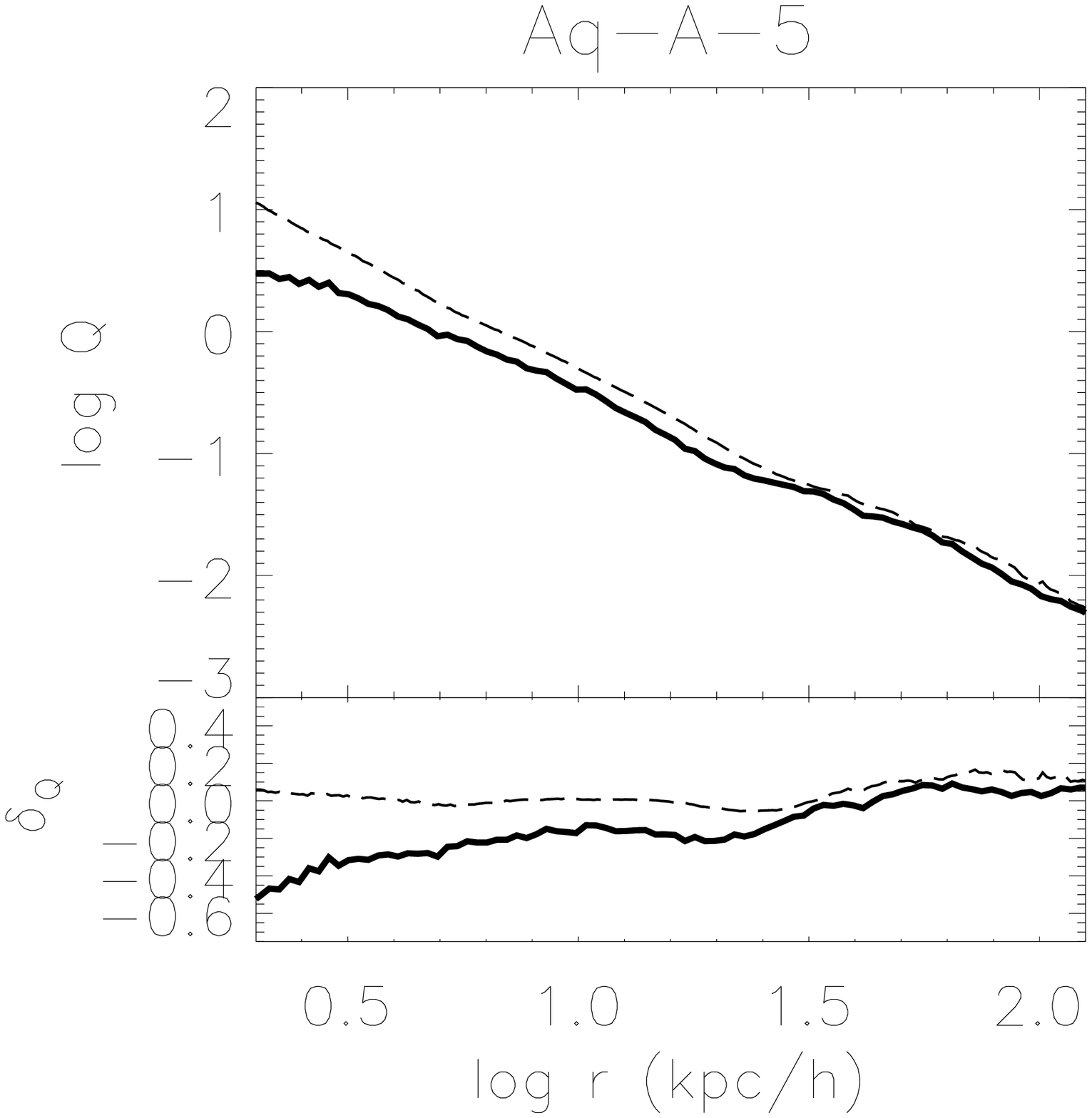}}
\resizebox{5cm}{!}{\includegraphics{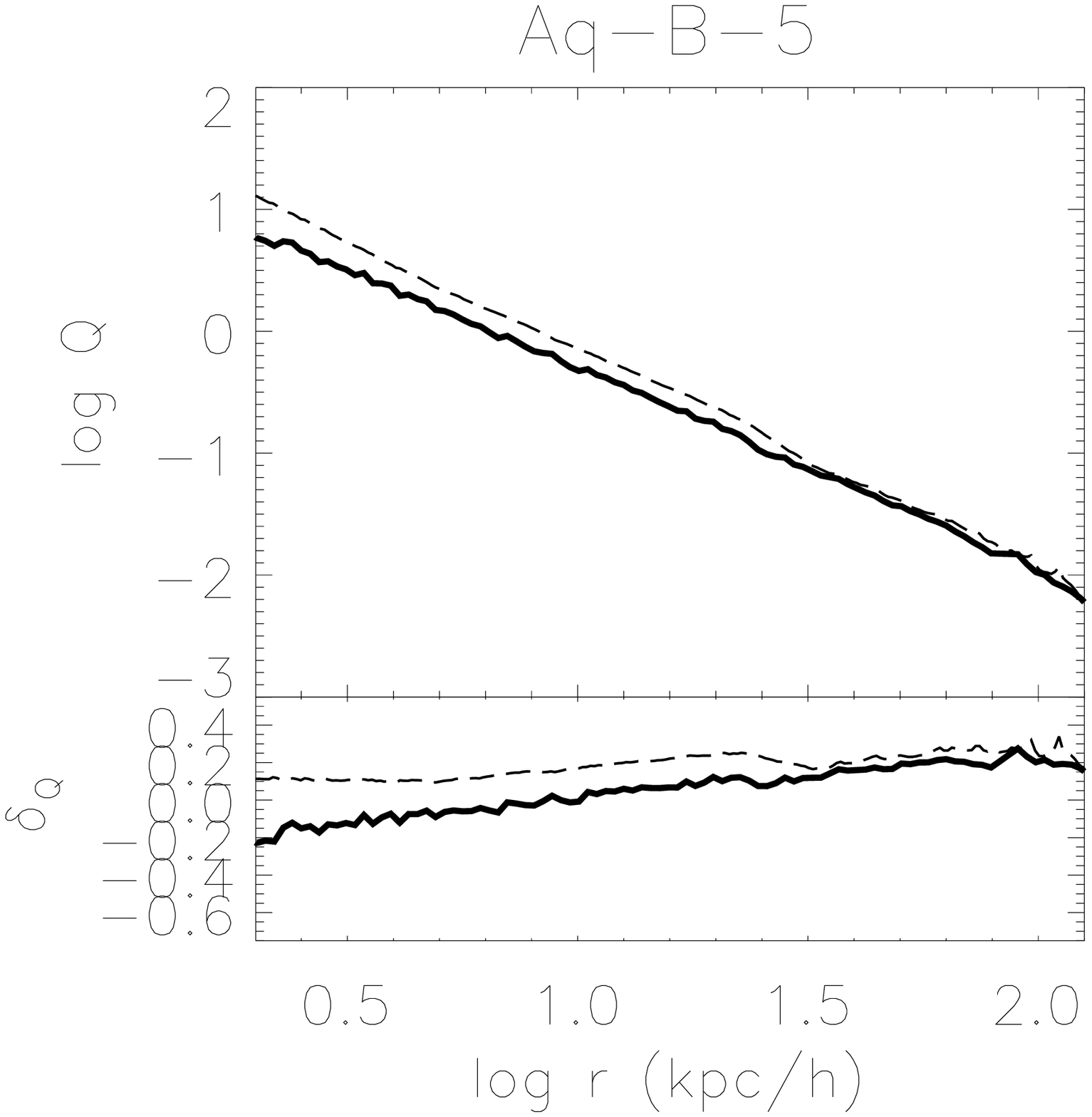}}
\resizebox{5cm}{!}{\includegraphics{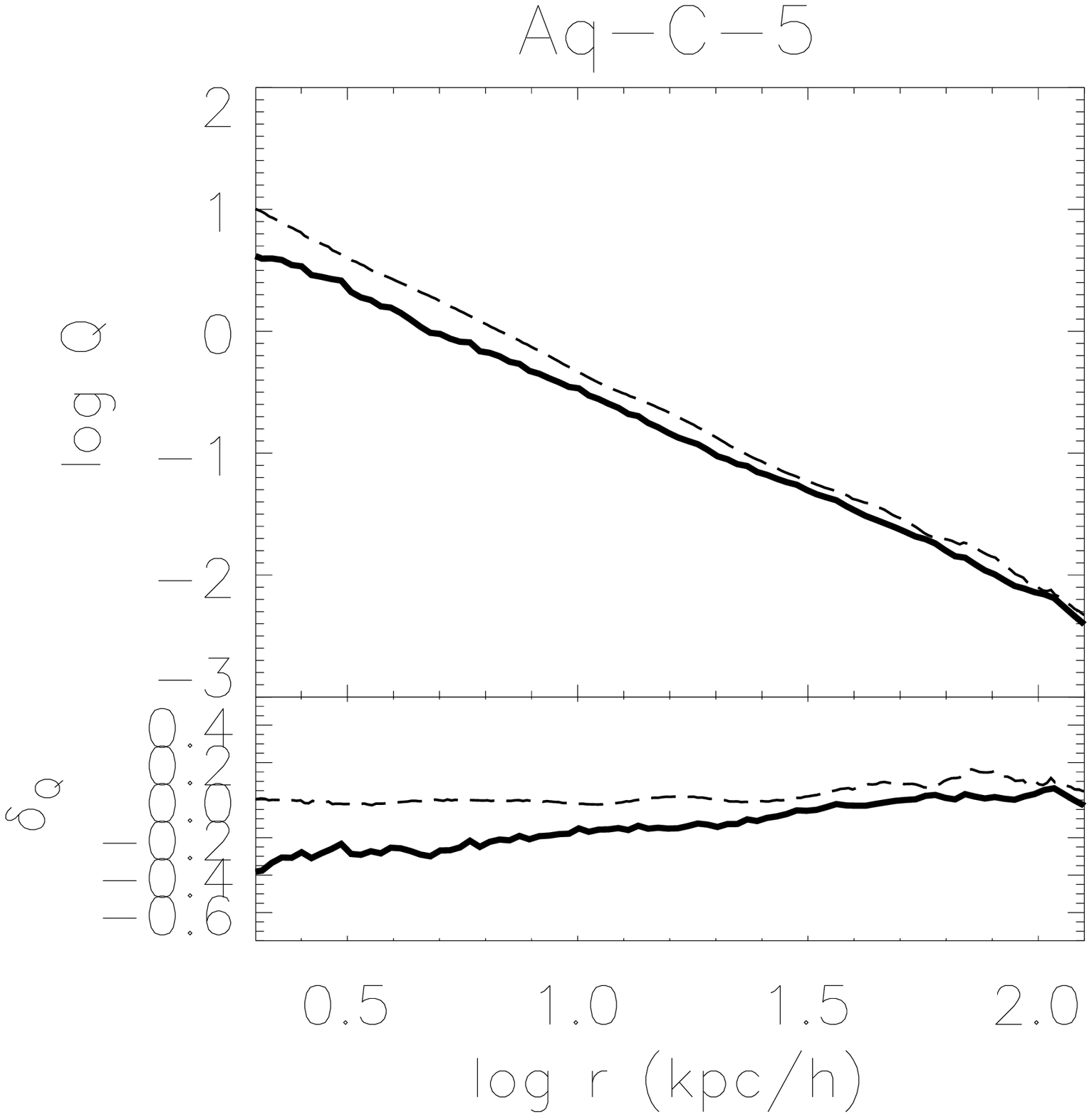}}\\
\resizebox{5cm}{!}{\includegraphics{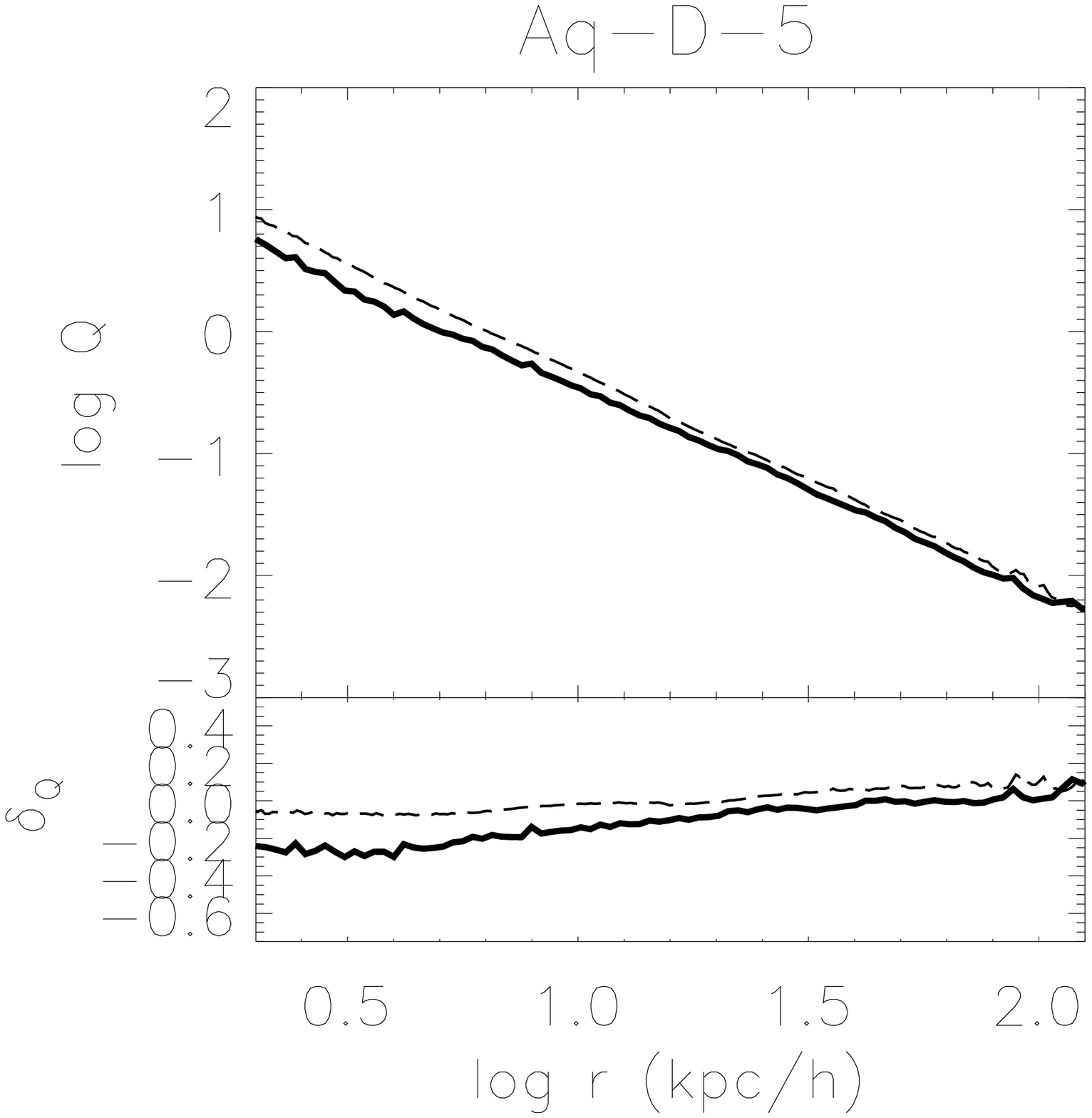}}
\resizebox{5cm}{!}{\includegraphics{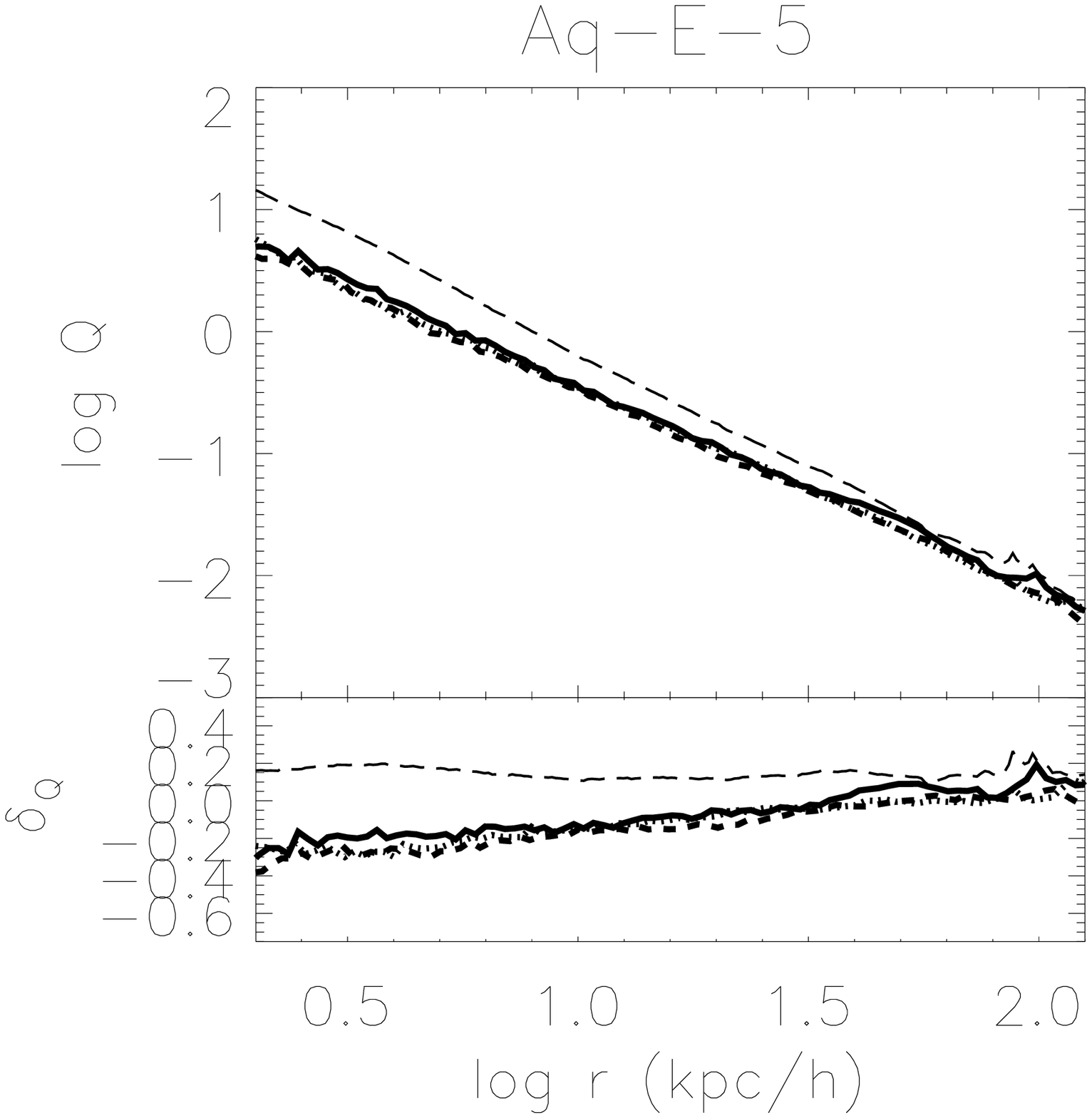}}
\resizebox{5cm}{!}{\includegraphics{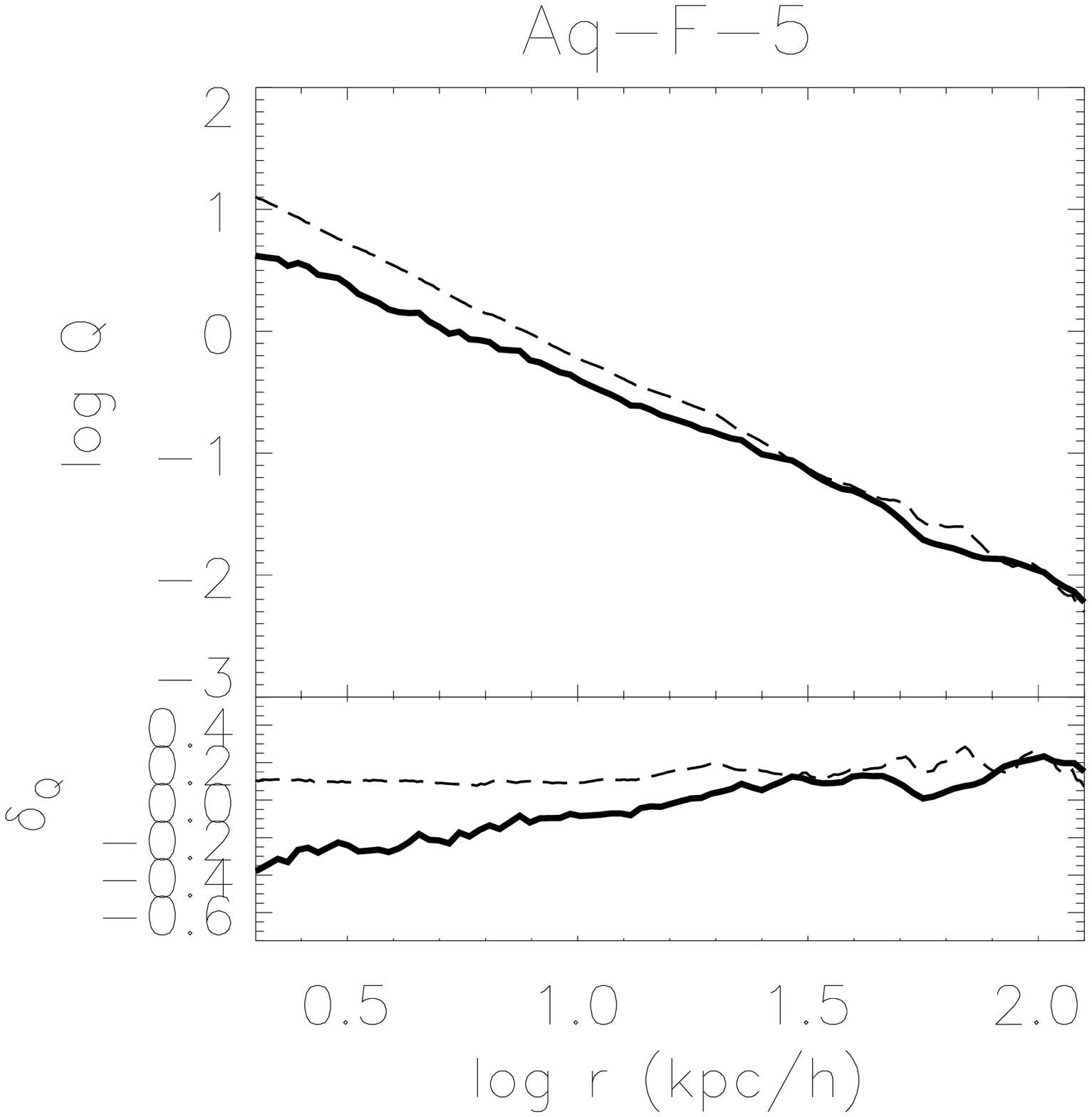}}
\hspace*{-0.2cm}
\caption{Pseudo-phase-space density profiles as a function of radius for
  haloes in the SPH (solid lines) and DM (dashed lines) runs ( Q is in units 
  of $h^{2} M_{\odot} {\rm kpc^{-3}}{\rm (km \ s^{-1})^{-3}} $). Residuals of the
  two relations from the power-law relation predicted in Bertschinger's (1985)
  similarity solution are shown in the narrow lower panels of each plot. We include the lower resolution versions of Aq-E: Aq-E-6 (short dashed lines) and Aq-E-7 (dotted lines).}
\label{bert}
\end{figure*}

As may be seen in Fig. ~\ref{bert}, where we show $Q(r)$ for both the SPH and
the DM runs, all our SPH haloes show a less steep relation than their DM
counterparts or predicted by Bertschinger's (1985) similarity solution.  The
residuals (small boxes in Fig. ~\ref{bert}) show that the modified profile is
still well fit by a power law at most radii although in a couple of cases (in
particular Aq-A-5) there is some indication for deviations in the innermost
region. In Table ~\ref{tab2} we give the best fit power-law indices for the
SPH haloes over the radial range $2\ h^{-1}{\rm kpc}< r < r_{200}$. 
Interestingly, but not surprisingly considering our previous findings, the
level of departure from the Bertschinger value depends on the halo.

\begin{figure*}
\resizebox{5cm}{!}{\includegraphics{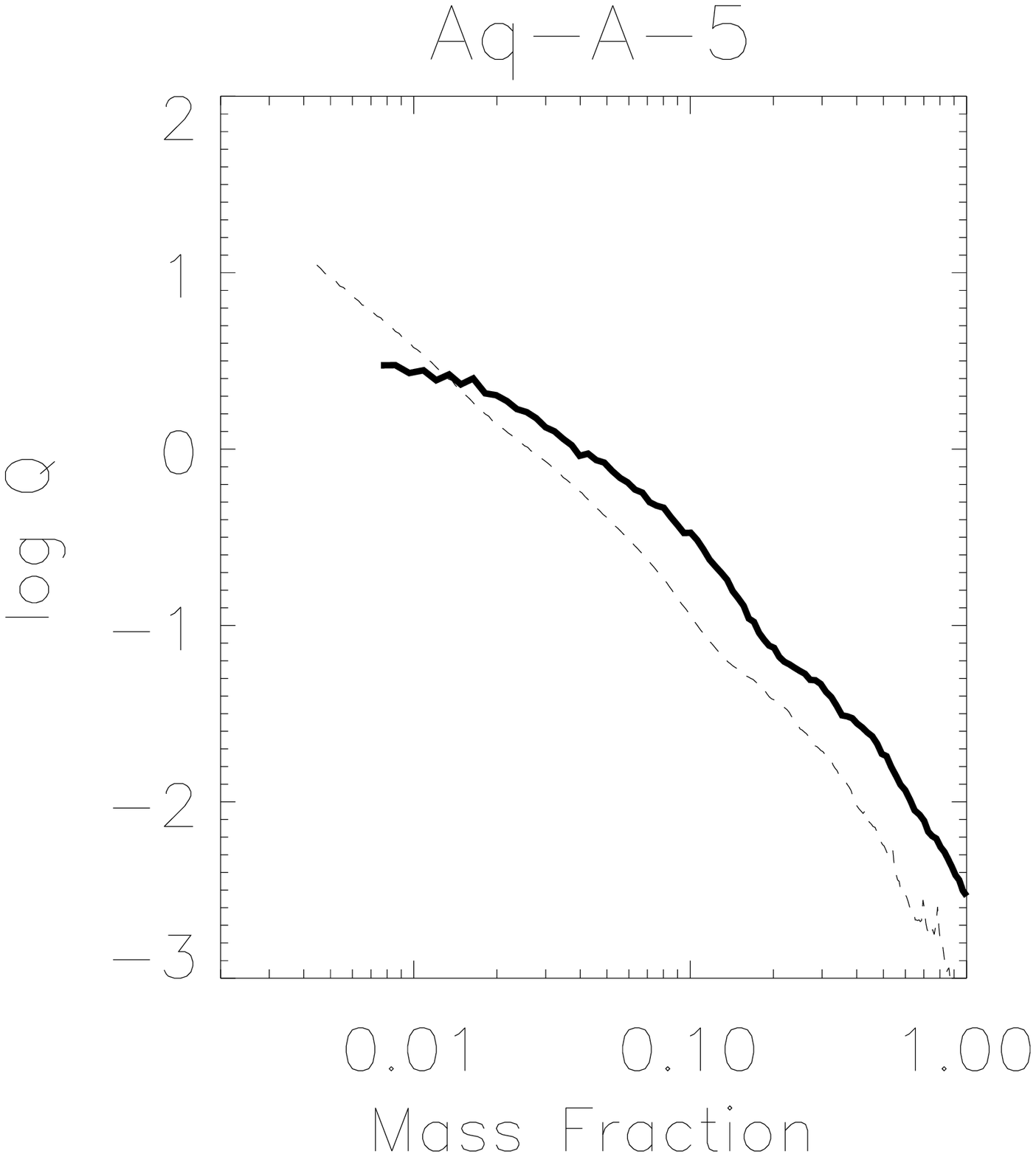}}
\resizebox{5cm}{!}{\includegraphics{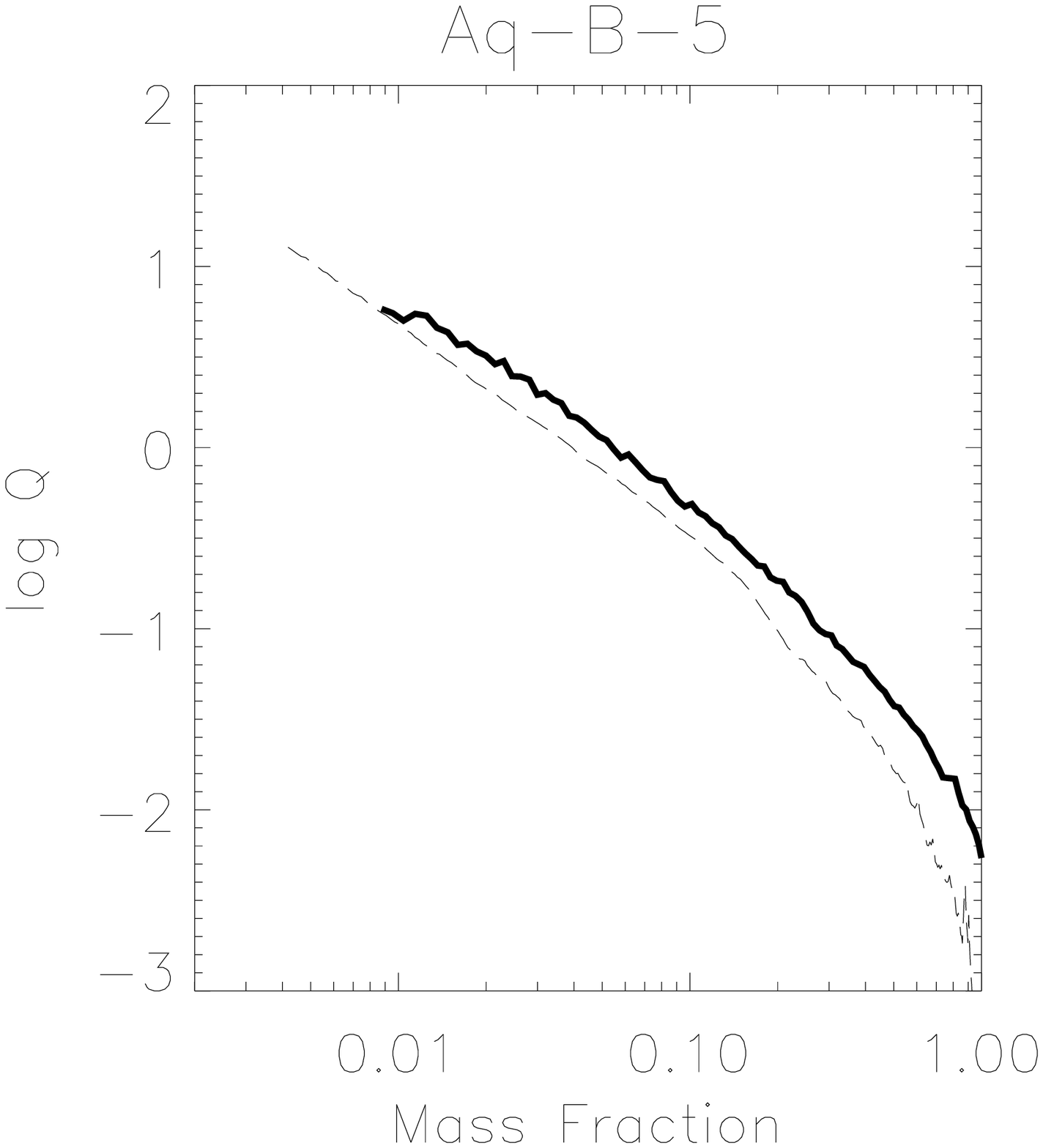}}
\resizebox{5cm}{!}{\includegraphics{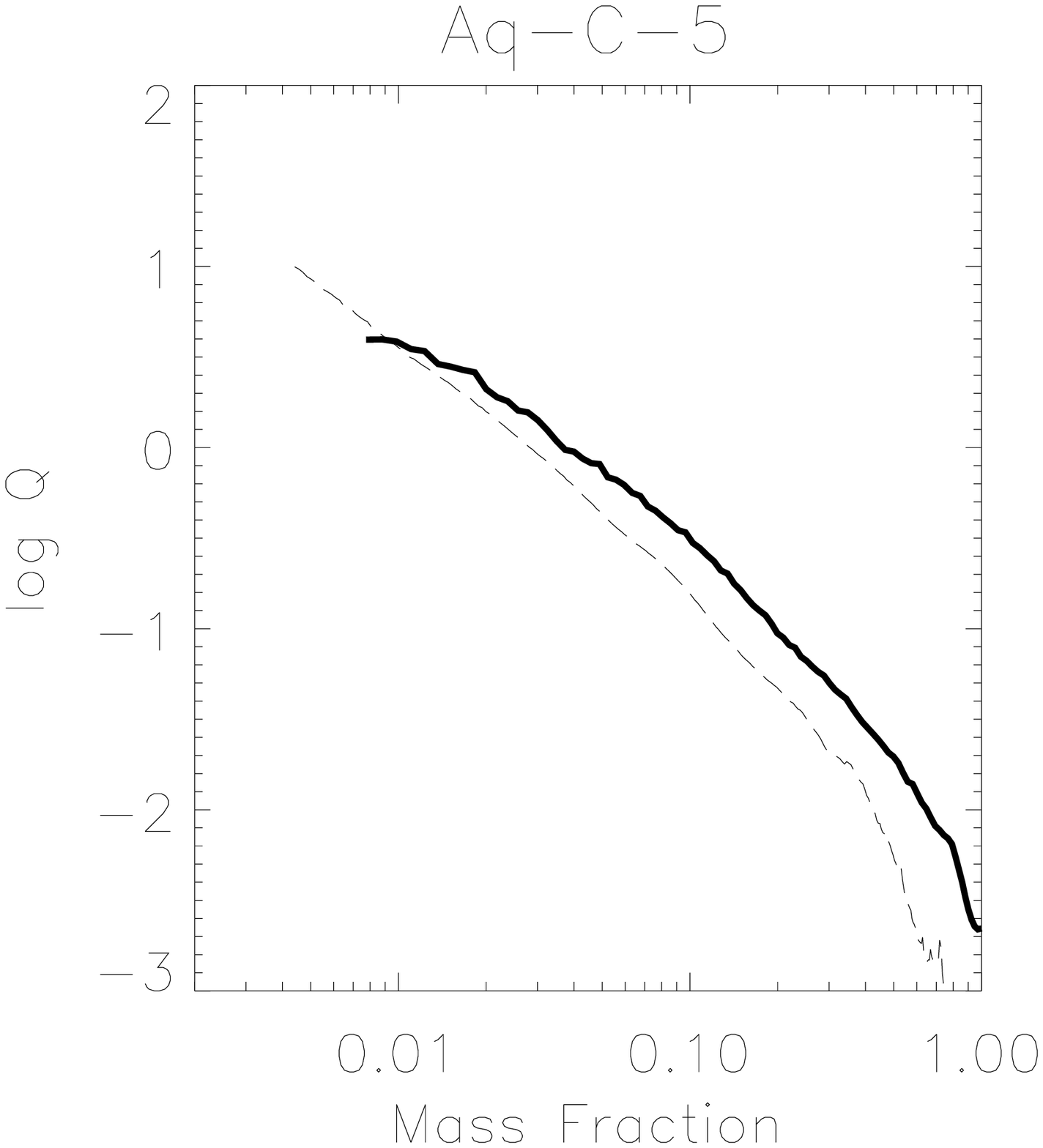}}\\
\resizebox{5cm}{!}{\includegraphics{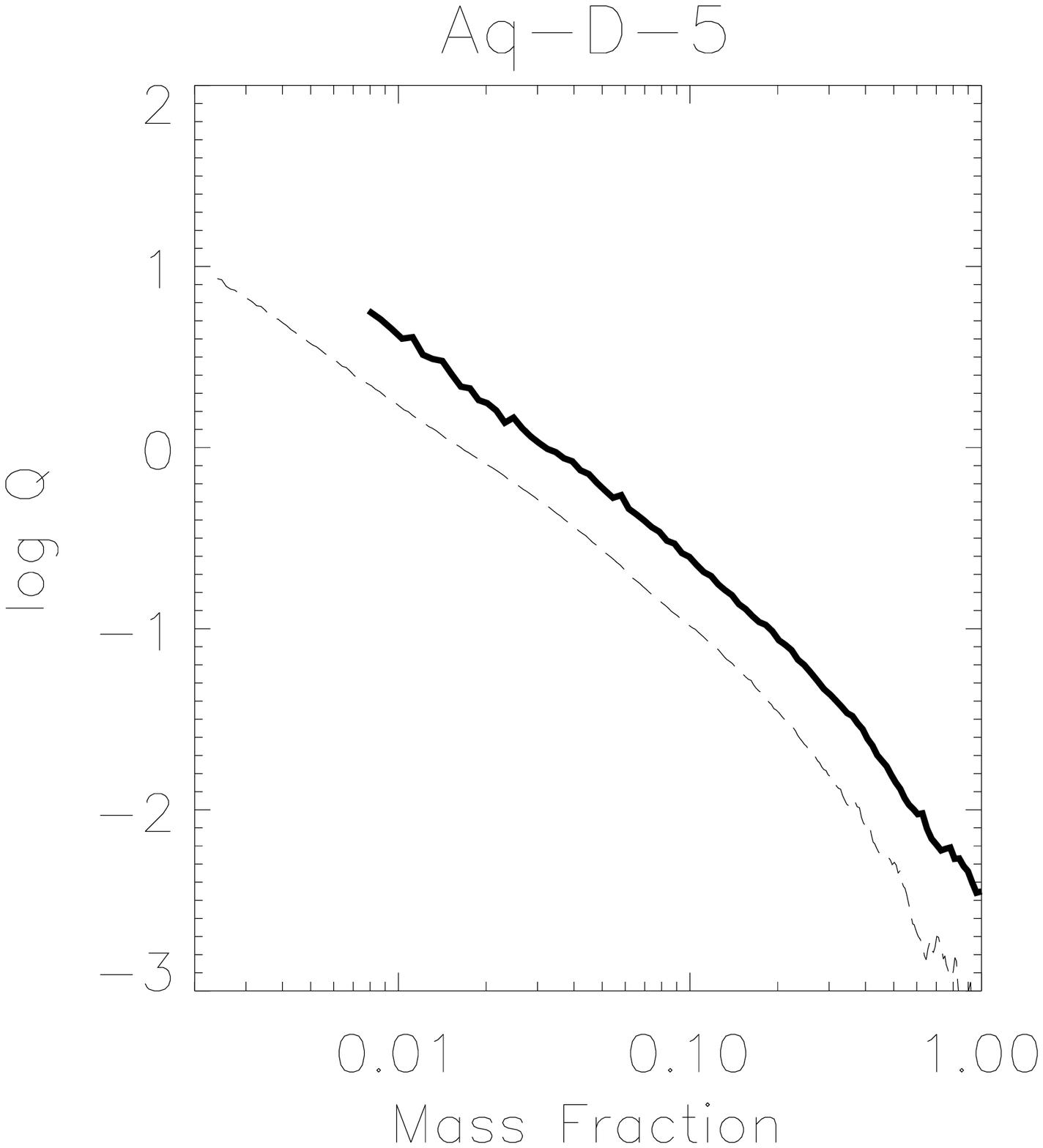}}
\resizebox{5cm}{!}{\includegraphics{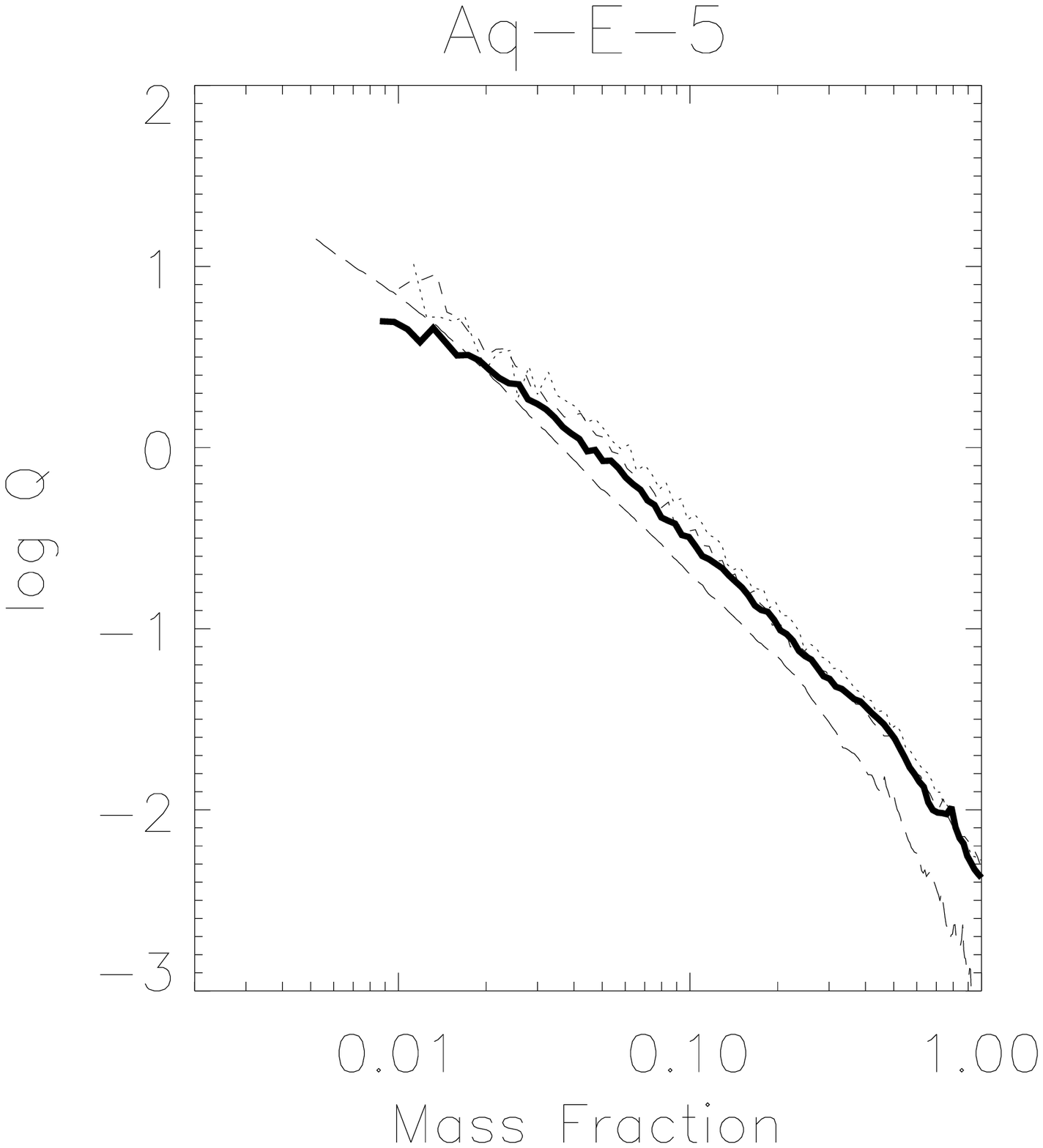}}
\resizebox{5cm}{!}{\includegraphics{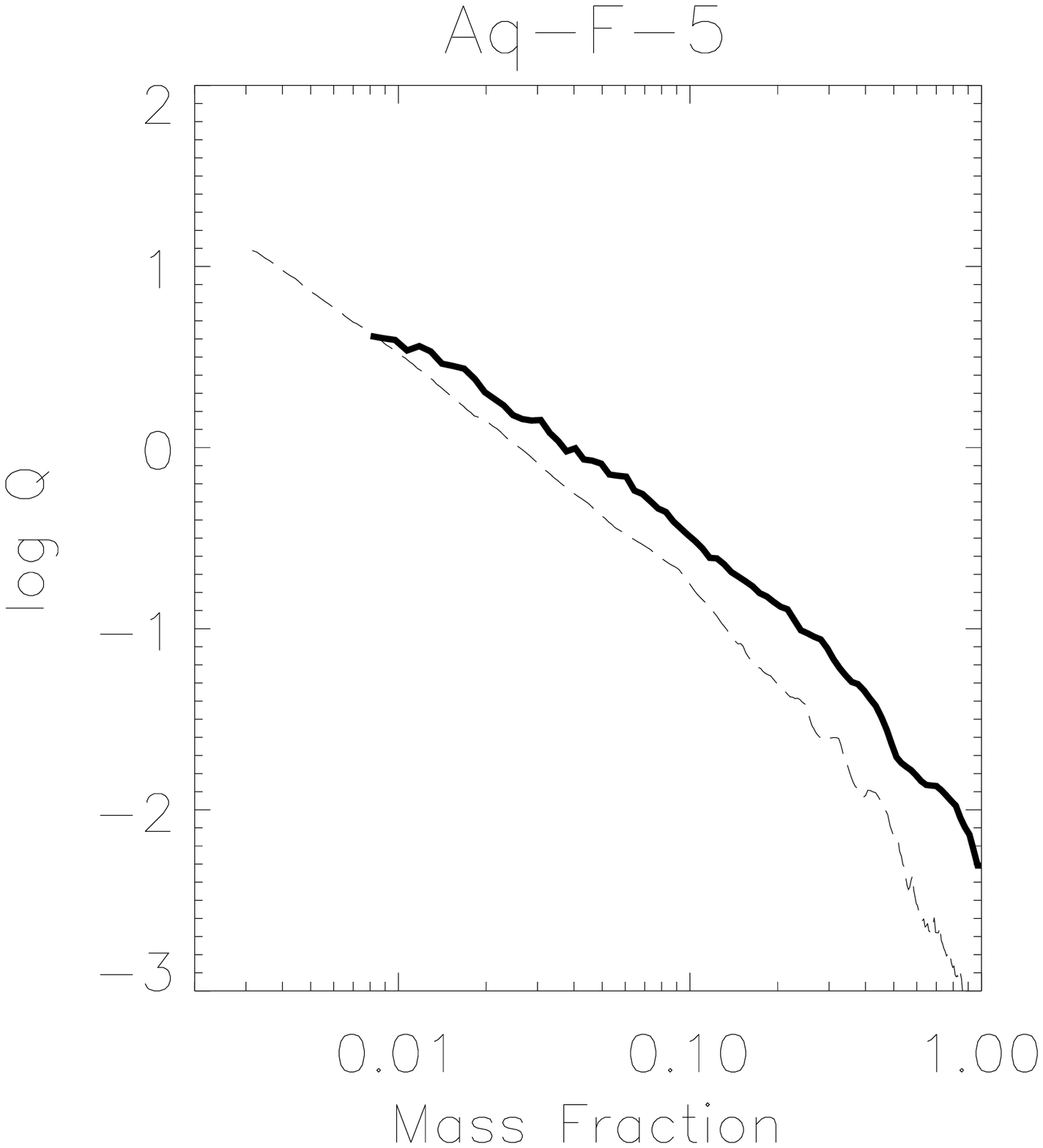}}
\hspace*{-0.2cm}
\caption{Pseudo-phase-space density profiles as a function of enclosed dark
  matter mass fraction for haloes in the SPH (solid lines) and DM (dashed
  lines) runs (Q is in units
    of $h^{2} M_{\odot} {\rm kpc^{-3}}{\rm (km \ s^{-1})^{-3}} $). The profiles are shown over the same radial range already shown
  in Fig.~\ref{bert}. We include the lower resolution versions of Aq-E: Aq-E-6 (short dashed lines) and Aq-E-7 (dotted lines).}
\label{bertmasas}
\end{figure*}

It could be possible that the contraction of the dark matter proceeds
adiabatically and that the detected increase in entropy at each radius merely
reflects the contraction itself.  To investigate this point, we plot $Q$ as a
function of enclosed dark matter mass (relative to the total dark matter
mass within $r_{200}$) rather than as a function of $r$. If the contraction
were purely adiabatic, the SPH haloes should show identical behaviour to
their DM counterparts when $Q(r)$ is plotted against $M(r)/M_{200}$ in this
way. As shown in Fig. ~\ref{bertmasas}, this is not the case in any of our haloes.

 The lower numerical resolution
versions of Aq-E  do not show significant differences in the pseudo-phase-space density profiles with respect to  Aq-E-5 as can be seen from Fig. ~\ref{bert}, because the numerical effects in the density and velocity profiles compensate each other. However, if this relation is plotted as a function of the
enclosed mass as seen in Fig. ~\ref{bertmasas} the differences are slightly larger.

\section{Shapes}

Haloes formed by hierarchical clustering of dark matter in CDM scenarios are
typically triaxial (e.g. Barnes \& Efstasthiou 1987; Frenk et al 1988; Jing \&
Suto 2002) with the inner regions being approximately prolate (e.g. Hayashi et
al. 2007). However, when baryons condense to make galaxies in these inner
regions they become rounder and approximately oblate (Dubinski 1994; Tissera
\& Dom\'{\i}nguez-Tenreiro 1998; Kazantzidis et la. 2004).  This latter result
seems in better agreement with a variety of observational shape estimates for
galaxy haloes (e.g., Sackett \& Sparke 1990; Kuijken \& Tremaine 1994;
Koopmans et al. 1998; Helmi 2004; Weijmans et al. 2008).  We measure the shapes
of our SPH and DM haloes following the Dubinski \& Carlberg (1991) method,
based on the eigenvalues of the moment of inertia tensor, $I$.  For each bin
of N particles we computed ellipsoidal radii
\begin{equation}
r=[ x^2 +\frac{y^2}{(b/a)^2} + \frac{z^2}{(c/a)^2}]^{1/2}
\end{equation}
The semi-axes of the triaxial ellipsoids ($a > b >c$) are calculated
iteratively as
\begin{equation}
\frac{b}{a}=(\frac{I_{22}}{I_{11}})^{1/2},
\frac{c}{a}=(\frac{I_{33}}{I_{11}})^{1/2}
\end{equation}
where $I_{11}>I_{22}>I_{33}$ are the eigenvalues of the tensor of inertia
($I_{jk}=\sum_i x^j_i x^k_i/r_i^2)$. In order to determine $a$, $b$ and $c$, an
iterative cycle is set up, starting with $b/a=c/a=1$.

\begin{figure*}
\resizebox{6cm}{!}{\includegraphics{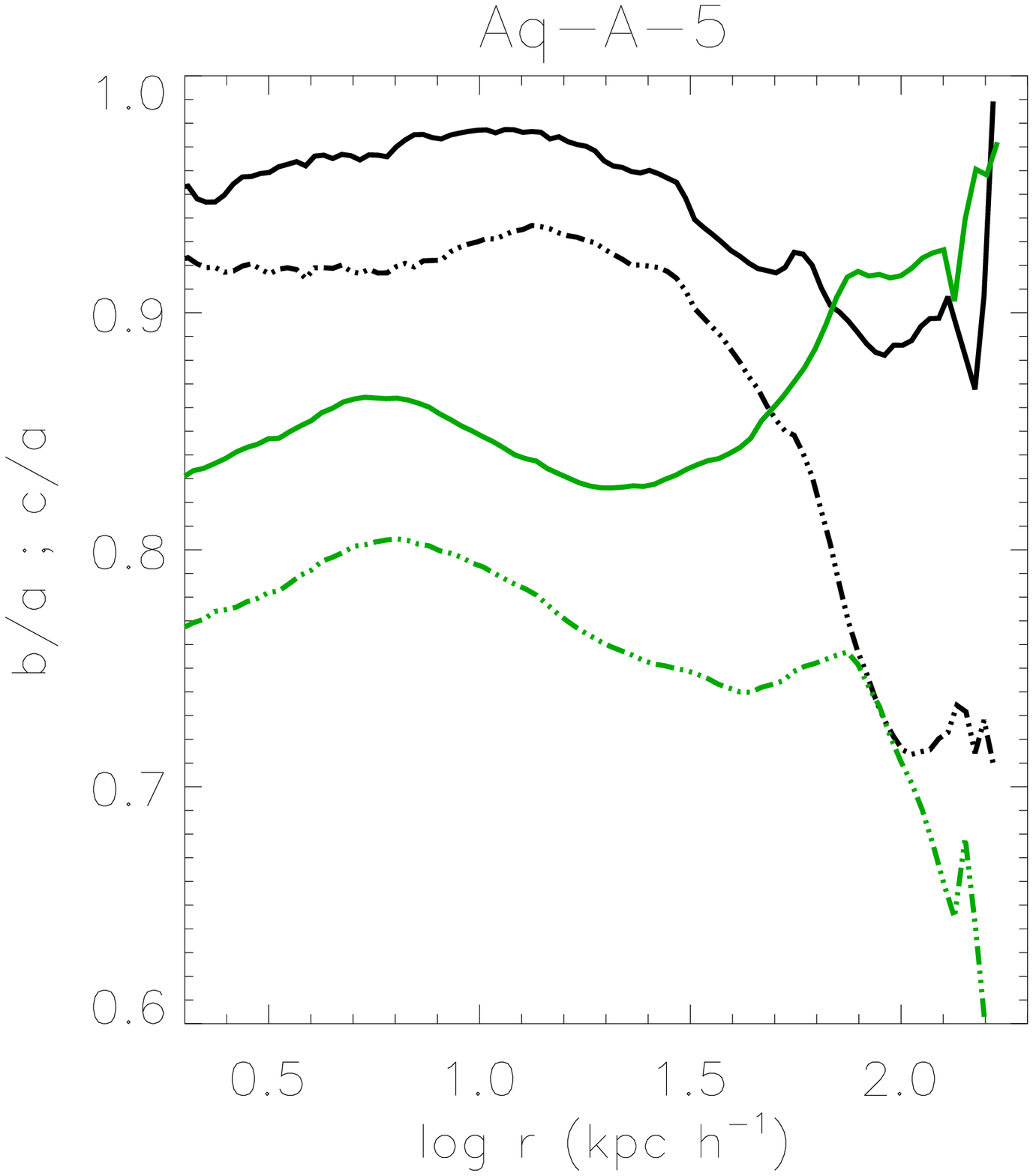}}
\resizebox{6cm}{!}{\includegraphics{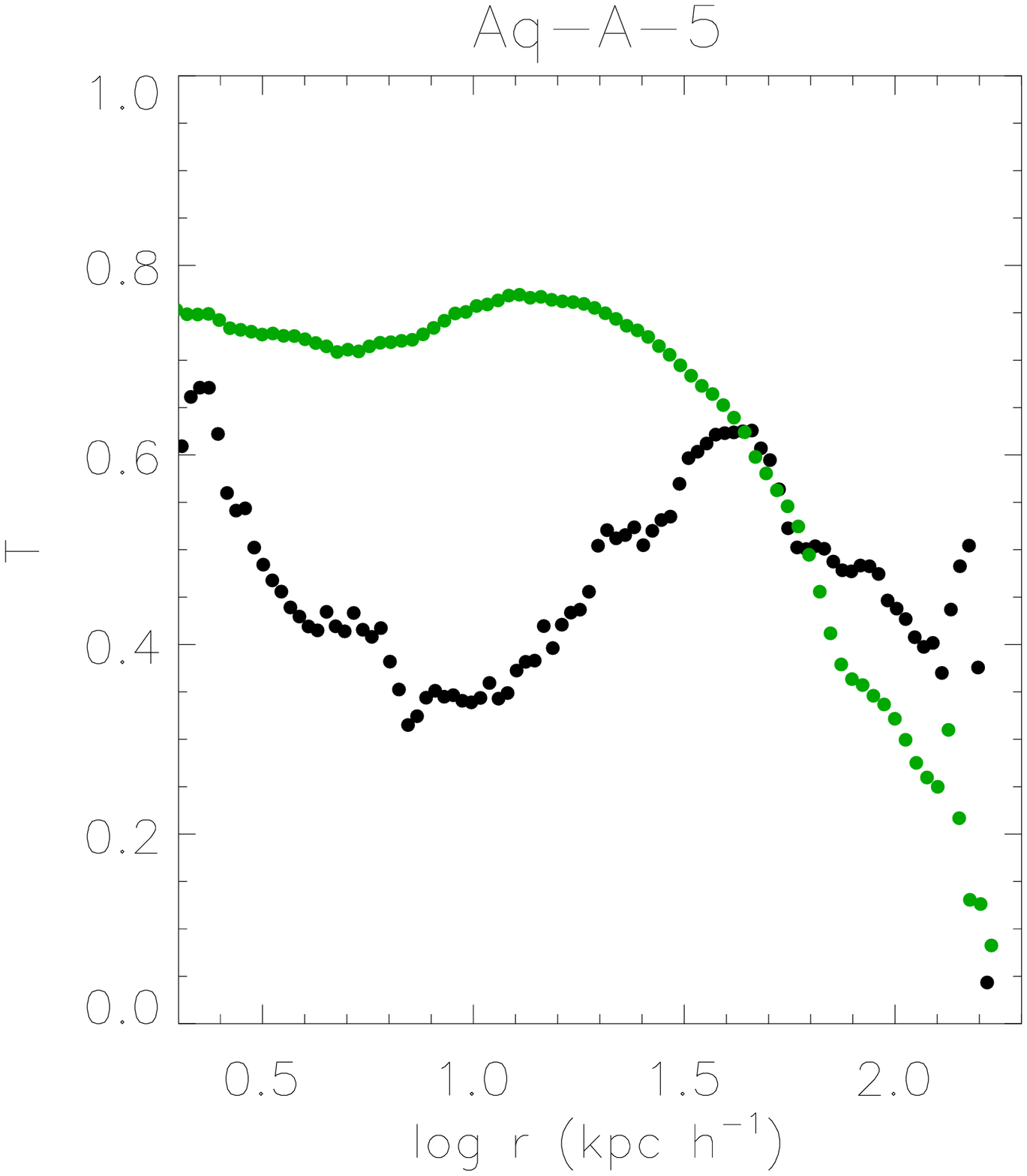}}\\
\resizebox{6cm}{!}{\includegraphics{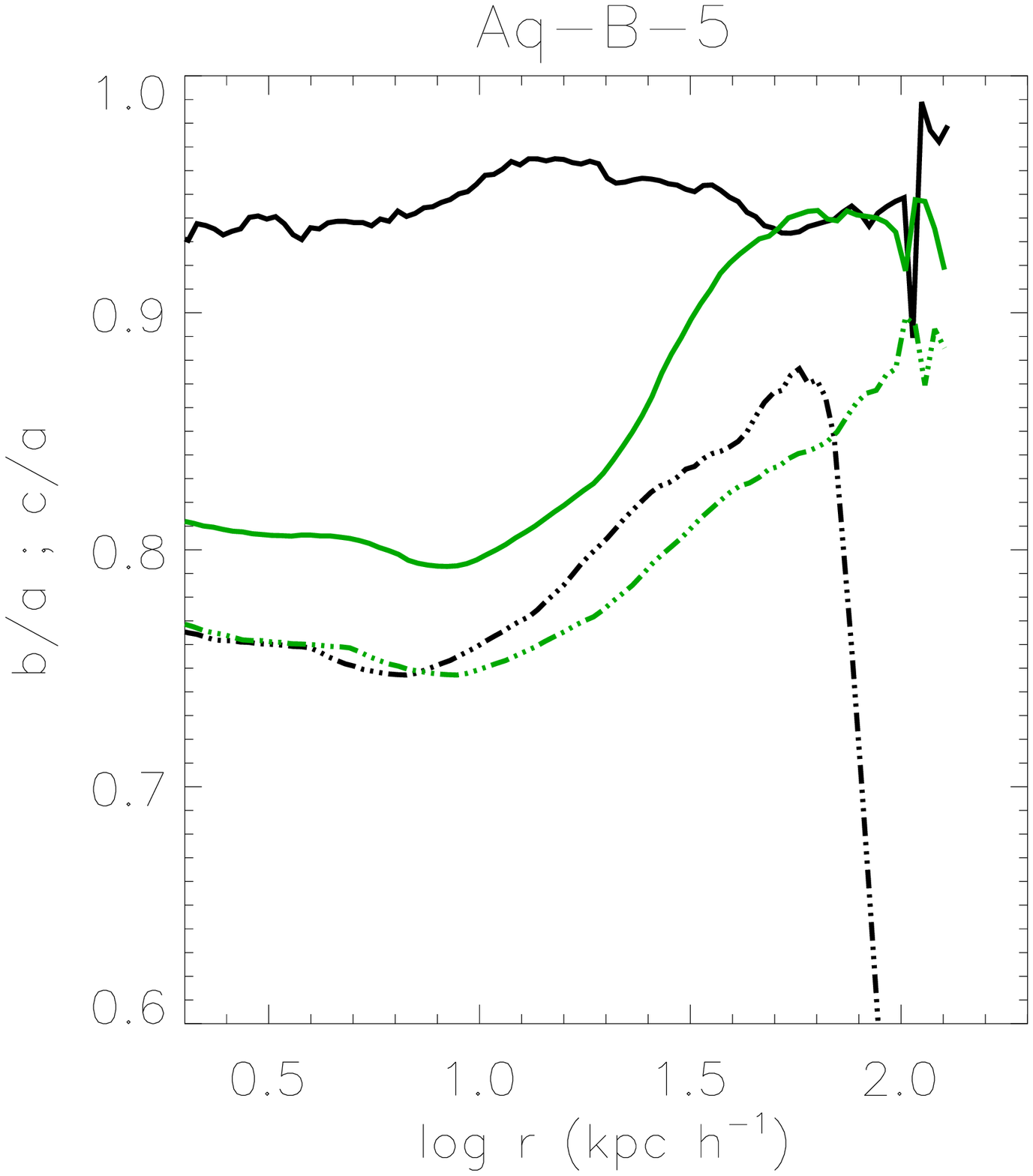}}
\resizebox{6cm}{!}{\includegraphics{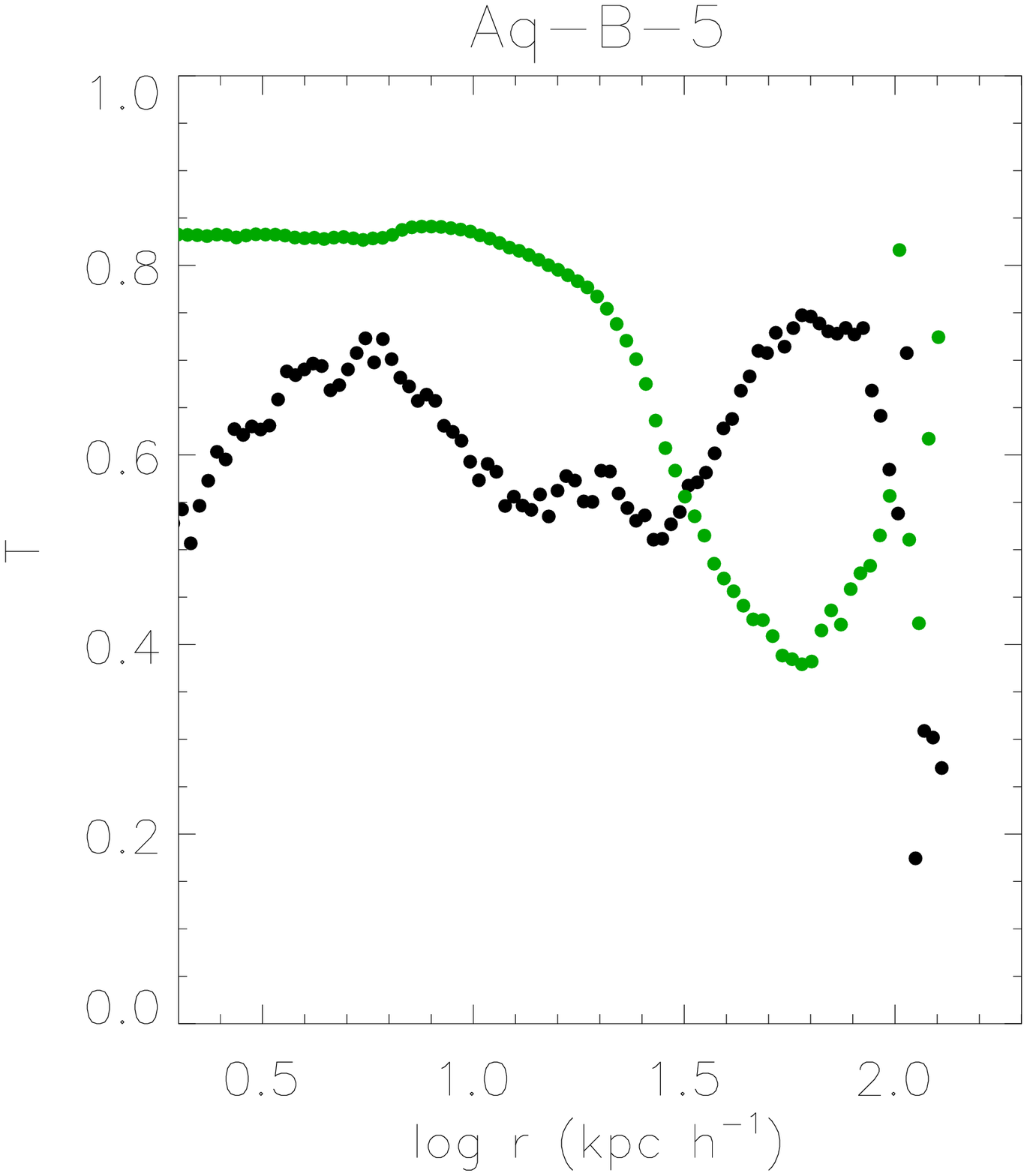}}\\
\resizebox{6cm}{!}{\includegraphics{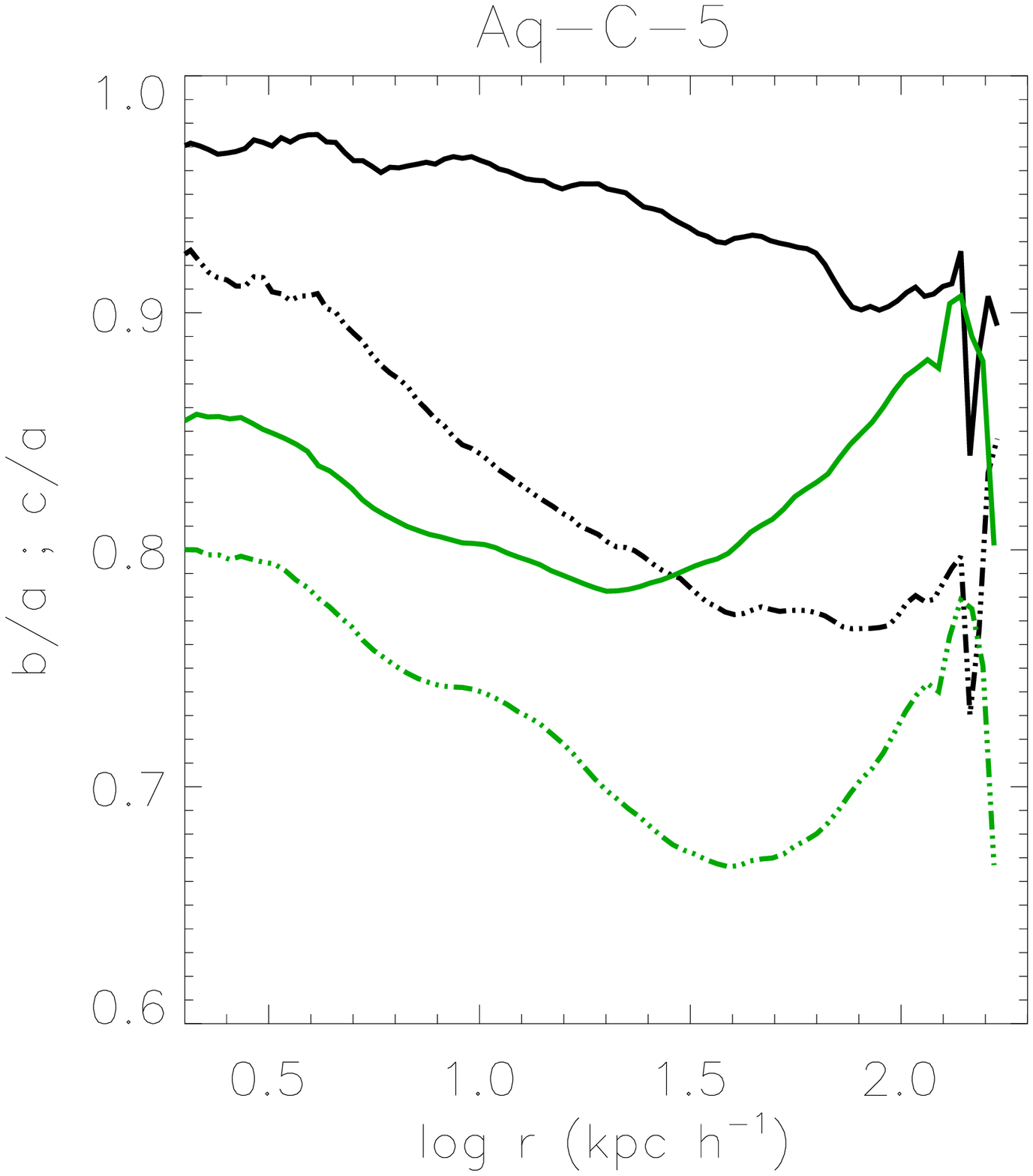}}
\resizebox{6cm}{!}{\includegraphics{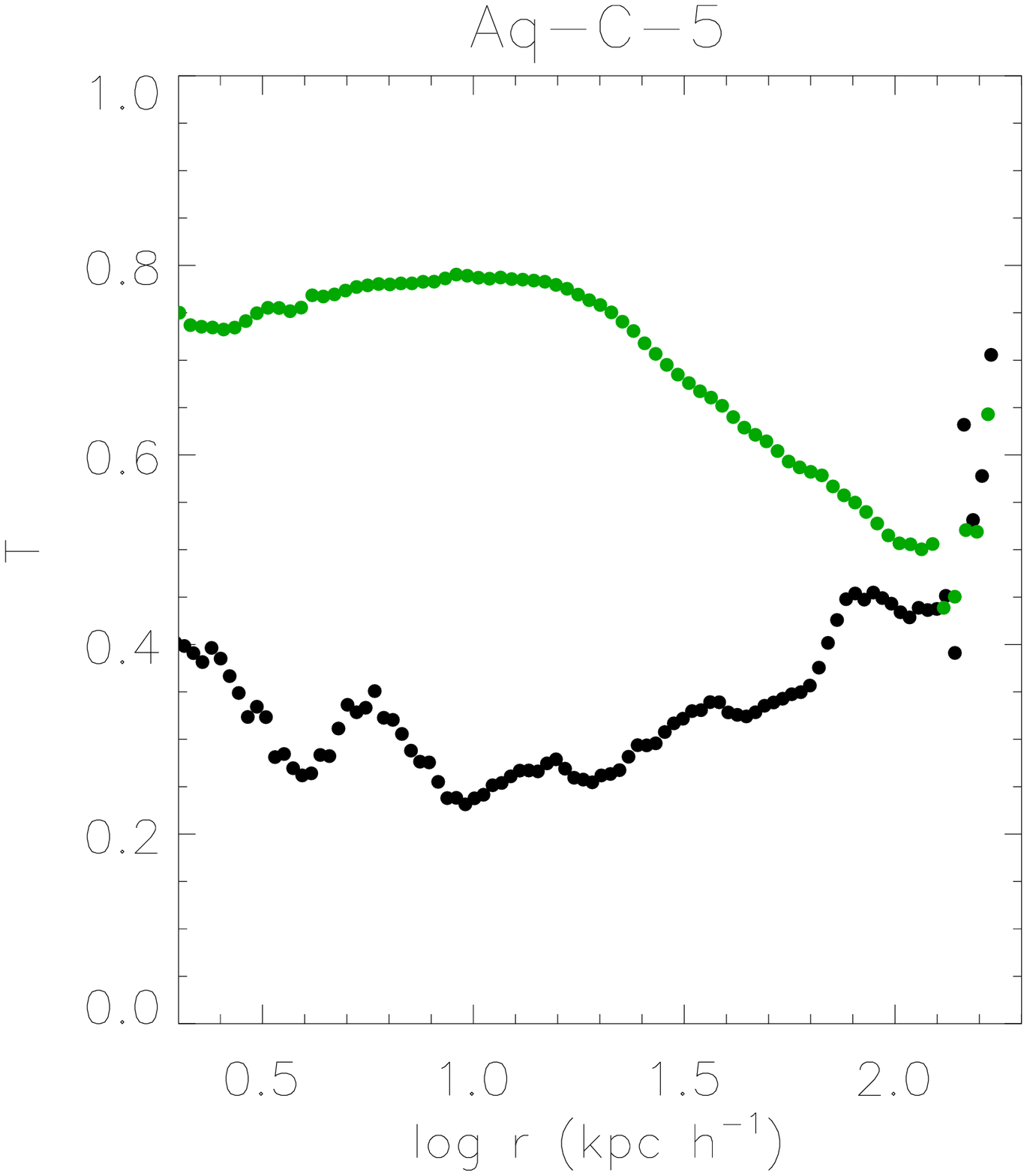}}\\
\hspace*{-0.2cm}
\caption{Axis ratios $b/a$ (solid lines) and $c/a$ (dotted-dashed lines) for
  haloes in our SPH (black) and DM (green) runs. In the right panels, the
  triaxiality parameter $T$ is also shown; $T=0$ for oblate systems while
  $T=1$ for prolate ones.  We include the lower resolution versions of Aq-E: Aq-E-6 (blue) and Aq-E-7 (red).
}
\label{sshi}
\end{figure*}

In Fig. ~\ref{sshi} we show the axis ratios and the triaxiality parameter $T$
defined as $ T=(a^2-b^2)/(a^2-c^2)$ as a function of radius. Purely prolate
objects have $T=1$, while purely oblate ones have $T=0$.  In agreement with
previous work, we find a general trend for the SPH runs to be more oblate than
their DM counterparts.  Although in all haloes both axis ratios increase at
most radii, each halo shows its particularities.  The largest change in shape
occurs in Aq-C-5 which becomes oblate with $ T < 0.4$ over the whole analysed
range.
The smallest change in shape is found in Aq-F-5 which only weakly modifies its
overall prolate shape, although the axis ratios do increase noticeably.  The
only galaxy without any disc component at all inhabits this halo.  Halo Aq-E-5
shows a strong dependence of $T$ on radius in the DM run, varying from
$T\approx 0.6$ to $T\approx 0.1$. This behaviour is preserved when baryons are
included although the axis ratios do increase slightly.  Interestingly, the
galaxy in this halo is the only one with a strongly rotating bulge.
Initially triaxial shapes become much more nearly oblate in the
baryon dominated regions for Aq-A-5, Aq-B-5 and Aq-D-5. These
haloes have galaxies with extended, diffuse surviving discs.
 For halo Aq-E we have included the analysis of the shapes for  the two lower resolution
versions in Fig. ~\ref{sshi}. In agreement with the other parameters, the shapes tend to converge as 
numerical resolution increases.
As expected, the dark matter halo in the lowest resolution version, Aq-E-7, has 
 the noisiest  shape.

\begin{figure*}
\resizebox{6cm}{!}{\includegraphics{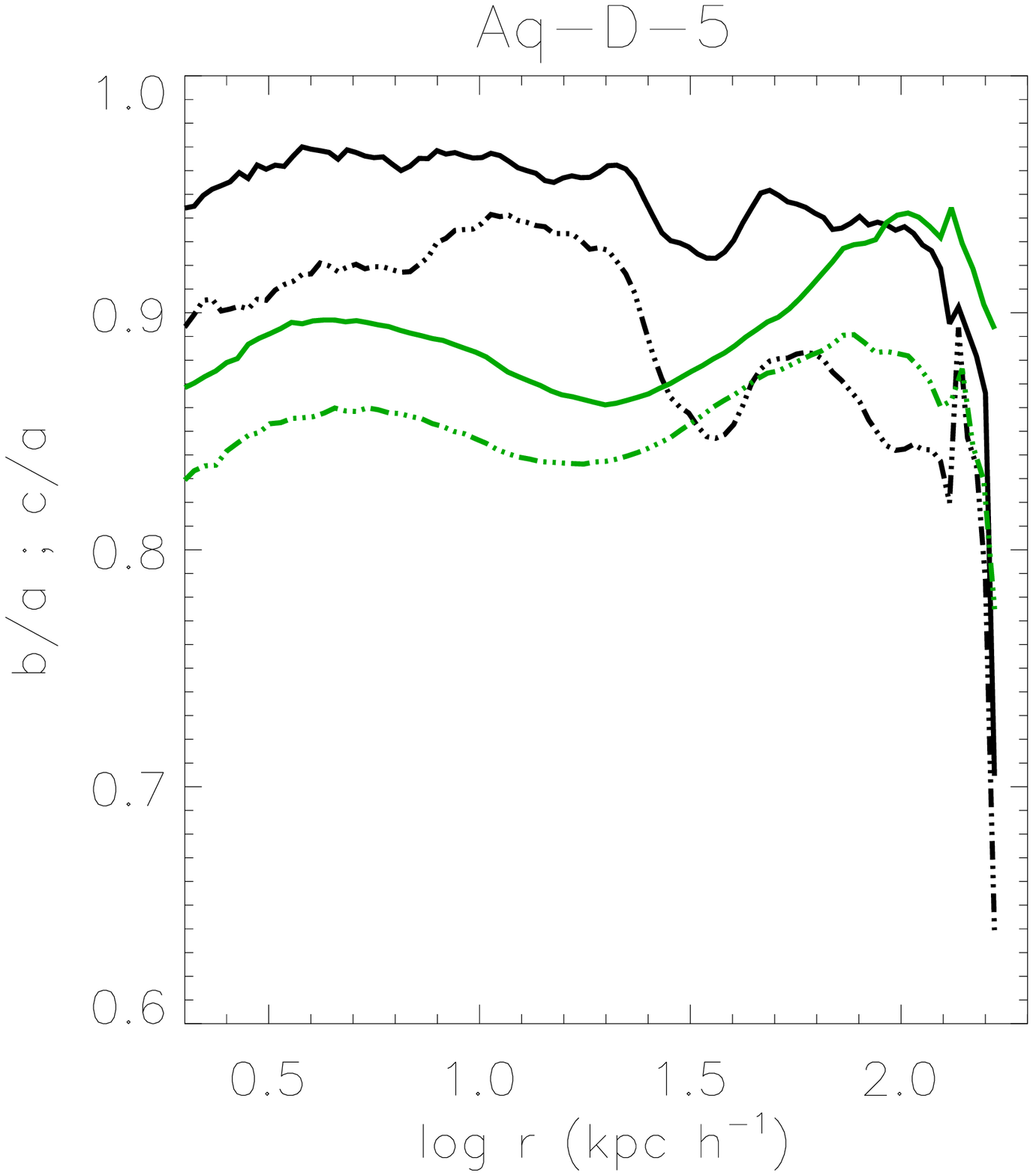}}
\resizebox{6cm}{!}{\includegraphics{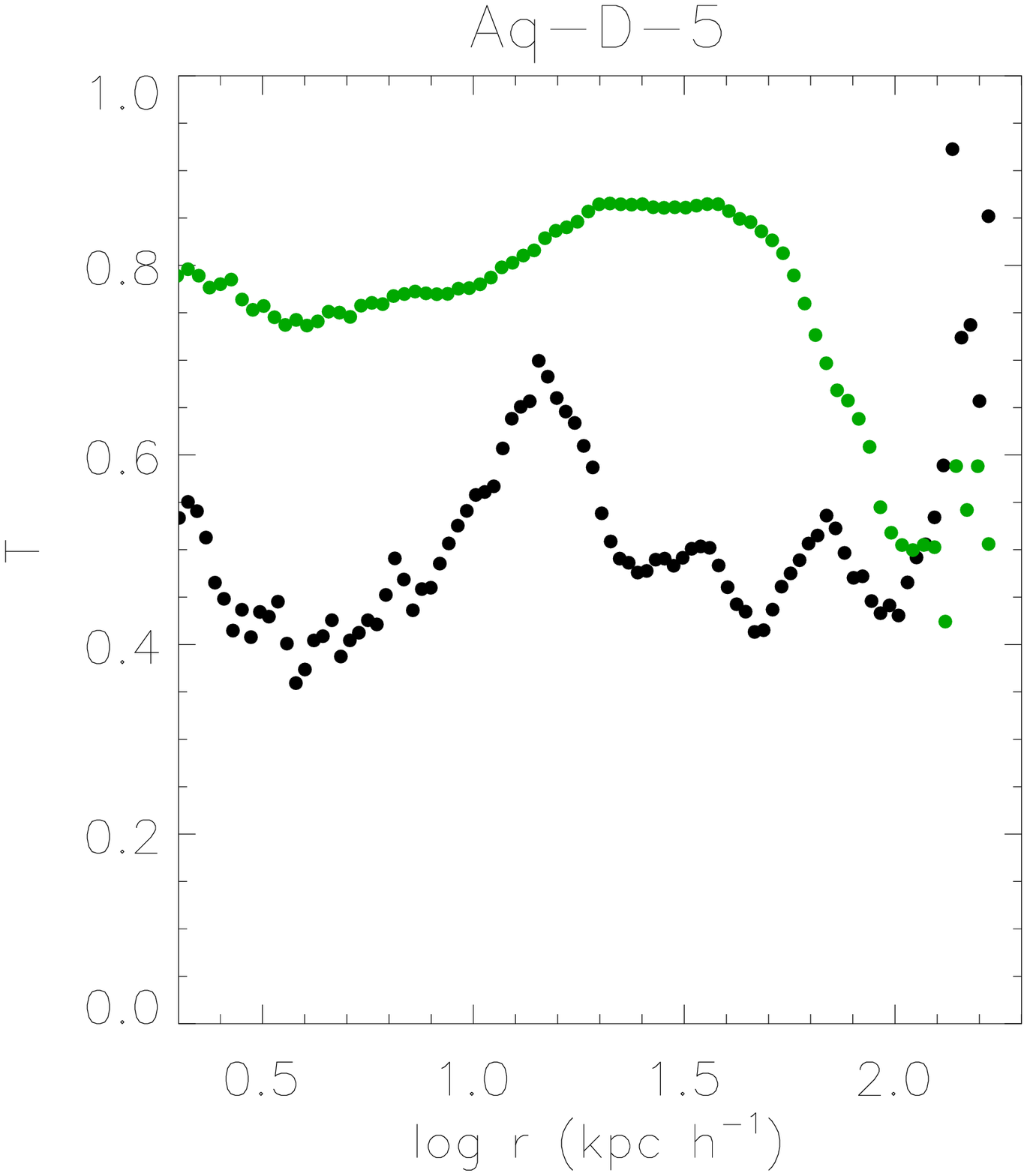}}\\
\resizebox{6cm}{!}{\includegraphics{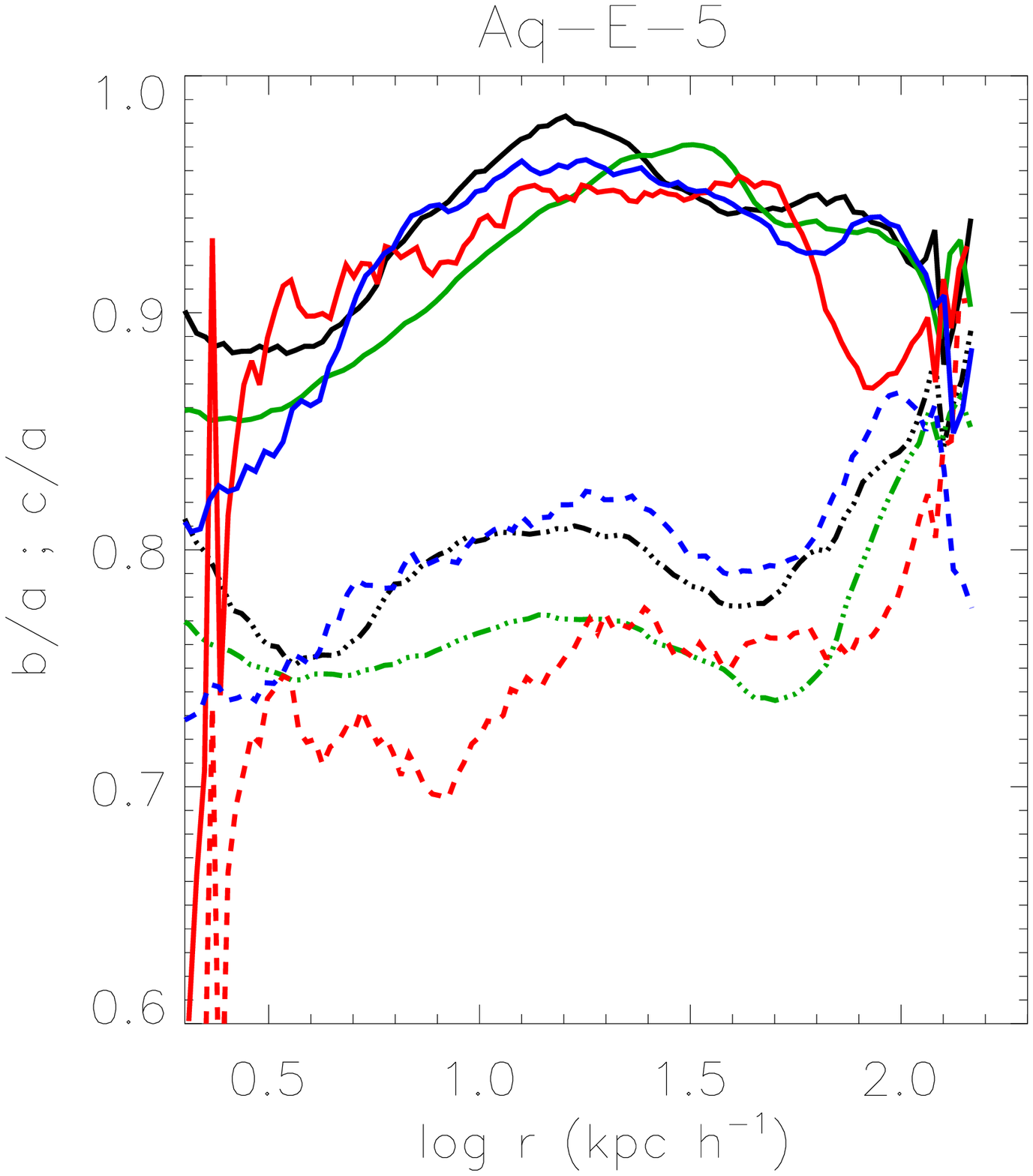}}
\resizebox{6cm}{!}{\includegraphics{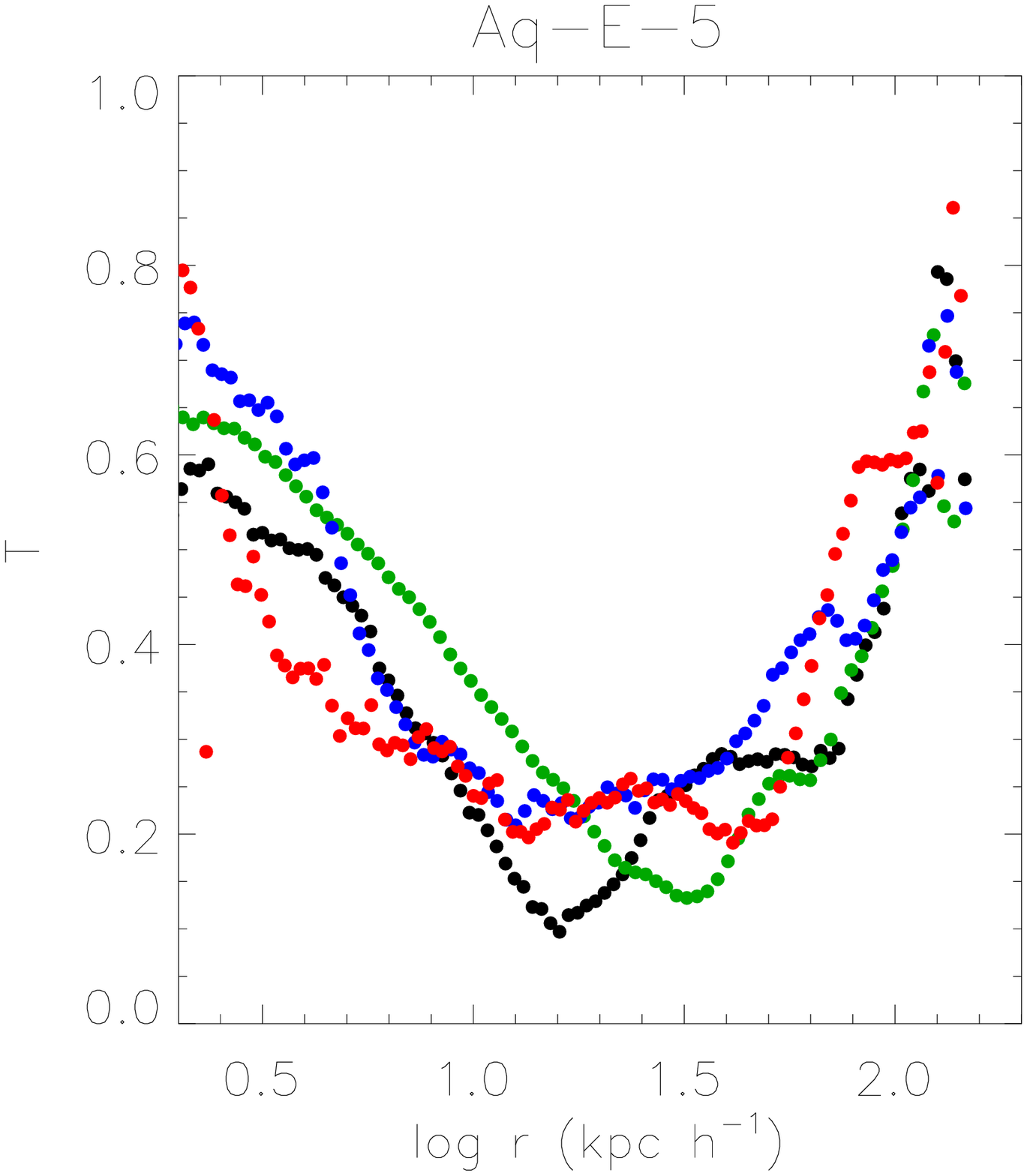}}\\
\resizebox{6cm}{!}{\includegraphics{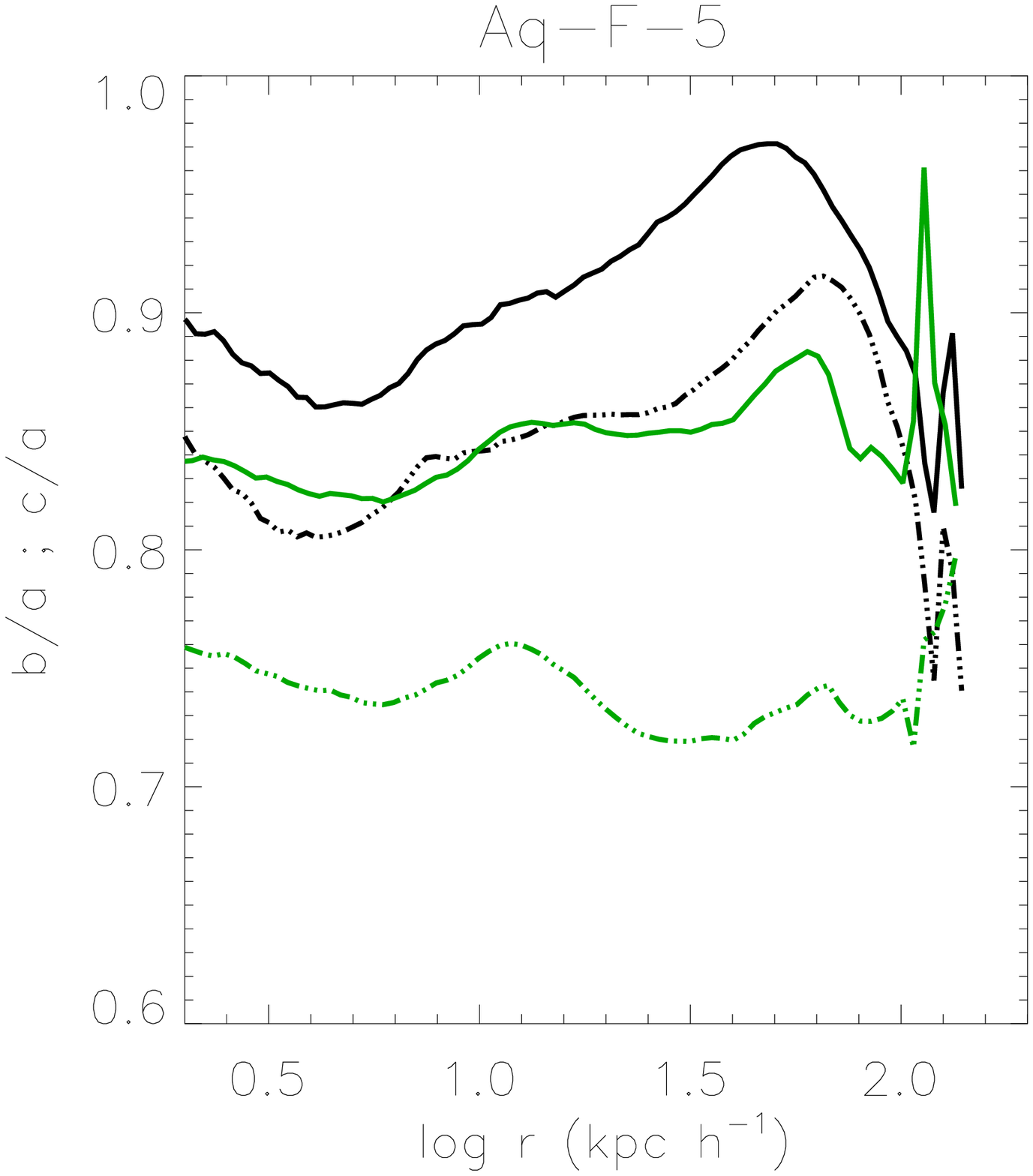}}
\resizebox{6cm}{!}{\includegraphics{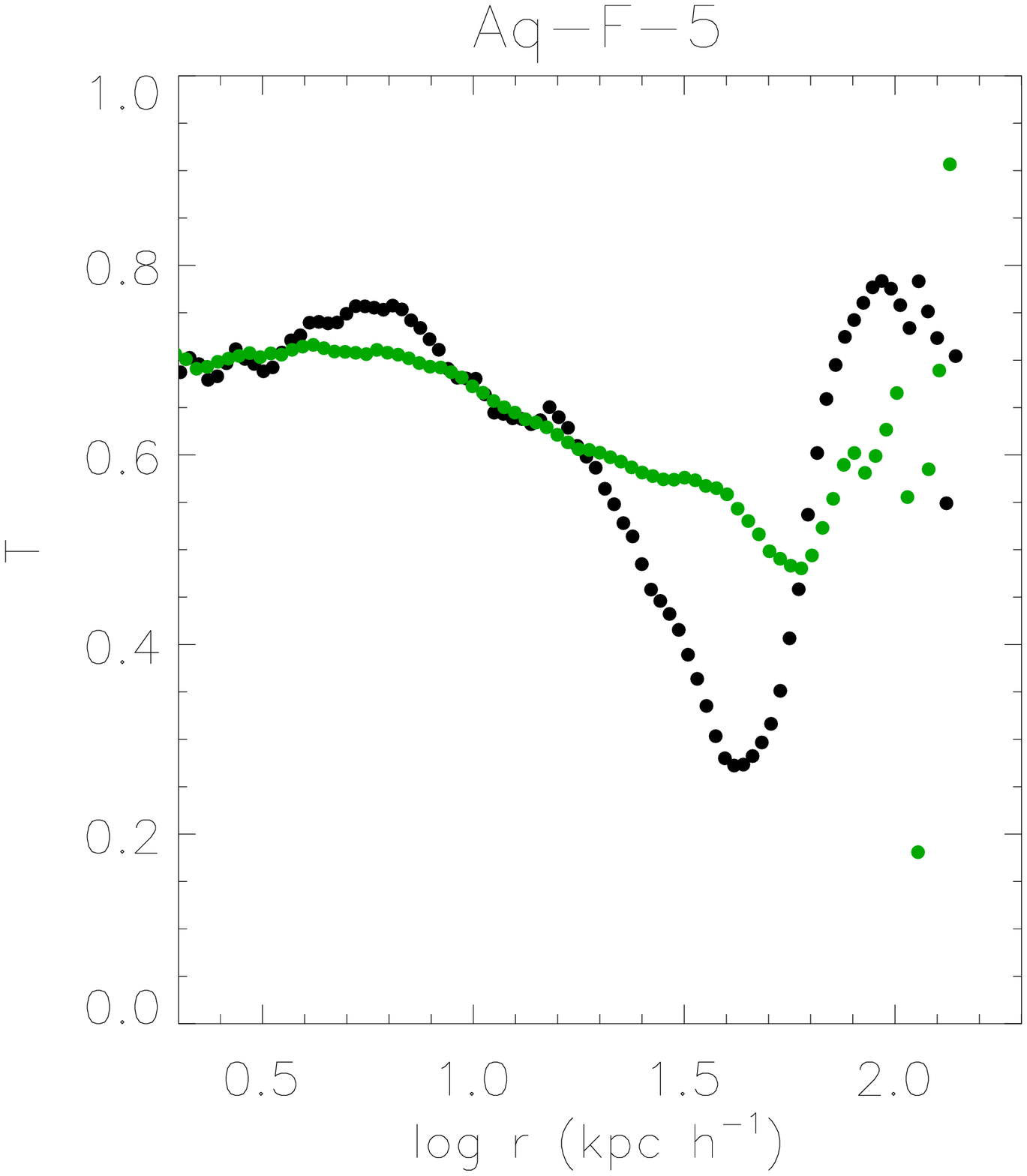}}\\
\hspace*{-0.2cm}
\centerline{{\bf Figure 8} --{\it continued}}
\end{figure*}

\section{Circular velocity and the adiabatic contraction models}

The adiabatic contraction (AC) hypothesis has been used extensively to model
the effects that baryon condensation produces on dark matter haloes. As
discussed in the Introduction, there is now a consensus that the simplest
implementation of this hypothesis overestimates the dark matter concentration
in the central region and makes it very difficult to reconcile observations
with CDM models.  This discrepancy may well reflect issues other than dark
matter compression by baryonic gravity, but adopting an inappropriate AC
hypothesis can certainly exacerbate the problem.  In this section, we will
apply the standard model of  B86, the modified
algorithm of  G04 and the statistical motivated relation
of  A09, comparing their predictions to our simulated dark
matter circular velocity profiles.

\begin{figure*}
\resizebox{7cm}{!}{\includegraphics{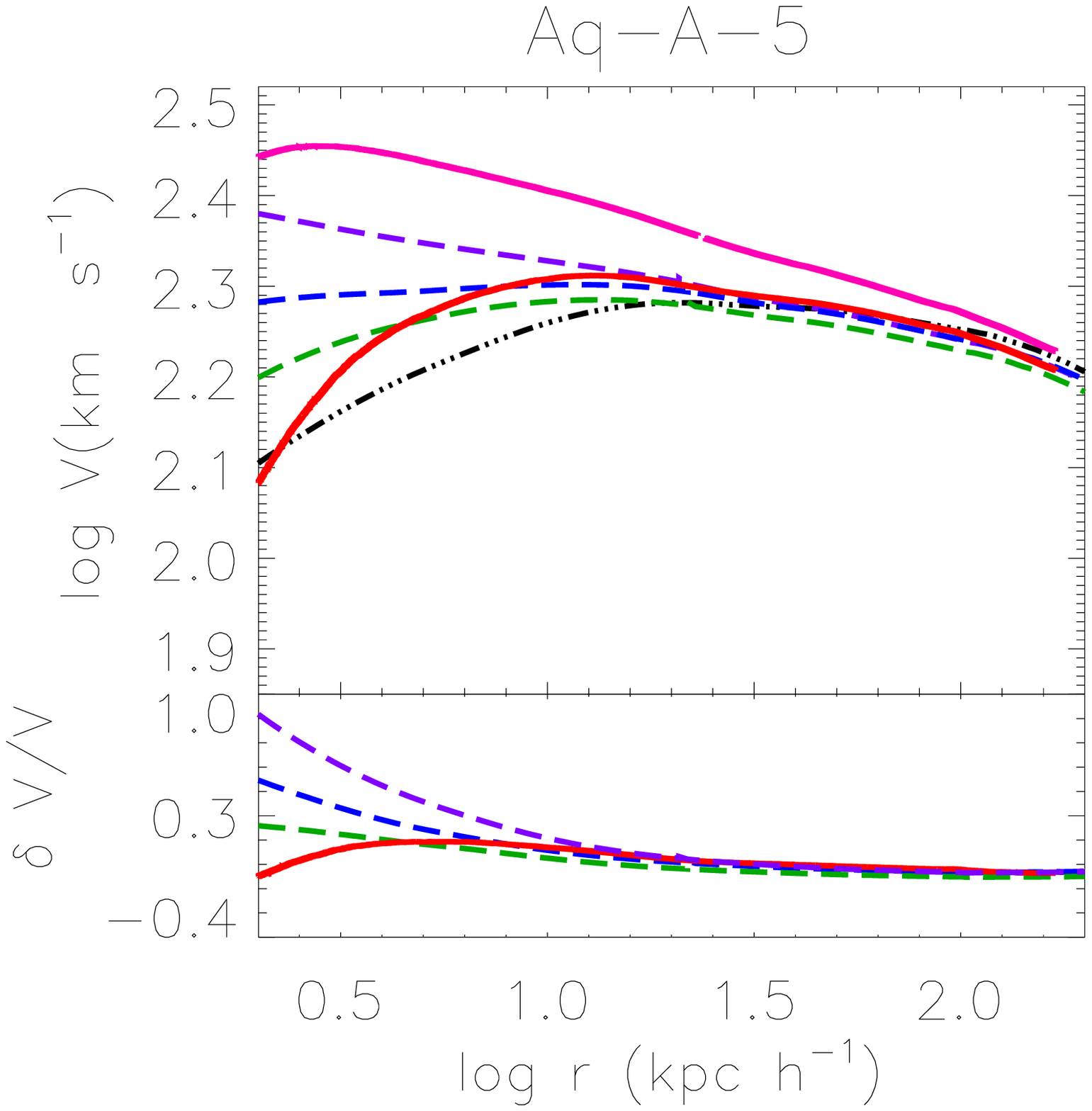}}
\resizebox{7cm}{!}{\includegraphics{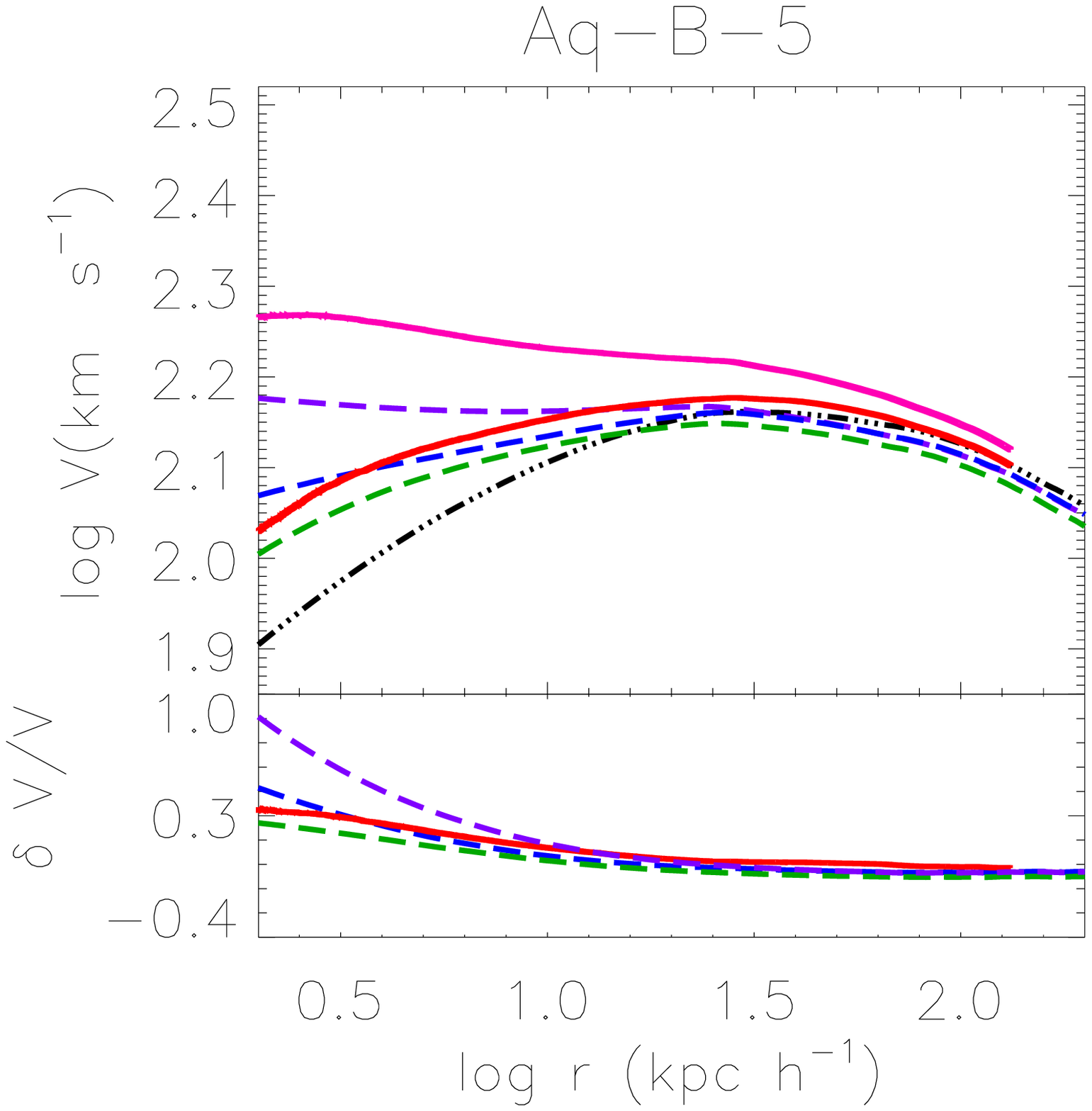}}\\
\resizebox{7cm}{!}{\includegraphics{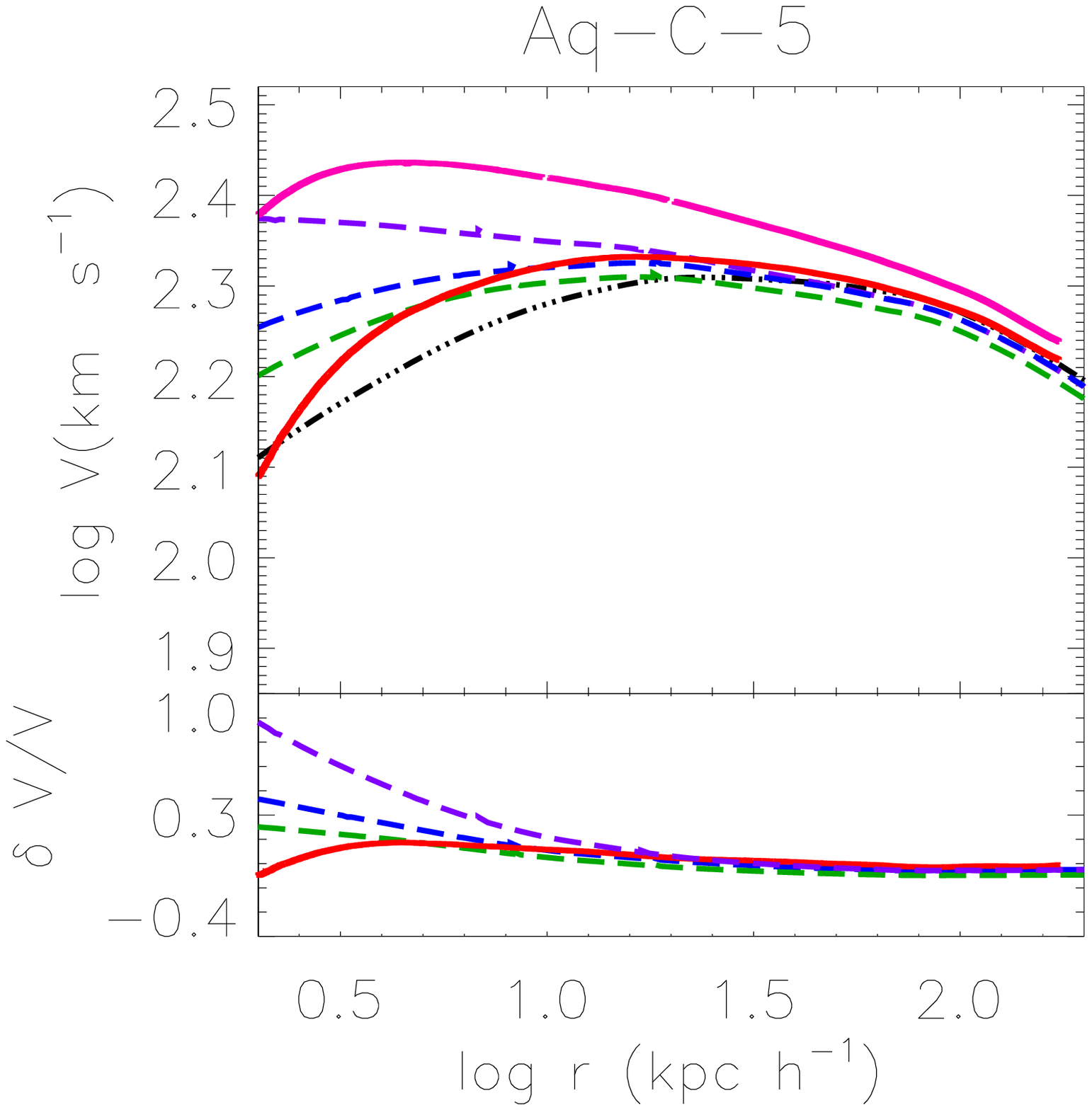}}
\resizebox{7cm}{!}{\includegraphics{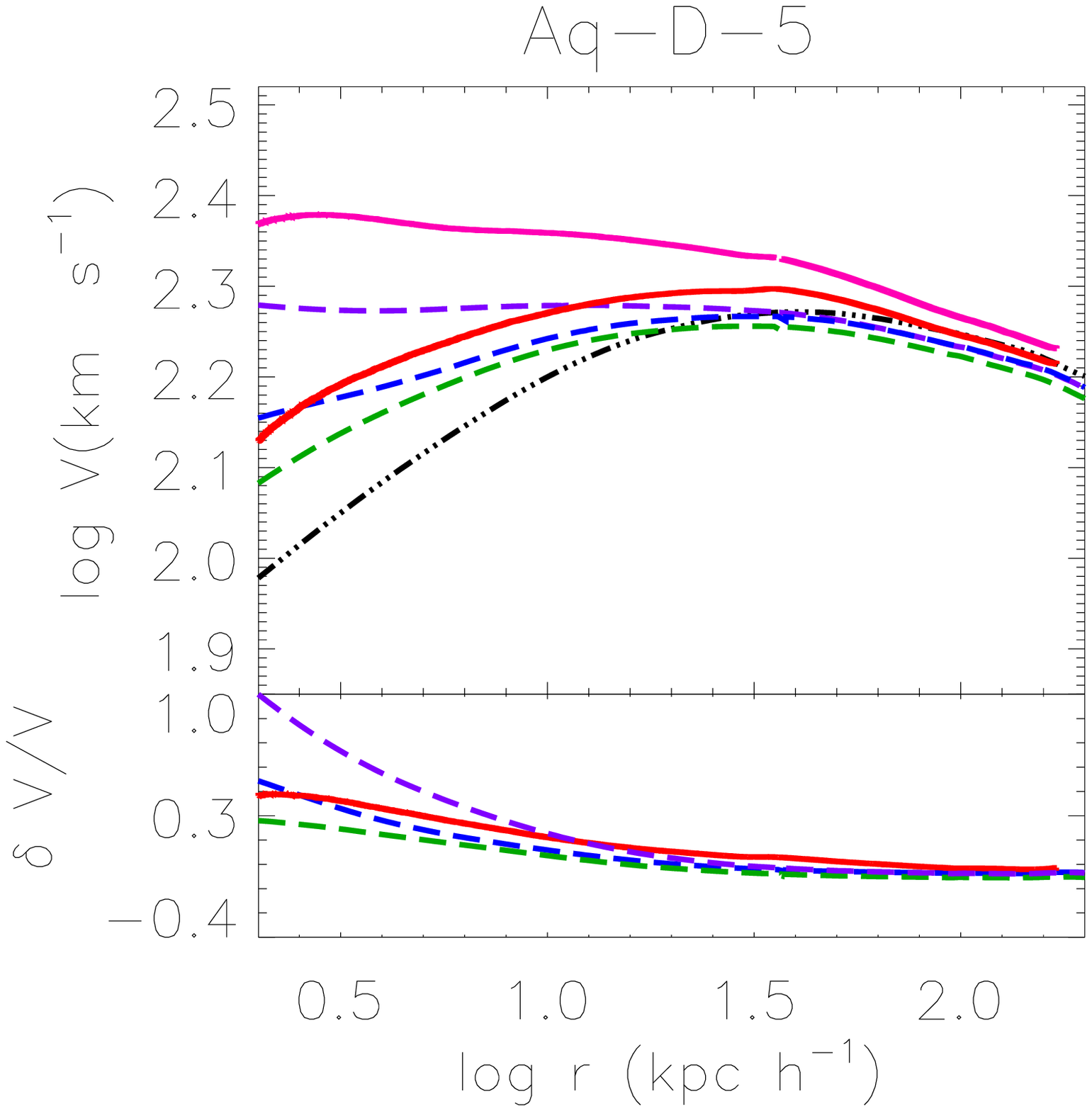}}\\
\resizebox{7cm}{!}{\includegraphics{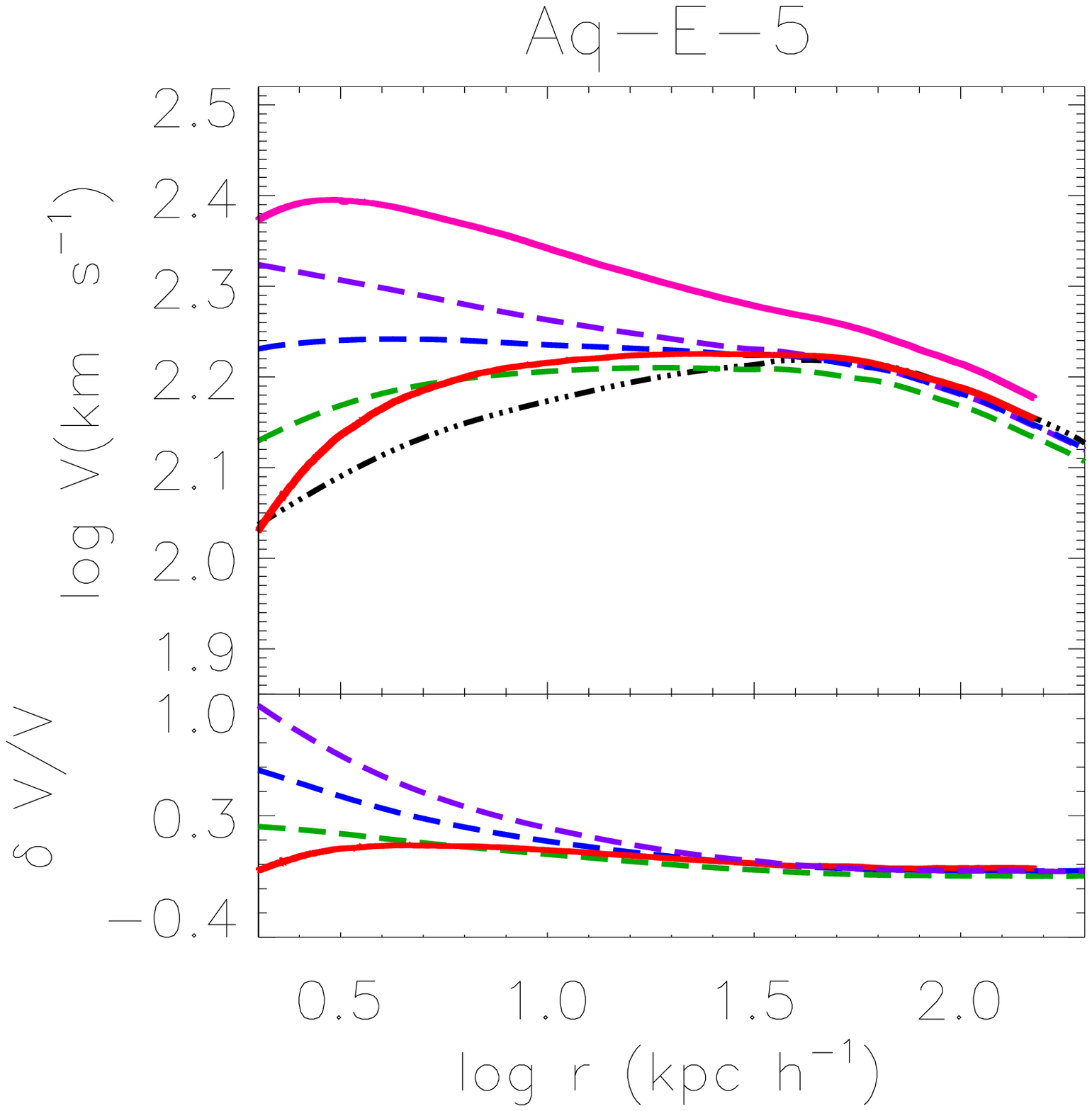}}
\resizebox{7cm}{!}{\includegraphics{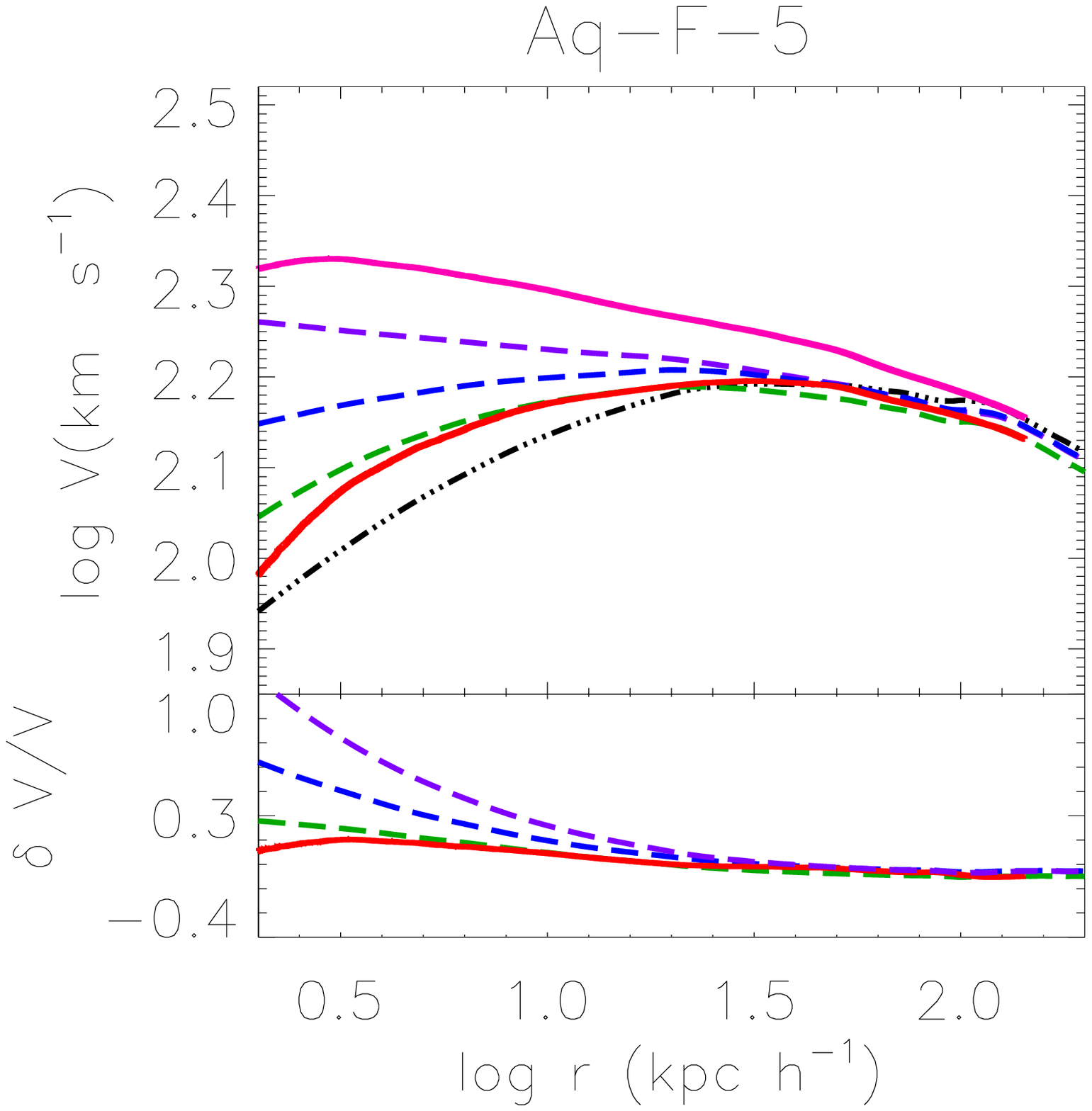}}
\hspace*{-0.2cm}
\caption{Dark halo circular velocity curves predicted by applying the AC
  models of Blumenthal et al. (1986, violet dashed lines), Gnedin et
  al. (2004; blue dashed lines) and Abadi et al. (2009, green dashed lines) to
  the circular velocity curves of our DM simulations assuming the central
  baryon distribution found in our SPH simulations.  We also show the directly
  measured circular velocity curves in the SPH (red solid lines) and DM (black
  dotted-dashed lines) simulations. Thus if the AC models worked perfectly
  their curves would match the red solid lines. In the lower panels we show
  residuals of the three model predictions from the original DM curves, and
  compare them with the residuals for the actual curves in the SPH
  simulations. For this comparison, the curves for the AC models and DM
  simulations have been scaled down by 8\% to account for the baryon fraction
  in the SPH simulations. We also show circular velocity curves for the total
  mass (DM $+$ baryons) in the SPH runs (magenta lines).}
\label{aida}
\end{figure*}

The differences between these various recipes can be seen in Fig. ~\ref{aida}.
These plots show the directly measured dark matter circular velocities ($V_c =
[G M_{dm}(r)/r]^{1/2}$ where $M_{dm}(r)$ is the dark matter mass enclosed within $r$) in our SPH (red lines) and DM (black lines)
simulations and compares them to the curves predicted by modifying the result
for the DM run according to each of the three recipes, assuming the baryon
distribution found in the SPH run. The DM profiles have been reduced by 8\% to
account for the global baryonic fraction of 0.16 adopted in the SPH
simulations.  In the narrow panels at the bottom of each plot we show the
residuals from the pure DM result for each of the models (the three dashed
curves) and for the dark matter distribution actually found in the SPH
simulation (the solid red curve).  As expected, the AC model of B86
overpredicts the amount of dark matter in the central regions by up to $50\%$
or so.  The G04 and A09 models provide a better representation with A09 being
closer to the ``true'' result in most cases.  However, for two haloes (Aq-B-5
and Aq-D-5), the A09 model underestimates the dark matter mass by several tens
of percent at some radii.  This is expected, since the Abadi et al (2009)
prescription is based on a large sample of dark matter haloes, and for
individual haloes one can expect overestimates and underestimates in roughly
equal numbers, just as we find.   Note that for Aq-B-5 and Aq-D-5 the G04 prescription actually gives the best description at all but the smallest radii.

In general terms, the dark matter mass in the innermost regions is usually
overpredicted and, more importantly, the shape of the circular velocity is
often poorly matched by all of these models. 
 At least in three of our haloes (Aq-A-5, Aq-C-5 and Aq-E-5), the  velocity curves rise 
 in the central region more gently  than  predicted by the models (in agreement with Fig.1).

\section{Conclusions}

We have analysed the dark matter distributions in six galaxy-sized haloes
belonging to the Aquarius Project, comparing results from the original dark
matter only simulations to those from re-simulations including baryonic
processes (S09).  In agreement with previous work, we find that dark matter
haloes become more concentrated when baryons condense at their centres, but
that the characteristics of the contraction do not correlate in a simple way
with the total amount of baryons.  Our main contribution here is to analyse
similar mass haloes with a variety of formation histories using high
resolution simulations with and without baryons.  Our results show that the
response of haloes to the presence of baryons is sensitive to the details of
halo assembly.  This set of six galaxy-sized haloes provides the opportunity to
study which properties can be considered common to such haloes and which
depend significantly on their particular formation history.

Our findings can be summarized as follows:
\begin{itemize}

\item In the regions dominated by baryons, haloes become significantly more
  concentrated than their dark matter only counterparts. The level of
  concentration varies significantly from object to object, however, it is
  not simply related to the total baryonic mass accumulated in the central
  galaxy.  For $r < r_{-2}$, the dark matter density profiles of many of our
  simulations are nearly isothermal except, possibly, very close to the
  centre.

\item The velocity dispersion structure is modified in all haloes, with
  velocity dispersion increasing monotonically to small radii in all cases.
  The temperature inversion of NFW profiles is no longer present once the
  galaxy has formed.  In some systems, the tangential dispersion increases
  more than the radial dispersion (but not all), causing them to become more nearly
  isotropic.

\item The $\beta$-$\gamma$ relation proposed by Hansen \& Moore (2006) is
  obeyed at most over a restricted radial range for some of our
  haloes. It works best in the central region of those systems where the
  isothermal behaviour extends over a relatively small range.  The departures
  from their predictions become large when the profile is isothermal or has
  constant anisotropy over an extended radial range.
 
\item Pseudo-phase-space density  no longer follows the same power law in
  radius as in Bertschinger's (1985) similarity solution once galaxy formation 
  is included. The profile differs from one halo to the next and in no halo is it consistent with the purely adiabatic contraction of the dark matter only case.

\item As in previous work, the condensation of baryons makes the central
  regions of all our haloes less aspherical and more nearly oblate, although in
  two cases the changes are small. One of these has no significant disk
  component and the other has a bulge with substantial nett rotation.

\item None of the simple adiabatic contraction models proposed in earlier work
  is able to describe how the radial density profiles of  our haloes are
  modified by baryon condensation. The scheme suggested by Abadi et al (2009; see also Pedrosa et al. (2010))
  is the most successful of those we consider, although it can significantly
  over- or underestimate the effects in individual cases. For the same haloes, 
 the central mass is  overestimated by more than $50\%$.   The circular
  velocity curves of our galaxy formation simulations are not as centrally
  peaked as predicted by such adiabatic contraction models, but are still more
  rapidly rising  than in many observed spirals.  Since none of
  our simulated galaxies is disk-dominated it is unclear whether this is a
  problem.

\end{itemize}

Our analysis show that haloes of similar virial mass respond in different ways
to baryon condensation depending on the details of their assembly history.
Consequently, a much more detailed understanding of galaxy formation is needed
before we can make reliable predictions for the detailed structure of the
dark matter haloes surrounding galaxies.

\section*{Acknowledgements}
We thank the referee of this paper for her/his useful comments that helped to improved this paper.
The simulations were carried out at the Computing Centre of the Max-Planck-Society in Garching.
This research was supported by the DFG cluster of excellence 'Origin and Structure of the Universe'. PBT thanks the Max Planck Institute for Astrophysics (Garching, Germany) for its hospitality during her visits. This work was partially funded by PROALAR 07 (DAAD-Secyt collaboration), PICT
32343 (2005) and PICT Max Planck 245 (2006) of the Ministry of Science and
Technology (Argentina).

\end{document}